\documentclass{article}
\usepackage{graphicx} 
\usepackage{setspace}
\usepackage{xcolor}
\usepackage{hyperref}
\usepackage{caption}
\usepackage{listings}
\usepackage{morefloats}
\usepackage{float}
\usepackage{fancyhdr}
\usepackage{mathtools}
\usepackage{natbib}
\usepackage{bm}
\usepackage{bbm}
\usepackage{amsfonts}
\usepackage{tabularx}
\usepackage{booktabs}
\usepackage[margin=1.25in]{geometry}
\usepackage[toc,page]{appendix}
\usepackage{comment}
\usepackage{placeins}
\usepackage{xargs}
\usepackage{makecell}

\definecolor{myblue}{RGB}{50, 0, 185}
\definecolor{harvardcrimson}{rgb}{0.60, 0.0, 0.04}
\definecolor{myred}{rgb}{200, 0.0, 0.25}
\definecolor{Odkaz}{rgb}{0, 0, 0.5}

\definecolor{upennblue}{RGB}{75,156,211}
\definecolor{upennred}{RGB}{153,0,0}

\newcommand{\E}[1]{\text{E}\left[#1\right]}
\newcommand{\Eind}[2]{\text{E}_{#1}\left[#2\right]}

\newcommand{\nn}{\mathcal{N}_{\bm{\rho}}}
\newcommand{\nnt}{\mathcal{N}_{\bar{\bm{\rho}}}}
\newcommand{\lr}{\alpha^{\text{learn}}}
\newcommand{\brho}{\bm{\rho}}
\newcommand{\Xst}{\mathbf{X}^{\text{state}}}
\newcommand{\Xse}{\mathbf{X}^{\text{seq}, T}}
\newcommand{\Xstagg}{\mathbf{X}^{\text{agg. state}}}

\hypersetup{colorlinks=true, urlcolor=upennblue, citecolor=upennblue, linkcolor = upennred}
\hypersetup{
 pdftitle={Deep Learning in the Sequence Space},
 pdfauthor={Azinovic-Yang, Marlon and Zemlicka, Jan}
}
\title{Deep Learning in the Sequence Space\thanks{This paper has greatly benefited from numerous discussions with Felix Kubler, Jonathan Payne, Simon Scheidegger and Yucheng Yang. We are also grateful for the comments of participants at seminars at the University of Zurich, UNC Chapel Hill, Princeton University, City University of New York, and the 2025 Torino Conference on Machine Learning for Economics and Finance as well as for helpful discussions with Jaroslav Borovi\v{c}ka, Anusha Chari, Jaden Chen, Jeppe Druedahl, Lutz Hendricks, William Jungerman, Yasutaka Koike-Mori, Lilia Maliar, Stanislav Rabinovich, Jacob R{\o}pke, and Can Tian. Žemlička gratefully acknowledges support from the Swiss National Science Foundation and the kind hospitality of the Department of Economics at Princeton University.}}

\author{Marlon Azinovic-Yang\footnote{Email: marlonay@unc.edu} \\ University of North Carolina at Chapel Hill \and Jan Žemlička\footnote{Email: jan.zemlicka@df.uzh.ch}\\
University of Zurich and Swiss Finance Institute}

\date{First version: September 9, 2025\\
This version: March 13, 2026}

\begin{document}

\maketitle

\begin{abstract}
\singlespacing
We develop a deep learning algorithm for constructing globally accurate approximations to functional rational expectations equilibria of dynamic stochastic economies in the sequence space. We use deep neural networks to parameterize key equilibrium objects, such as policies or prices, as functions of truncated histories of exogenous shocks. We train the neural networks to satisfy equilibrium conditions along simulated paths of the economy. We illustrate the performance of our method in three environments: (i) a high-dimensional overlapping generations economy with multiple sources of aggregate risk; (ii) an economy with heterogeneous households and firms facing uninsurable idiosyncratic risk and large shocks to idiosyncratic and aggregate volatility; and (iii) a stochastic life-cycle economy with a continuous asset choice and a discrete early-retirement choice that induces local convexities in the continuation values of working-age cohorts. 
We also propose practical neural policy architectures that guarantee monotonicity of predicted policies, enabling the endogenous grid method to simplify parts of the algorithm. We achieve high precision throughout, with the mean error in equilibrium conditions below $0.2\%$.
\end{abstract}


\onehalfspacing
\section{Introduction}
We bring deep learning solution methods to the sequence space and propose a new global method for computing equilibria in economies with aggregate risk.\footnote{We use the expression ``sequence space'' liberally, referring to the space of truncated sequences of aggregate shocks, not (only) the space of perfect foresight sequences. To be more precise, one could alternatively refer to our approach as a moving-average rather than a sequence-space approach.} Exploiting the ergodicity property of a large class of dynamic economies, we approximate the aggregate state vector using a truncated sequence of aggregate shocks. We use deep neural networks to parameterize the mapping from truncated histories of aggregate shocks to equilibrium objects of interest, and optimize them using stochastic gradient descent to minimize errors in the equilibrium conditions over the simulated ergodic set of the economy.

We make four distinct contributions to the literature. First, we show how deep learning can be used to compute global approximations to equilibria of dynamic stochastic economies in the sequence space. Second, we show how to construct neural network architectures that predict approximate policy functions, which are guaranteed to satisfy ex-ante known monotonicity and concavity properties. Third, based on our monotonicity-preserving architectures, we show how to obtain supervised learning targets for policy and price functions, using the endogenous gridpoint method of \cite{carroll_2006}. Fourth, we illustrate the rich applicability and accuracy of our method by solving three models, each featuring a different computational challenge.
We first solve an overlapping generations model with portfolio choice and regime switching.
We then proceed to compute a global solution to an economy with a continuum of heterogeneous firms and a continuum of heterogeneous households. To our knowledge, we are the first to do so in an economy featuring incomplete markets and household heterogeneity in the spirit of \cite{imrohorouglu1989cost}-\cite{bewley_1976}-\cite{huggett1993risk}-\cite{aiyagari_1994}-\cite{krusell1998income}, together with firm heterogeneity in the spirit of \cite{bloometal_2018} and \cite{khan2008idiosyncratic}. Lastly, we apply our method to solve an overlapping generations model with intra- and intergenerational risk and a discrete early-retirement choice. The discrete choices lead to local convexities in the continuation values of working-age cohorts, rendering their Euler equations insufficient for uniquely characterizing their optimal savings policies.

To describe the main idea of our method, we now take the canonical \cite{krusell1998income} model as an example. 
Let $z_t$ denote the realization of stochastic TFP in period $t$, and let $z^t :=(z_0, z_1, \dots, z_t)$ denote the history of realized shocks. In an infinite horizon model the history of shocks becomes infinitely long. The state of the economy is the current level of TFP $z_t$, together with the cross-sectional distribution of households over wealth and productivity. Let $e^i_t$ denote the idiosyncratic productivity of the household $i$ in period $t$, let $k_t^i$ denote their capital holdings, and let $\mu_t = \mu_t (e, k)$ denote the cross-sectional distribution. Since the distribution, even when discretized into a histogram as in \cite{young2010solving}, is a high-dimensional object, the presence of the distribution in the state vector poses a long-standing challenge for solution methods.\footnote{See, for example, \cite{moll2025trouble}.} The approach of \cite{krusell1998income} is to approximate the cross-sectional distribution $\mu_t$ with a low-dimensional statistic such as aggregate capital $K_t$.
Instead of approximating the state vector using its moments or a histogram, we build on the seminal perturbation approaches in the sequence space developed in \cite{boppartetal_2018} and \cite{auclertetal_2021} and approximate the aggregate state vector using \emph{sequence-space} objects. In particular, we propose to approximate the aggregate state of the economy by a truncated history of exogenous aggregate shocks. In the context of the \cite{krusell1998income} model, for example, let $z^t_T = (z_{t-T}, z_{t-T+1}, \dots,z_{t-1}, z_t)$ denote the last $T$ realizations of TFP. Instead of solving for equilibrium functions $f^{\text{approx}}(z_t, K_t) \approx f(z_t, \mu_t)$, we propose to solve for equilibrium functions $f^{\text{approx}}(z^t_T) \approx f(z_t, \mu_t)$. For any $T<\infty$ our approach implies a truncation error. However, in economies featuring ergodic dynamics, this truncation error converges to zero as $T \to \infty$. Hence, we can approximate the information contained in the aggregate state vector arbitrarily well by choosing $T$ sufficiently large. Consequently, our function approximators need to be able to handle large $T$, implying a high-dimensional input $z_t^T$, potentially even higher dimensional than a histogram of the wealth distribution. We will refer to our approach, which uses a truncated history of aggregate shocks as input as \emph{sequence-space approach} and to the standard approach (using, for example, the wealth distribution or a lower dimensional sufficient statistic such as standard moments, as input to the equilibrium functions) as \emph{state-space approach}. 

We use neural networks to approximate all equilibrium functions that need to be solved for. Like polynomials, neural networks are universal function approximators \citep{hornik1989multilayer}. More importantly however, neural networks have shown great promise in ameliorating the curse of dimensionality \citep{bellman_1961} associated with high-dimensional function-approximation problems, such as solving partial differential equations \citep[see, \emph{e.g.}][]{grohsetal_2018, jentzenetal_2018} or computing equilibria in dynamic economies \citep[see, \emph{e.g.}][]{azinovic2022deep, maliar2021deep, kahou2021exploiting, guetal_2023}.\footnote{Section \ref{sec:literature} provides a more detailed discussion of the related literature in economics.} 
The sequence-space formulation has two main advantages in the context of deep learning solution methods. First, the sequence-space formulation is potentially more parsimonious, especially in economies with rich cross-sectional heterogeneity.\footnote{e.g. The state-space formulation of the workhorse HANK economy features a state vector that includes a cross-sectional distribution of households over idiosyncratic shocks as well as liquid and illiquid assets. A tensor-product discretization of such a distribution might easily generate tens of thousands of aggregate state variables, whereas a highly accurate sequence-space representation could be obtained using a few hundred of shock lags at quarterly frequency.} Second, using truncated shock histories as the only network input mitigates a dangerous feedback loop present in simulation-based deep learning solution methods. While the distribution of shock histories is exogenous and fixed,\footnote{Because the distribution of fundamental shocks is just a primitive of the economic model at hand.} the distribution of endogenous states moves during the training as the algorithm updates the approximate policy function. As discussed in \cite{azinovic2022deep}, a large update to the policy function might shift the distribution of training states into areas that have been previously only sparsely covered, causing large sample error, and correspondingly large gradients, potentially inducing further jumps in the policy function. While our algorithm still relies on simulating endogenous states as an input for evaluating loss function\footnote{e.g. for evaluation of marginal product of capital} those do not enter as an input into neural networks parameterizing the core equilibrium objects of the economy.

The second contribution of our paper is that we show how to construct neural network architectures that guarantee monotonicity or even concavity of the resulting approximator with respect to a chosen set of inputs. Rather than directly approximating individual policy function using a neural network that takes the aggregate and idiosyncratic state as input,\footnote{As in \cite{azinovic2022deep} or \cite{maliar2021deep}.} we build on the operator learning approach of \cite{zhong_2023}. An operator network maps the sequence of aggregate shocks to policy \emph{functions}. The predicted policy functions then map idiosyncratic state variables into the individual policy choices. Therefore, instead of having to repeatedly evaluate the neural network for many combinations of aggregate and idiosyncratic states, the neural network predicts a single policy function for each aggregate state. The predicted policy function is then evaluated for all idiosyncratic states of interest.
For the \cite{krusell1998income} economy, for example, the neural network predicts a consumption \emph{function} based on the realized history of aggregate shocks, which then takes only individual productivity and asset holdings as input. 
Additionally, ensuring monotonicity allows us to implement parts of the algorithm using the endogenous grid method (EGM) \citep{carroll_2006}. The use of EGM is particularly advantageous when tackling models where agents face a mix of discrete and continuous choices since it enables the use of an upper envelope algorithm \citep{druedahl2017general, dobrescu2023fast}. 

\section{Literature\label{sec:literature}}
The idea of approximating the endogenous aggregate state by a truncated sequence of aggregate shocks was first introduced by \cite{chienetal_2011} in the context of an endowment economy with heterogeneous agents and with a single aggregate shock that takes two discrete values. In their setting, they find that considering the history of the past five aggregate shocks is enough to yield a high-accuracy approximation. In the context of production economies with aggregate risk \cite{Lee_RTM_2025} introduces the repeated transition method (RTM). The underlying idea of the RTM is that, in an ergodic economy, all possible states will be realized along a sufficiently long simulation path. \cite{Lee_RTM_2025} uses this fact to construct the continuation value function for the Bellman equations of the agents in the economy. The RTM iterates over simulating a long path of the economy. In each iteration it constructs the continuation values by using the previously simulated path as a look-up table.
Our contribution relative to this literature is generality. While those two papers also develop sequence-based global solution methods, the applicability of these methods has limitations that do not pose a significant challenge for our method. Namely, the algorithm of \cite{chienetal_2011} struggles to resolve economies featuring stronger persistence, such as production economies, because strong persistence implies the need to keep track of a long history of shocks. The RTM approach of \cite{Lee_RTM_2025} can handle a rich set of environments, including production economies; however, its main bottleneck lies in handling economies with a large number of different aggregate shocks. At each iteration the RTM algorithm has to construct the continuation value function for every \emph{period}. To do so, the algorithm has to split the simulated path into partitions sorted by the realized value of aggregate shock, and then to search through each partition for a realization where the aggregate endogenous state (e.g. wealth distribution) is closest to the \emph{next period} distribution with respect to some norm. This structure implies that the RTM algorithm works best in environments with a small number of discrete aggregate shocks. Complex stochastic processes that cannot be approximated using a parsimonious Markov chain pose a challenge for RTM, as the length of simulated path required for ergodic representation grows with the cardinality of shock discretization. In contrast, we demonstrate that our method can efficiently solve rich economies featuring production and complex driving stochastic processes. Besides that, our algorithm can handle discrete as well as continuous shocks. Furthermore, neural networks can learn from a large number of simultaneous trajectories, allowing for significant acceleration on modern massively parallel computing architectures.\footnote{Global solution methods to solve models on the simulated paths using state-space approaches include \cite{brumm2025tensor}, \cite{levintal2018taylor}, and \cite{azinoviccolekubler_2023}}

Instead of using the truncated history of exogenous shocks as a summary statistic for the aggregate state, recent work by \cite{yang2025structural} and \cite{guarda2025narrow} uses short histories of endogenous aggregate variables in the spirit of \cite{krusell1998income}. \cite{yokomoto_2025} implements the \cite{krusell1998income} algorithm using neural networks as function approximators. 
\cite{yang2025structural} and \cite{yokomoto_2025} use prices as their summary statistic for the endogenous aggregate state, allowing for fast computation of market clearing prices in the simulation step. \cite{yang2025structural} update policies with a policy gradient algorithm to maximize simulated lifetime rewards, while \cite{yokomoto_2025} updates policies using the Bellman equation. Households' decisions depend on the endogenous aggregate state only through prices. Hence, households directly track the aggregate variable that enters into their optimization problem. In \cite{yang2025structural} very short histories are sufficient to forecast conditional future prices well, keeping the state-space for households low-dimensional and thereby addressing the challenges raised in \cite{moll2025trouble}.\footnote{Applications in \cite{yang2025structural} include the \cite{krusell1998income} model, a \cite{huggett1993risk} model with aggregate risk, and a one asset HANK model along the lines of \cite{auclertetal_2021}.} Although we focus on histories of exogenous shocks, our method can easily be extended to additionally use histories of endogenous aggregate variables, such as prices.
In contrast to \cite{yang2025structural} our approach works directly with equilibrium conditions and hence allows us to obtain supervised learning targets for household policies. Furthermore, our approach applies to models where the individual problem depends directly on a high dimensional aggregate state and not only through a few low-dimensional aggregate statistics, such as prices.\footnote{See, for example, \cite{payneetal_2024}.}

Perturbation methods in the sequence space have become the method of choice for solving heterogeneous agent models following the game-changing work by \cite{boppartetal_2018} and \cite{auclertetal_2021}. \cite{boppartetal_2018} and \cite{auclertetal_2021} introduce local methods, which linearize heterogeneous agent models with respect to aggregates, in the sequence space.\footnote{See also \cite{reiter2009solving} and \cite{bayerluetticke_2020} for linearization in aggregate variables in the state space.} 
These methods, together with the accompanying libraries,\footnote{See \url{https://github.com/shade-econ/sequence-jacobian} for a user friendly library for the method in \cite{auclertetal_2021}.} have substantially increased the tractability of heterogeneous agent models with aggregate risk, rendering a wide set of previously intractable models computationally tractable. Our method complements this literature by providing a global method, which is hence applicable for models with larger shocks and without certainty equivalence with respect to aggregate risk. The cost of having a global method is that the runtime is substantially longer in comparison to local methods, such as the sequence-space Jacobian method by \cite{auclertetal_2021}.

The toolbox that enables us to approximate functions on the very high dimensional domain of long histories of economic shocks is deep learning. Deep learning as a tool to compute equilibria in economic models dates back to \cite{duffy_mcnelis_2001}, who use a neural network to solve a stochastic growth model using a parameterized expectations algorithm. \cite{norets_2012} makes use of neural networks in the context of discrete-state dynamic programming. Closer to this paper \cite{azinovic2022deep} and \cite{maliar2021deep} use neural networks to approximate equilibrium price and  policy functions, and train them to satisfy equilibrium conditions, such as first-order optimality conditions, Bellman equations, and market clearing conditions along the simulated paths of the economy. \cite{azinoviczemlicka_2024} extend these methods by introducing market clearing layers\footnote{Market clearing layers are neural network architectures, designed such that a prediction by the neural network is always consistent with market clearing.} and step-wise model transformations to robustly solve models with portfolio choice. \cite{kase2022estimating} show how to use deep learning to compute the solutions of economic models for ranges of economic parameters, speeding up the maximum likelihood estimation of these parameters. \cite{han2021deepham} and \cite{kahou2021exploiting} introduce symmetry preserving neural network architectures and \cite{valaitis2024machine} use deep learning within a parameterized expectations algorithm. \cite{fernandezetal_2023} use neural networks for a nonlinear forecasting rule within a \cite{krusell1998income} algorithm. \cite{kahou2022spooky} investigate the relationship between transversality conditions and solutions found by a deep-learning based algorithm. \cite{druedahlropke_2025} use deep learning to compute optimal choices in finite-horizon life-cycle models with up to eight durable goods. \cite{druedahletal_nonconvex_2026} show how the methodology of \cite{druedahlropke_2025} can be applied to solve non-convex life-cycle models featuring a mixture of discrete and continuous choices. We view our discrete-choice algorithm introduced in section \ref{sec:retirement_model} as complementary to \cite{druedahletal_nonconvex_2026}: we focus on general equilibrium setups while keeping the optimization problem of an individual agent comparatively simple, whereas \cite{druedahletal_nonconvex_2026} solve complex individual optimization problems with many states and choices in partial equilibrium.
In recent work \cite{druedahletal_2026} build on our approach to construct a global solution to a canonical HANK economy with aggregate risk using deep learning in the sequence space. They study the effect of aggregate uncertainty on fiscal multipliers and impulse responses to aggregate shocks.

\cite{zhong_2023} introduces operator learning, which we build on in section \ref{sec:HA}. While previous deep learning approaches used neural networks to directly parameterize the mapping from a combination of an aggregate and idiosyncratic state to individual choices, \cite{zhong_2023} uses neural networks to approximate the mapping from an aggregate state into a function which then returns individual choices as a function of individual state. We extend the operator learning method of \cite{zhong_2023} by showing how to encode monotonicity or concavity of the predicted functions into the network architecture. \cite{sun_2025_cont} uses neural networks to approximate continuation values in combination with otherwise standard methods. Neural network based solution methods in continuous time are developed or applied in \cite{duarte2018machine}, \cite{sauzet2021projection}, \cite{gopalakrishna_2021}, \cite{guetal_2023}, \cite{gopalakrishna2024asset} and \cite{payneetal_2024}.\footnote{Additional applications of deep learning based solution methods in variety of settings include \cite{folini2025climate} and \cite{friedl2023deep} (climate economics), \cite{carvalho2025planning} (production networks),  \cite{kase2025generative} (HANK with financial frictions), \cite{duarte2021simple} (partial equilibrium finite horizon model with rich asset choice), \cite{bretscher2022ricardian} (international business cycle), \cite{sun_2025_dist} (spatial economics), \cite{jungerman_2023} (monopsony power in the labor market), \cite{adenbaum2024deep} (\cite{bewley_1976} type economy), and \cite{reiter2025neural} (hybrid solution approach for OLG).}

Our contribution relative to the existing literature on deep learning solution methods is twofold. First, we show that deep learning can be efficiently used to construct global approximations to the functional rational expectations equilibria in the sequence space. Second, we construct shape-preserving neural network operator network architectures that allow us to incorporate known shape properties of certain equilibrium functions (e.g. monotonicity and concavity of consumption function in canonical consumption-savings problems) directly into the neural network structure.\footnote{The idea of shape-preservation in functional approximation methods for dynamic economies was introduced by \cite{juddsolnick_1994}, and further developed by \cite{cai_2012}.} The ability to guarantee monotonicity of the consumption function allows us to apply the method of endogenous gridpoints of \cite{carroll_2006} and hence simplify the solution algorithm for a large class of problems featuring optimization problems that are amenable to EGM.

\section{Algorithm and illustrative application\label{sec:illustration}}
\subsection{Model\label{sec:RA}}
We first explain our proposed algorithm by applying it to the simplest workhorse economy with aggregate risk, the stochastic growth model of \cite{brock_1972}.
This model can be accurately solved using conventional methods, so this section does not aim to claim a computational achievement of our method. Instead, it serves to lay out the proposed algorithm as transparently as possible. In sections \ref{sec:OLG}, \ref{sec:HA}, and \ref{sec:retirement_model}, we then apply our algorithm in settings that pose a substantial challenge for existing numerical methods.

Time is infinite and discrete, $t = 0, 1, \dots$. We consider an infinitely lived representative household that derives utility $u(C_t)$ from consumption $C_t$. The representative household owns the capital stock $K_t$ and inelastically supplies a constant amount of efficient units of labor $L = 1$ at the equilibrium wage $w_t$. The single good in the economy is produced by a representative firm endowed with a Cobb-Douglas production technology
\begin{align}
Y_t = A_t K_t^{\alpha} L^{1-\alpha},
\end{align}
where $A_t$ denotes stochastic total factor productivity, that evolves according to an AR(1) process
\begin{align}
\log(A_{t}) = \rho^A \log(A_{t-1}) + \sigma^A \epsilon^A_t,
\end{align}
where $\epsilon^A_t \sim \mathcal{N}(0, 1)$.
The firm rents capital $K_t$ and efficient units of labor $L$ on competitive spot markets and pays a rental rate $r^K_t = \alpha A_t K_t^{\alpha -1} L^{\alpha}$ on capital and a wage $w_t = (1 - \alpha) A_t K_t^{\alpha} L^{-\alpha}$ for efficient units of labor.
In each period, the representative household chooses how much to consume and how much to invest in capital. Capital evolves according to $K_{t+1} = (1 - \delta) K_t + I_t$, where $I_t$ denotes the investment in capital and $\delta$ denotes its depreciation rate. The households' problem is given by
\begin{align}
    &\max_{\{K_{t}\}_{t=1}^{\infty}}\E{ \sum_{t=0}^{\infty} \beta ^ t u(C_t)} \\
    \text{subject to}:& \nonumber \\
    C_t &= L w_t + (1 - \delta + r^K_t) K_t - K_{t+1}. \nonumber
\end{align}
The Euler equation, which characterizes the equilibrium together with a transversality condition, is given by
\begin{align}
u'(C_t) &= \E{\beta u'(C_{t+1})(1 - \delta + r_{t+1}^K)} \\
\Leftrightarrow 0 &= \frac{(u')^{-1}\left( \E{\beta u'(C_{t+1})(1 - \delta + r_{t+1}^K)}\right)}{C_t} - 1.\label{eq:ree_bm}
\end{align}
Where the Euler equation \eqref{eq:ree_bm} is rearranged, such that errors in the optimality condition can be interpreted as relative consumption errors. This allows for an implicit, yet interpretable accuracy measure \citep[see][]{judd1998numerical}.
Traditionally, the model would be solved using recursive methods, where the state of the economy $\Xst _t := [A_t, K_t] \in \mathbb{R}^2$ is given by the current level of TFP together with the current capital stock.
The solution to the model is a policy function $\pi(\Xst_t) = K_{t+1}$, which maps the state of the economy to the endogenous variables of interest, in this case the capital stock for the next period.\footnote{There are other equivalent policies, like a policy for investment or consumption, which together the budget constraint and the law of motion for capital lead to the same choice for capital stock in the next period.}
\subsection{Algorithm\label{sec:algorithm}}
\subsubsection{Main idea}
Our algorithm uses a deep neural network to approximate the policy function. Let $\nn$ denote a neural network with trainable parameters $\brho$. Following the \emph{Deep Equilibrium Nets} algorithm (or similar algorithms often used in the literature) the goal is to approximate the policy function by a neural network. This would mean to find neural network parameters $\brho$, such that for all states
\begin{align}
\nn(\Xst_t)\approx \pi(\Xst_t) = K_{t+1}.
\end{align}
The main idea of the algorithm we propose in this paper is to train the neural network to predict policies not based on $\Xst = [A_t, K_t]\in \mathbb{R}^2$ but instead purely based on a truncated history of shocks, $\Xse := [A_{t - T + 1}, A_{t - T + 2}, \dots, A_{t-1}, A_t]\in \mathbb{R}^T$, such that
\begin{align}
\nn(\Xse_t)\approx \pi(\Xst_t) = K_{t+1}.
\end{align}
At first sight, using the history of shocks as an input may look like a step in the wrong direction for two reasons: first, for $T<\infty$ the sequence of shocks is only an approximately sufficient statistic for the state of the economy. Hence, by truncating the history, we introduce a limit to the precision that even a perfectly trained neural network could attain. Second, the dimensionality of $\Xse$ is sometimes larger than the dimensionality of $\Xst$.\footnote{For example in this simple stochastic growth economy.} Therefore, the sequence space approach would not work well together with approximation methods, for which a low-to-medium dimensional domain is pivotal, such as, for example, (adaptive) sparse grids.\footnote{See \cite{kruegerkubler_2004} for sparse grids and \cite{brumm2017using} for adaptive sparse grids.}
As we illustrate in this paper, however, the sequence approximation to the endogenous aggregate state is promising when used together with deep neural networks as function approximators. 

\subsubsection{Sufficiency of the truncated history of shocks} For our method to work, it must be that the truncated sequence of shocks is an approximate sufficient statistic for the endogenous aggregate state of the economy, in this case capital $K_t$. This can only be the case if the influence of capital at period $t-T$, $K_{t-T}$, on capital in period $t$, $K_t$, is vanishing as $T\rightarrow\infty$, \emph{i.e.} our method relies on the ergodicity property of the underlying economy.

To see that this condition indeed holds in our stochastic growth economy, we now consider a special case with full depreciation $\delta = 1$ and logarithmic utility $u(C) := \log(C)$. In this case, the model admits a closed form solution with $K_{t+1} = \alpha \beta A_t K_t^\alpha$. Taking log on both sides, we obtain $\log(K_{t+1}) = \log(\alpha \beta) + \log(A_t) 
 + \alpha \log(K_t)$. Iterating forward, we obtain
\begin{align}
\log(K_t) &= \log(\alpha \beta) + \log(A_{t-1}) 
 + \alpha \log(K_{t-1}) \nonumber \\
 &= \log(\alpha \beta) + \log(A_{t-1}) 
 + \alpha (\log(\alpha \beta) + \log(A_{t-2}) 
 + \alpha \log(K_{t-2})) \nonumber \\ 
 &= \log(\alpha \beta) + \log(A_{t-1}) 
 + \alpha (\log(\alpha \beta) + \log(A_{t-2}) 
 + \alpha (\log(\alpha \beta) + \log(A_{t-3}) 
 + \alpha \log(K_{t-3}))) \nonumber \\
 &= \dots = \underbrace{f(\alpha, \beta, A_{t-1}, A_{t-2}, A_{t-3}, \dots, A_{t-T})}_{\text{function of $\alpha$, $\beta$ and the last $T$ productivity values}} + \underbrace{\alpha ^ T \log(K_{t-T})}_{\text{truncation error}}, \label{eq:bm_truncerror}
\end{align}
where the function $f$ depends only on the parameters $\alpha$, $\beta$ and the truncated history of the last $T$ productivity values. The value of capital at $t-T$ only affects the value of capital with a coefficient of $\alpha^T$, where $\alpha < 1$ is the share of capital in production, typically around $0.3$. The truncation error therefore vanishes exponentially at the rate $\alpha$.

As an alternative to using lagged productivity levels, we could use a truncated history of innovations as the input to the neural network.\footnote{Since the innovations are i.i.d.\ by construction, they may be easier for the network to process than highly persistent levels.}
Iterating the AR(1) implies
\begin{align}
\log(A_t) 
&= \rho^T \log(A_{t-T}) + \sigma \sum_{j=0}^{T-1} \rho^{j}\,\epsilon_{t-j} \\
&=: g\!\left(\rho,\sigma,\epsilon_{t-T+1},\ldots,\epsilon_t\right) + \underbrace{\rho^T \log(A_{t-T})}_{\text{truncation error}}.
\end{align}
Hence, for large enough $T$ and $|\rho| < 1$ the truncated history of innovations is an approximate sufficient statistic for the history of productivity values, which in turn are an approximate sufficient statistic for the endogenous aggregate state.

The same argument can also be made in a single step by observing that
\begin{align}
\begin{bmatrix}
\log(A_t) \\
\log(K_t)
\end{bmatrix}
= 
\underbrace{
\begin{bmatrix}
\rho & 0 \\
1 & \alpha
\end{bmatrix}}_{=: M} 
\begin{bmatrix}
\log(A_{t-1}) \\
\log(K_{t-1})
\end{bmatrix}
+  
\begin{bmatrix}
\sigma\epsilon_{t} \\
\log(\alpha \beta)
\end{bmatrix}
\end{align}
The two eigenvalues of $M$ are $\rho$ and $\alpha$, implying that the impact of $\log(A_{t-T})$ and $\log(K_{t-T})$ on $(\log(A_t),\log(K_t))$ decays geometrically at rate $\max\{|\rho|,|\alpha|\}$.
This also implies that in order to accurately predict the policy functions based on a truncated history of innovations in a setting where $\rho$ is close to 1, the history needs to be very long, leading to high dimensional state $\Xse_t$.\footnote{With a slight abuse of notation, we use $\Xse$ to indicate the approximate state in sequence space based on either sequences of innovations ($\epsilon$) or values ($A$).}

\subsubsection{Remaining parts of the algorithm\label{sec:remaining}}
The remaining parts of the algorithm follow \cite{azinovic2022deep} (DEQN) and are summarized here for completeness.

\paragraph{Loss function} 
We train the neural network by minimizing a loss function. 
Specifically, we use the mean squared error in the household optimality condition, \eqref{eq:ree_bm}, expressed in terms of relative consumption error.
By minimizing the (weighted) Euler equation errors, we train the neural network policy function to take shape that is consistent with the optimal choices of the representative household.

In order to evaluate the households optimality condition implied by the policy encoded the neural network, $\nn(\Xse_t) = K_{t+1}$, we need access to the current value of capital and productivity\footnote{Because we need to evaluate objects like current period output, or marginal product of capital next period.} $\Xst_t = [A_t, K_t]$. Hence, even though the neural network's input does only consist of $\Xse_t$, we need to also keep track of the associated state vector $\Xst_t$ in order to be able to evaluate the loss function. The distribution over possible  $t+1$ histories, $\Xse_{t+1}$, is implied by the current history $\Xse_t$, together with the stochastic process for the exogenous variable. The distribution over possible $t+1$ states, $\Xst_{t+1} = [A_{t+1}, K_{t+1}] = [A_{t+1}, \nn(\Xse_t)]$, is given by the current state $\Xst_t$, the policy function encoded by the weights of the neural network, and the stochastic processes for the exogenous variables.

We construct the relative Euler equation error by plugging the neural network approximation of the history-based policy function into equation
\eqref{eq:ree_bm}.
\begin{align}
\text{ree}(\Xse_t, \Xst_t, \brho) &:= \frac{(u')^{-1}\left( \E{\beta u'(C_{t+1})(1 - \delta + r_{t+1}^K)}\right)}{C_t} - 1 \label{eq:Rep:relconserr}\\
\text{where}&\nonumber\\
C_t &= A_t K_t^\alpha + (1 - \delta)K_t - K_{t+1} \\
C_{t+1} &= A_{t+1} K_{t+1}^\alpha + (1 - \delta)K_{t+1} - K_{t+2} \\
K_{t+1} &= \nn(\Xse_t) \\
K_{t+2} &= \nn(\Xse_{t+1}) \\
\Xse_{t+1} &=[A_{t+1-T},\dots, A_{t+1}].
\end{align}

We define the loss function as a simple average of the square of the relative Euler equation error evaluated across a dataset $\mathcal{D}$ of state-histories pairs.
\begin{align}
\mathcal{D} &:= \{(\Xse_1, \Xst_1), (\Xse_2, \Xst_2), \dots, (\Xse_N, \Xst_N) \} \\
\mathcal{L}(\mathcal{D}, \brho) &:= \frac{1}{N}\sum_{(\Xse_t, \Xst_t)\in \mathcal{D}}\text{ree}(\Xse_t, \Xst_t, \brho) ^ 2.\label{eq:loss_RA}
\end{align}

\paragraph{Updating the neural network parameters}
We update the parameters using the ADAM optimizer \citep[][]{kingma2014adam}, which is a variant of gradient descent. The update rule implied by vanilla gradient descent is given by
\begin{align}
\brho^{\text{new}} = \brho^{\text{old}} + \alpha^{\text{learn}} \nabla_{\brho}\mathcal{L}(\mathcal{D}, \brho) \label{eq:sgd_vanilla}
\end{align}
The ADAM optimizer modifies the equation \eqref{eq:sgd_vanilla} by introducing parameter-specific adaptive learning rates rather than using a single constant learning rate for all parameters.

\paragraph{Sampling the most relevant state-history pairs}
To generate a dataset $\mathcal{D}$ of state-history pairs, we closely follow the ergodic simulation scheme used in the DEQN algorithm. We start the simulation with a random sample of initial exogenous and endogenous states, and then simulate exogenous states using their stochastic process, and evolve endogenous states using the current approximation of the policy functions. Relative to the original DEQN method, our algorithm additionally keeps track of truncated shock histories. This is important because we use the truncated histories of shocks as neural network inputs. A key advantage of using truncated shock histories as network input is that their distribution does not change during the training, as it is pinned-down by model primitives, \emph{i.e.} shock distributions, rather than by policy functions.

\subsection{Parameterization}
We set relative risk aversion $\gamma = 2$, depreciation $\delta = 0.1$, and patience $\beta = 0.95$. The capital share in production is set to $\alpha = \frac{1}{3}$, and the process for logarithm of TFP has persistence $\rho^A = 0.8$ and $\sigma ^A = 0.03$.
The model parameters are summarized in table \ref{tab:RA_calib}.

\subsection{Implementation}
We train a densely connected feed-forward neural network to predict the savings rate $s_t$ in period t as a function of the last $T$ realizations of innovations to aggregate productivity, \emph{i.e.}, $\Xse_t = [\epsilon_{t-T}, \dots, \epsilon_{t-1}, \epsilon_t] \in \mathbb{R}^T$ and $s_t = \nn(\Xse_t)$. Approximating the savings rate as opposed to directly parameterizing the savings function $K_{t+1}$, has the advantage that we can ensure that savings rate is bounded in $(0, 1)$ by applying a sigmoid activation in the output layer.\footnote{The sigmoid function is defined as $\text{sigmoid}(x):=\frac{1}{1+e^{-x}}$.} Because of that, our policy function is guaranteed to satisfy the budget constraint and also guaranteed to ensure non-negativity of consumption. 
Given the current capital stock and productivity level, we compute the quantity of resources on hand $M_t = A_t K_t^\alpha L^{1 - \alpha} + (1 - \delta) K_t$. The savings rate and resources on hand then imply the current consumption $C_t = (1 - s_t) M_t$ and the next period's capital $K_{t+1} = s_t M_t$. This illustrates why our algorithm still needs to keep track of values of state variables. Even though states are not used as network input, they are still required to construct objects like resources at hand, marginal products, etc.

We organize the training procedure into a sequence of \emph{episodes}. An episode starts by inheriting a neural policy function and a dataset $\mathcal{D}_j$ consisting of state-history pairs from the previous episode. Then we use the approximate policy function and a pseudo-random number generator to simulate a new training dataset of $N^{\text{data}}$ state-history pairs $\mathcal{D}_j = \{(\Xst_{j, i}, \Xse_{j, i})\}_{i = 1}^{N^{\text{data}}}$:
\begin{align}
&\text{obtain savings rate for old state}: &s_{j, i} = \nn(\Xse_{j, i}) \\
&\text{compute implied capital in the next period}: &K_{j+1, i} = s_{j, i} (A_{j, i} K_{j, i}^\alpha + (1-\delta)K_{j, i})  \\
&\text{draw a new random shock}: &\epsilon_{j+1, i}\sim\mathcal{N}\left(0, 1\right)\\
&\text{update the truncated history:} &\Xse_{j+1, i} = \left[[\Xse_{j,i}]_{2 : T}, \epsilon_{j+1, i} \right]\\
&\text{update the agg. state:} &\Xst_{j+1, i} = [K_{j+1, i}, A_{j+1, i}].
\end{align}
Generating new training data in this way is computationally cheap and does not pose a bottleneck to our algorithm. This has the advantage that we can generate new training dataset after every episode, such that no datapoint is used twice during the training. Therefore, overfitting is not a concern. A key difference of our approach, relative to previous deep learning based approaches is that the distribution of neural network inputs, $\Xse$, is exogenous and hence remains stationary throughout the training. The state-space representation $\Xst$ on the other hand, is converging to the ergodic set of states as the neural network learns the equilibrium policies.

After obtaining a fresh dataset, we update the policy function by running a short training loop using the Adam optimizer \citep[see][]{kingma2014adam} with learning rate $\alpha^\text{learn}$ and mini-batch size $N^{\text{mini-batch}}$. We evaluate the gradient of the loss function using standard reverse mode automatic differentiation.

\subsection{Training}
\paragraph{Hyperparameters}
We use a history of $T = 100$ previous productivity innovations as the approximate sufficient statistic replacing the state vector as a network input. We parameterize the history-based policy function using a densely-connected feed-forward neural network with three hidden layers. Each layer consists of 128 \emph{gelu} activated neurons.\footnote{The gelu activation function is given by $f(x) = x \Phi(x)$, where $\Phi(x)$ denotes the Gaussian cumulative distribution function.} The output layer is sigmoid activated, ensuring that the predicted savings rate lies between 0 and 1.
To approximate the conditional expectation in the Euler equation we use Gauss-Hermite quadrature with $N^{\text{quad}}=8$ integration nodes.\footnote{An alternative would be to discretize the AR(1) process, as in \cite{rouwenhorst_1995}.} 
For minimization of the loss function, we rely on the basic ADAM optimizer with a learning rate of $10^{-5}$ and mini-batch size of $256$. In each episode, our algorithm simulates $4096$ state-history pairs, implying $16$ gradient descent steps per episode.
The hyperparameters are summarized in table \ref{tab:RA_hp}.

\begin{figure}
    \centering
    \includegraphics[width=0.6\linewidth]{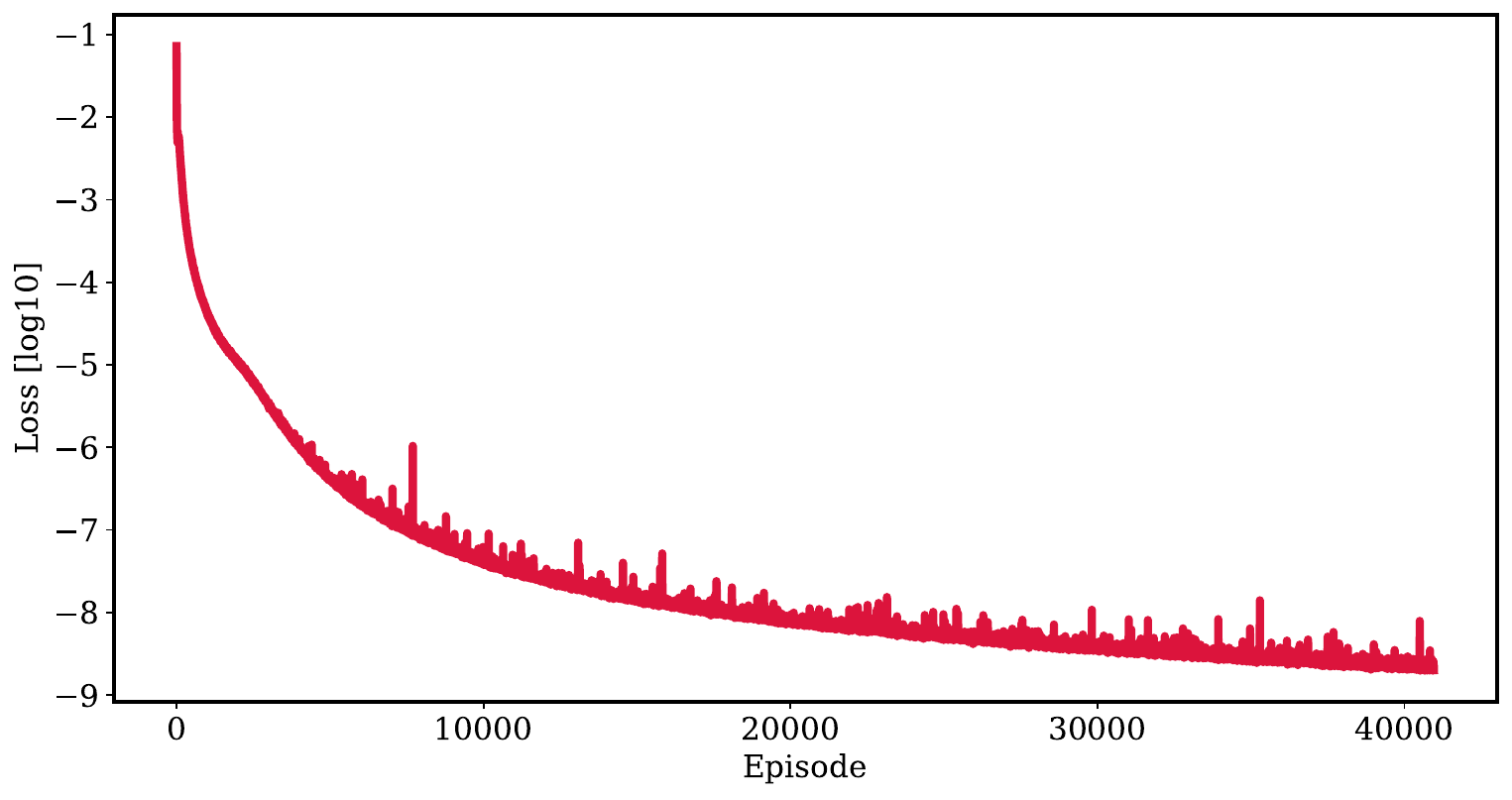}
    \caption{Loss function when training the neural network for $\approx$40{,}000 episodes to solve the \protect\cite{brock_1972} model. The loss function is the mean squared error in the equilibrium conditions of the model, each episode consists of 4096 simulated states.\label{fig:RA_loss_training}}
\end{figure}

\paragraph{Training progress}
Figure \ref{fig:RA_loss_training} shows the loss function during training. The loss function decreases steadily during the training and reaches a value well below $10^{-8}$ by the end of training.\footnote{The training run takes roughly 4 minutes on a Tesla T4 GPU.}

\subsection{Accuracy}
\begin{figure}
\centering
    \includegraphics[width=0.32\linewidth]{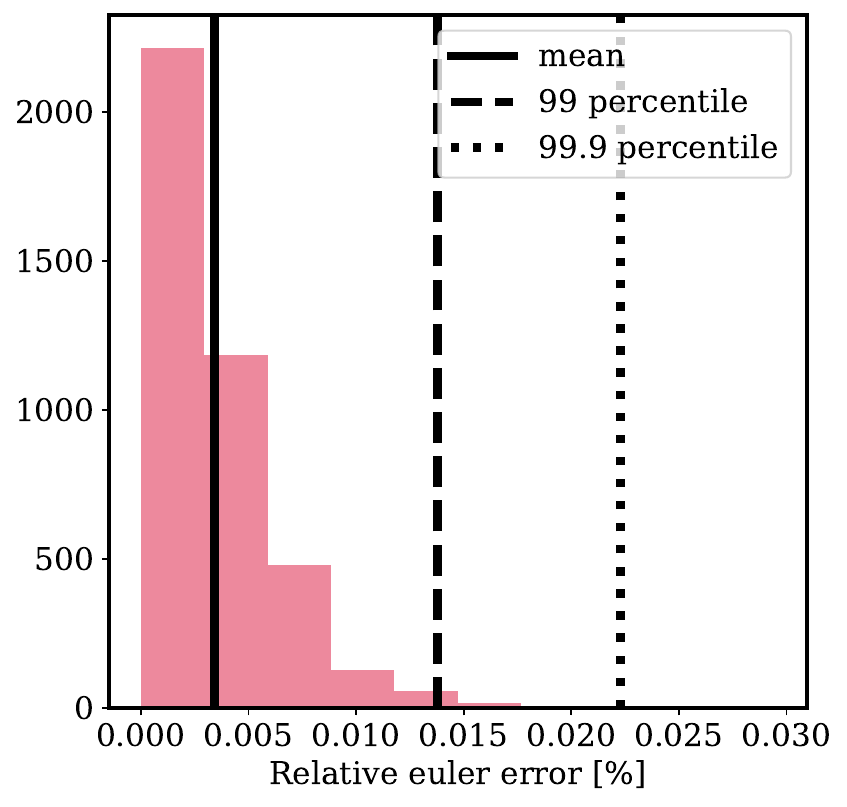}
    \includegraphics[width=0.32\linewidth]{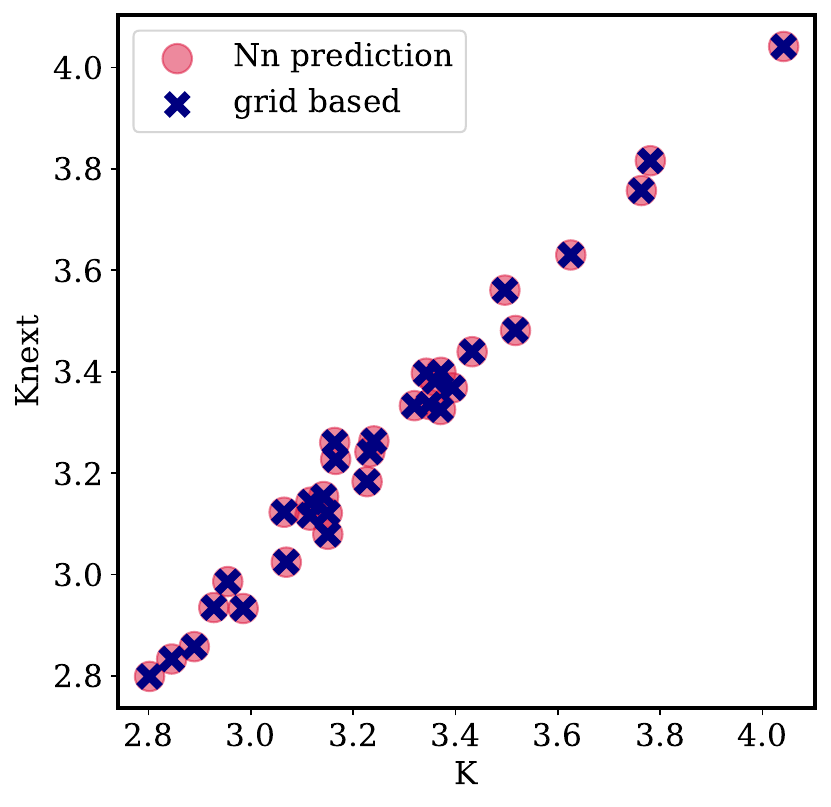}
    \includegraphics[width=0.315\linewidth]{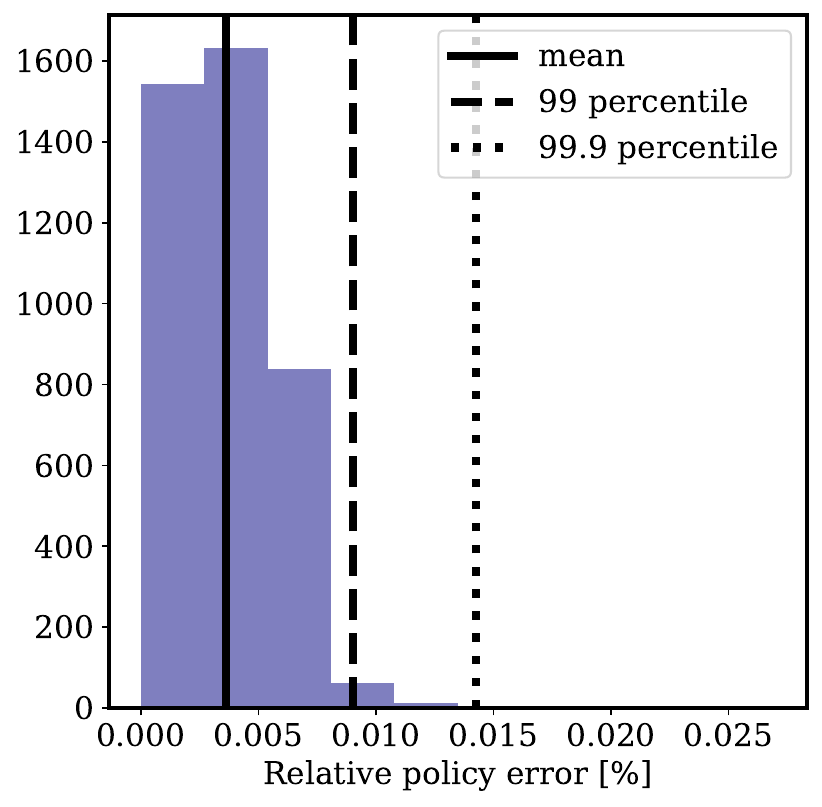}
    \caption{The left panel shows the distribution of errors in the optimality condition (equation \eqref{eq:ree_bm}) in \% on 4096 simulated states after training the neural network. The solid vertical line shows the mean error, the dashed vertical line shows the 99th percentile of errors and the vertical dotted line shows the 99.9th percentile of errors. 
    The middle panel compares the policy learned by the neural network (round dots) to the policy solved for with a conventional grid-based method. The right panel shows the distribution of errors when comparing the policy learned by the neural network to the policy solved with a conventional method.\label{fig:RA_acc}}
\end{figure}

We now turn to assessing the accuracy of our solution. First, we inspect the error in the equilibrium conditions on the simulated ergodic set of states.
The single equilibrium condition that characterizes the solution to the model is the Euler equation \eqref{eq:ree_bm}. The left panel in figure \ref{fig:RA_acc} shows the distribution of the absolute value of equilibrium condition error expressed in units of relative consumption errors. The vertical lines show the mean, the 99th percentile as well as the 99.9th percentile of the error distribution. As shown in the figure, the errors in the equilibrium conditions are low, with a mean error below $0.005\%$ and the 99.9th percentile of errors below $0.025\%$. 

The error in the equilibrium conditions is an implicit measure, and depending on the model at hand, there might not be an obvious mapping between the magnitude of this implicit error measure and magnitude of actual policy error. For a complicated model, equilibrium conditions error might be the only available error measure. However, in this simple model, we have access to a high-quality reference solution provided by a standard grid-based solution methods. With a dense enough grid, the present model can be solved to very high accuracy, such that we can regard the obtained solution as a good proxy to the \emph{true} policy function. Exploiting this classical solution, in figure \ref{fig:RA_acc}, we show a comparison of the policy learned by the neural network, to the policy computed with standard policy time iteration \citep[see, \emph{e.g.},][]{judd1998numerical}. 
The comparison shows two things: first, the difference between the two policies is small, the mean relative difference is below $0.005\%$, and the 99.9th percentile is below 0.015\%. This illustrates that the neural network is able to approximate the equilibrium policies to very high accuracy. Second, when comparing the \emph{explicit} policy error (right panel) to the \emph{implicit} error in the equilibrium conditions (left panel), we can observe that both errors are of the same magnitude. This indicates that the implicit error in the equilibrium conditions is a good proxy for the policy error. Although this is a reassuring finding, it is important to point out that this finding depends on the model, so we cannot generalize this conclusion to other models.

\section{Application to an OLG model\label{sec:OLG}}
In this section, we test the performance of our algorithm in a more challenging setup. We compute the approximate equilibrium in an overlapping generations economy featuring portfolio choice, non-trivial market clearing, and more complex stochastic processes for aggregate shocks. In particular, we model 72 overlapping generations of households who consume and save. Households allocate their savings into physical capital and risk-free bonds. While bonds are fully liquid, trading capital is subject to convex portfolio adjustment costs, and return to capital is exposed to aggregate risk.
There are three sources of uncertainty: as in the \cite{brock_1972} model, the logarithm of TFP follows an AR(1) process.
In addition, the economy switches between two regimes: normal times and disaster times. In disaster times, the depreciation to capital is stochastic and on average higher than during normal times. The regime process evolves according to a Markov transition matrix. Since capital is predominantly held by older households, the disaster leads to swings in the intergenerational wealth distribution.

We train the neural network to predict equilibrium prices and policies as a function of truncated histories of aggregate shocks. In this economy, our algorithm keeps track of the last $T$ realizations of each of the aggregate shocks, which include $T$ innovations to the logarithm of TFP, $T$ innovations to the depreciation of capital, and the last $T$ regime realizations. The purpose of this section is to demonstrate that our sequence space method is capable of obtaining highly-accurate approximations to equilibria in economies driven by multivariate stochastic processes. The overlapping generations setup furthermore generates an obvious long-range dependence of current outcomes on the realized sequence of past shocks.

\subsection{Model}
\subsubsection{Technology}
A representative firm operates a Cobb-Douglas technology and produces a perishable consumption good using labor and capital
\begin{align}
Y_t = A_t K_t^\alpha L_t^{1 - \alpha}.
\end{align}
$A_t$ denotes the stochastic total factor productivity (TFP). The logarithm of TFP evolves according to an AR(1) process
\begin{align}
\log(A_t) = \rho \log(A_{t-1}) + \sigma^A \epsilon_t^A,~\epsilon_t^A\sim \mathcal{N}\left(0, 1\right).
\end{align}
As in the stochastic growth economy of \cite{brock_1972}, the firm faces competitive input and output markets. We choose the consumption good as the num\'eraire, hence price of consumption is normalized to unity. We denote the wage by $w_t$ and rental rate of capital by $r^K_t$.

\subsubsection{Demographics}
Each period a new representative cohort of mass 1 enters the economy and lives for $H$ periods. We denote the age of a cohort by $h \in \{1, \dots, H\}$. Households are endowed with $l^h$ efficiency units of labor, which they supply inelastically at equilibrium wage. The labor endowment is normalized such that $L = \sum_{h=1}^H l^h = 1$. The age dependence of efficient labor supply serves as a device to model a life-cycle profile of household labor income. We abstract from bequest motives, taxes and social security.

\subsubsection{Preferences}
Households have time separable expected utility over consumption streams. We assume that the felicity function takes the constant relative risk aversion form with coefficient $\gamma$. Hence, households rank stochastic consumption stream according to
\begin{align}
\E{\sum_{i = 0}^{H - h} \beta^i  u(c_{t+i}^{h+i})},
\end{align}
where $\beta$ denotes their patience.

\subsubsection{Asset markets}
Households are able to re-allocate their consumption across time and states by trading in two assets. Two assets and infinitely many possible shock realizations next period\footnote{We assume innovations to the logarithm of TFP and capital depreciation to be Gaussian.} imply an incomplete-markets environment. Furthermore, there are two additional financial frictions. First, households are subject to short sale limits on capital and bonds. Second, while the bond is a liquid asset, capital is an illiquid asset and households have to pay adjustment costs when adjusting their capital holdings. 
The adjustment costs are given by
$\psi(k_{t+1}^{h+1}, k_t^h) = \xi^{\text{adj}}\left( k_{t+1}^{h+1} - k_t^h \right)^2$.
Short sale constraints on both assets are specified as exogenous limits $b_{t+1}^{h+1}\geq \underline{b}$ and $k_{t+1}^{h+1}\geq \underline{k}$.
The net supply of bonds is fixed at $B$. Bonds can be purchased or sold at equilibrium price $p_t$, and they deliver a risk-free payoff of 1 in period $t+1$. Purchasing a unit of capital in period $t$ promises a risky payout of $1 - \delta_{t+1} + r_{t+1}^K$ in period $t+1$. There are two sources of risk in returns to capital: first, the rental rate of capital $r_{t+1}^K$ depends on the total factor productivity $A_{t+1}$ in period $t+1$. Second, the depreciation rate of capital in period $t+1$ is stochastic as well and given by
\begin{align}
\delta_{t+1} &= \delta \frac{2}{1+D_{t+1}}
\text{, where  } \log(D_{t+1}) = 
\begin{cases}
\rho^{\delta} \log(D_{t}) & \text{normal times in $t+1$} \\
\mu^{\delta} + \rho^{\delta} \log(D_{t}) + \sigma^\delta \epsilon_{t+1}^\delta
 & \text{disaster in $t+1$}
\end{cases}
\end{align}
with $\mu^\delta<0$ and $\epsilon_{t+1}^\delta \sim \mathcal{N}(0, 1)$.
Normal and disaster times are modeled as two discrete Markov regimes with transition matrix $\Pi^{\text{regime}}=\begin{bmatrix}
 \pi^{n\rightarrow n} & 1 - \pi^{n\rightarrow n} \\
 1 - \pi^{d \rightarrow d} & \pi^{d \rightarrow d}
\end{bmatrix}$, where $\pi^{n \rightarrow n}$ and $\pi^{d \rightarrow d}$ denote the persistence of the normal and disaster regimes, respectively.

\subsubsection{Household problem}
Recursively formulated, the households' problem is characterized by the following Bellman equation
\begin{align}
V_t^h &= \max_{k_{t+1}^{h+1}, b_{t+1}^{h+1}} u(c_t^h) + \beta \E{V_{t+1}^{h+1}} \label{eq:OLG_Bellman}\\
\text{subject to}:&  \nonumber \\
c_t^h &= \underbrace{l^h w_t}_{\text{lab. inc.}} + \underbrace{b_t^h + (1 - \delta_t + r_t^K) k_t^h}_{\text{payout of assets}} - \underbrace{(p_t b_{t+1}^{h+1} + k_{t+1}^{h+1})}_{\text{savings}} - \underbrace{\psi(k_{t+1}^{h+1}, k_t^h)}_{\text{adj. costs}} \nonumber \\
\underline{b} &\leq b_{t+1}^{h+1}  \nonumber \\
\underline{k} &\leq k_{t+1}^{h+1}.  \nonumber 
\end{align}
In every period, households receive labor income and the payoff of their asset holdings, and decide how much to consume and how much to invest in each of the two assets, subject to adjustment costs and short-sale constraints.
The time index denotes the dependence on the aggregate state variables. 

The solution to the household problem is characterized by a set of Karush-Kuhn-Tucker (KKT) conditions. For each age-group $h=1,\dots, H-1$ and each asset, we have two sets of KKT conditions
\begin{align}
p_t u'(c_t^h) &= \beta\E{u'(c_{t+1}^{h+1})} + \lambda_t^{b, h} \\
0 &= \lambda_t^{b, h}(b_{t+1}^{h+1} - \underline{b}) \\
0 &\leq (b_{t+1}^{h+1} - \underline{b}) \\
0 &\leq \lambda_t^{b, h} \\
\left(1 + \frac{\partial}{\partial k_{t+1}^{h+1}}\psi(k_{t+1}^{h+1}, k_t^h)\right) u'(c_t^h) &= \beta\E{u'(c_{t+1}^{h+1})\left(1 - \delta_{t+1} + r_{t+1}^ k -\frac{\partial}{\partial k_{t+1}^{h+1}}\psi(k_{t+2}^{h+2}, k_{t+1}^{h+1})\right)} + \lambda_t^{k, h} \\
0 &= \lambda_t^{k, h}(k_{t+1}^{h+1} - \underline{k}) \\
0 &\leq (k_{t+1}^{h+1} - \underline{k}) \\
0 &\leq \lambda_t^{k, h}. 
\end{align}
For each age-group and each asset we use the Fischer-Burmeister function \citep[see, ][]{jiang_1999} $\psi^{FB}(a, b) := a-b - \sqrt{a^2 + b^2}$ to collapse the set of KKT conditions into a single equation. The above KKT conditions are fully characterized by
\begin{align}
0&=\psi^{FB}\left(p_t u'(c_t^h) - \beta\E{u'(c_{t+1}^{h+1})}, b_{t+1}^{h+1} - \underline{b}\right) \\
0&=\psi^{FB}\left(\left(1 + \frac{\partial}{\partial k_{t+1}^{h+1}}\psi(k_{t+1}^{h+1}, k_t^h)\right) u'(c_t^h) \right. \nonumber \\
&\left.~~~- \beta\E{u'(c_{t+1}^{h+1})\left(1 - \delta_{t+1} + r_{t+1}^ k -\frac{\partial}{\partial k_{t+1}^{h+1}}\psi(k_{t+2}^{h+2}, k_{t+1}^{h+1})\right)}, (k_{t+1}^{h+1} - \underline{k})\right)
\end{align}
Analogously to equation \eqref{eq:Rep:relconserr} for the representative agent model, we rearrange the equations so that the deviations from zero are expressed in units of relative consumption error. 

\subsubsection{Equilibrium}
Now, we define the recursive equilibrium of the economy
\paragraph{State space}
The state of the economy 
\begin{align}
 \Xst_t = [\chi_t, A_t, \delta_t,\mu_t]\in \{0, 1\} \times \mathbb{R}^+ \times \mathbb{R}^+\times \mathbb{R}^{H-1 + H - 2} \label{eq:OLGstate}   
\end{align}
is given by the regime indicator, $\chi_t$, total factor productivity $A_t$, capital depreciation $\delta_t$, and the distribution of assets across the age groups $\mu_t = \{(b_t^h, k_t^h)\}_{h = 1}^H$.\footnote{Strictly speaking, because we assume that households enter the economy without assets, and because of market clearing, a sufficient statistic for the asset distribution consists of $H-1$ free parameters for capital and $H-2$ for the bond, because the bond supply is fixed.} 
\paragraph{Functional rational expectations equilibrium}
The functional rational expectations equilibrium of our economy consists of $H$ policy functions $\pi^{k, h}(\Xst_t) = k_{t+1}^{h+1}$ for investment in capital, $H$ policy functions $\pi^{b, h}(\Xst_t) = b_{t+1}^{h+1}$ for investment in the bond, and a price function $\pi^p(\Xst_t) = p_t$, such that the KKT conditions hold and the bond and capital markets clear.\footnote{We use the first-order conditions of representative firm directly to compute the equilibrium wage and rental rate as a function of the state of the economy. The goods market clears by Walras' law.} The aim of our algorithm is to compute approximations to these equilibrium functions.

\subsection{Parameterization}
We set the model parameters to standard values in the literature. One model period corresponds to one year, and we model $H  = 72$ overlapping generations. The model ages $h = 1$ to $72$ correspond to adult life from $21$ years to $92$ years. 
The age dependent efficient units of labor are chosen to generate a life-cycle profile of labor income and are shown in figure \ref{fig:OLG_laborendowment}.
We set the coefficient of relative risk aversion at $\gamma = 2$ and patience to $\beta = 0.96$. We assume strict borrowing constraints with $\underline{b}=\underline{k}= 0$. The adjustment costs for capital are set to $\xi^{\text{adj}}=0.1$. The persistence of the logarithm of TFP is set to $\rho=0.85$, and the standard deviation of innovations to the logarithm of TFP is set to $\sigma^A=0.03$. The long-run depreciation of capital in normal times is given by $\delta = 0.1$. The parameters governing the depreciation of capital in the disaster regime are given by $\rho^\delta=0$, $\sigma^\delta=0.2$ and $\mu^\delta  = -1.10$, implying 50\% higher depreciation in the disaster when $\log(D_t) = \mu^\delta$. The per-period probability to enter into the disaster state is 6\% and the per-period probability to exit the disaster state is 33.33\%, implying $\pi^{n\rightarrow n} = 0.94$ and $\pi^{d \rightarrow d} = 2/3$. 
The parameter values are summarized in Table \ref{tab:OLG_params}.

\subsection{Implementation}
We use a densely-connected feed-forward neural network to parameterize a key set of functions and then obtain the remaining equilibrium objects in closed form. These key functions consist of the bond price, $p_t$, as well as the investment in capital, $k_{t+1}^{h+1}$, and investment in the bond, $b_{t+1}^{h+1}$, for each age-group.\footnote{To be precise, we only need to predict the asset choice for $H-1$ age groups, since, for our calibration, it is always optimal to consume everything in the last period of life. Similarly to our approach in the \cite{brock_1972} model, the neural network predicts the share of cash-at-hand invested in each of the two assets, which we then use to construct the prediction for the asset choices.}
We use the market clearing layers approach \citep[as in][]{azinoviczemlicka_2024}, so the neural network architecture ensures that the predicted policies are always consistent with bond market clearing.

Following our sequence space approach, the neural network predicts the policy and price functions as a function of truncated history of aggregate shocks. Hence, the input to the neural network is given by 
\begin{align}
\Xse_t = [
\underbrace{\chi_{t-(T - 1)}, \dots, \chi_{t-1}, \chi_{t}}_{T \text{ last regime indicators}}, \underbrace{\epsilon^A_{t-(T - 1)}, \dots, \epsilon^A_{t-1}, \epsilon^A_t}_{T\text{ last innovations to TFP}}, \underbrace{ \epsilon^\delta_{t-(T - 1)}, \dots, \epsilon^\delta_{t-1}, \epsilon^\delta_{t}}_{T \text{ last innovations to deprec.}}
] \in \mathbb{R}^{3T}.
\end{align}
As before, we train the neural network to minimize the errors in all equilibrium conditions. The resulting loss function consists of the $2\times (H-1)$ optimality conditions for the capital and bond policies for each age group, except the last who consumes everything. 
As in the previous example, the training data in episode $j$ are given by state-history pairs $\mathcal{D}_j = \{(\Xst_{j, i}, \Xse_{j, i})\}_{i = 1}^{N^{\text{data}}}$ and use the Adam optimizer \citep[see][]{kingma2014adam}. After each training episode we simulate states and histories forward to generate a new training dataset $\mathcal{D}_{j+1}$. We approximate the expectations over the continuous shocks $\epsilon^A_{t+1}$ and $\epsilon^\delta_{t+1}$ using Gauss-Hermite quadrature.

\subsection{Training}
Our training procedure follows the stepwise approach to solve models with multiple assets outlined in \cite{azinoviczemlicka_2024}. We proceed along four steps:

First, we set the net supply of bonds to zero. Given a strict no-short sale constraint the market clearing condition implies that all equilibrium bond holdings are zero. Hence we can first restrict our attention to solving for capital savings policy functions while keeping bond savings fixed at zero.

Second, we train the neural network price function to learn the bond price implied by consumption allocations of the capital-only economy. We plug the set of consumption functions obtained by solving the capital-only economy into bond Euler equations. We invert the Euler equations to recover implied bond prices. The largest of the prices would be the price of the first $\epsilon$ of bond supply introduced into this economy. We train the price network to replicate this price as a function of truncated histories of aggregate shocks.

Third, the previous two steps provide us with a good initial guess for the policy and price functions in an economy with a comparatively small bond supply. We now proceed to slowly increase the bond supply. After every small increase in the bond supply, we retrain the neural network using the previous solution as a starting point. We proceed until the full bond supply is reached.

Fourth, after reaching the full supply, we continue training the neural network on the final model specifications until we reach desired level of accuracy.
The hyperparameters for our implementations are summarized in table \ref{tab:OLG_hp}.

\subsection{Accuracy}
\begin{figure}
    \centering
    \includegraphics[width=0.33\linewidth]{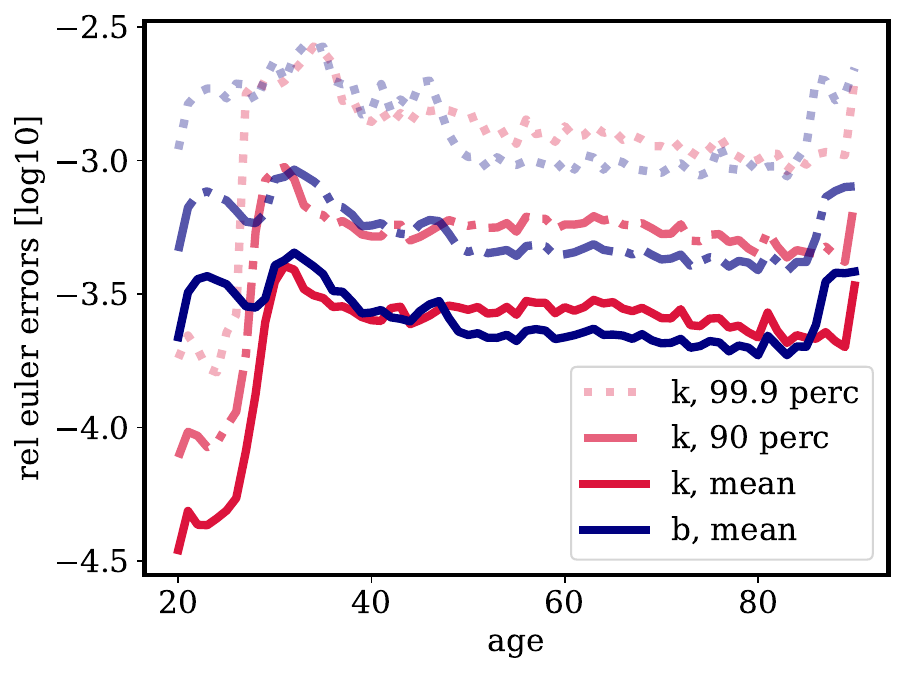}
    \includegraphics[width=0.31\linewidth]{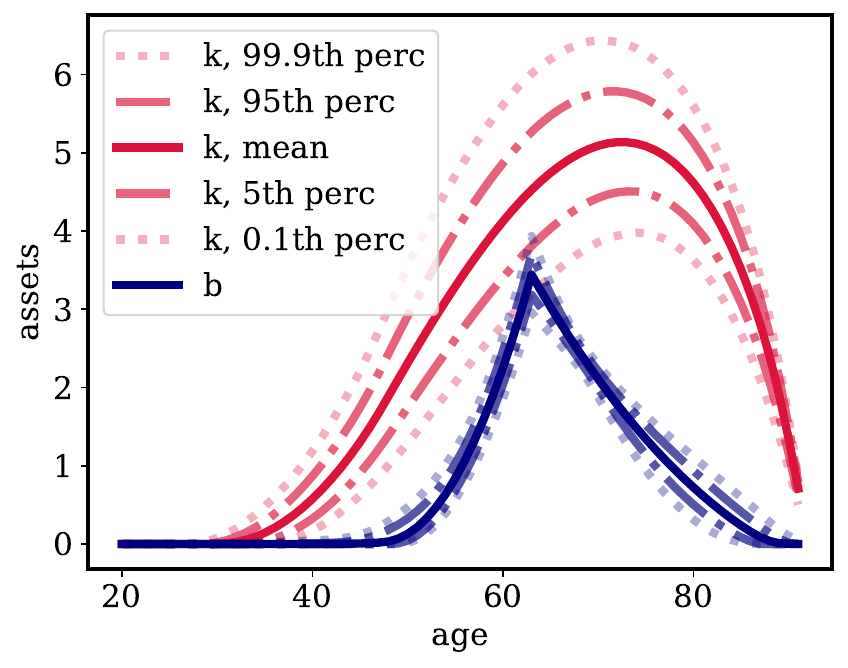}
    \includegraphics[width=0.32\linewidth]{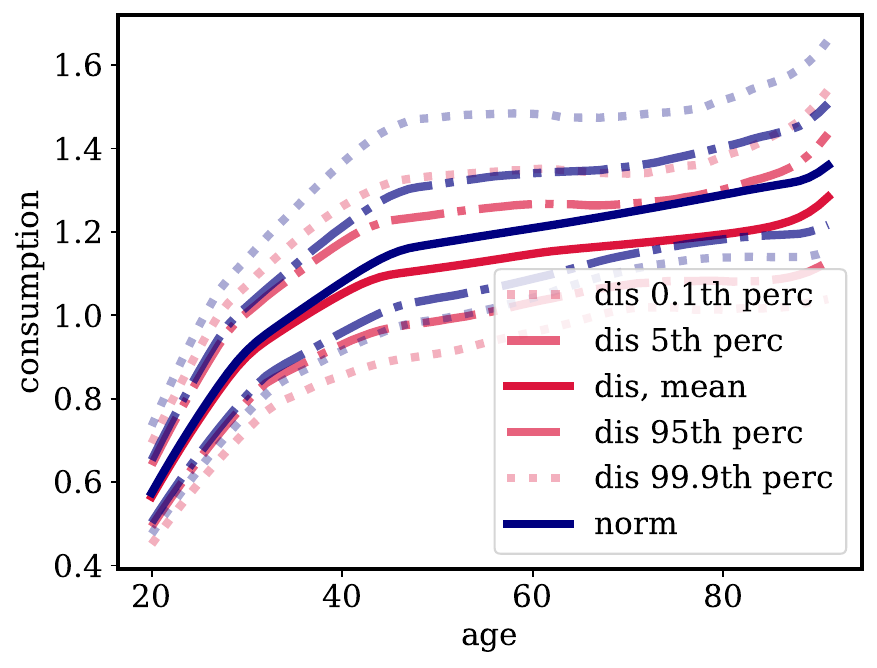}
    \caption{The left panel shows the final level of errors in the equilibrium conditions achieved by our training procedure. The dotted line shows the 99.9th percentile, the dash-dotted line the 90th percentile, and the solid line shows the mean. The red lines show the errors in the optimality conditions for capital for each age group, and the blue lines show the errors in the optimality conditions for the bonds. The middle panel shows the distribution of assets over the simulated ergodic set of states. The right panel shows the resulting consumption by age group. The blue lines in the right panel show consumption conditional on the economy being in the normal state, and the red lines show consumption statistics conditional on the economy being in the disaster state.\label{fig:OLG_finalerrors}}
\end{figure}

We assess the accuracy of our neural network solution by investigating the errors in the equilibrium conditions over the simulated ergodic set.
Figure \ref{fig:OLG_finalerrors} shows the errors in the equilibrium conditions for each age group along the simulated paths of the economy. The errors are expressed in units of relative consumption errors.
The model is solved accurately, the mean and 90th percentile of errors are below 0.1\% and the 99.9th percentile below 0.32\% for both assets and all the age groups. The mean error is around 0.03\%.

In the middle panel in figure \ref{fig:OLG_finalerrors} we show the distribution of asset holdings over the age groups. The savings behavior of households is driven by the life-cycle forces. Facing an increasing labor income profile, young households are borrowing constrained. Later in life, they start to accumulate savings to make sure they are able to finance their later-age consumption after the drop in their labor endowment after the age of $62$.\footnote{While we abstract from modeling retirement explicitly, the reduction in labor efficiency units around the age of $60$ serves as a proxy for the income decline associated with retirement.}

The right panel in figure \ref{fig:OLG_finalerrors} shows the distribution of consumption across the age groups, conditional on the regime of the economy. The disaster, modeled as an increase in the capital depreciation and uncertainty, leads to the largest consumption decline for households around the age 80.

\section{Application to heterogeneous households and heterogeneous firms\label{sec:HA}}
In this section, we apply our algorithm to compute an approximate equilibrium in an economy with heterogeneous households and heterogeneous firms. The households and the firms are both subject to uninsurable idiosyncratic risk, as well as to aggregate productivity shocks, and swings in the level of aggregate and idiosyncratic volatility.
Firms operate a decreasing returns to scale technology that produces the final good using capital and labor. The firms own their capital, hire workers on a Walrasian labor market, and pay dividends to their shareholders. Firms make their decisions to maximize the present discounted value of their dividends. We assume that firms use a weighted average of households' marginal rates of substitution to discount future cash flows. The weighting is proportional to households' stock holdings. Households supply labor inelastically at equilibrium wage, consume, and invest in the aggregate stock market index. Besides the fluctuations in wages and stock prices induced by aggregate shocks, households are subject to uninsurable idiosyncratic labor endowment shocks.   
The state space of the economy includes two endogenously moving distributions: the firm distribution over capital and idiosyncratic firm productivity, as well as the household distribution over stock holdings and idiosyncratic labor productivity. The distribution of firms is relevant to households because it determines the dividends paid on stock holdings, and the wage rate on supplied labor. In turn, the stochastic discount factor used by firms to make their investment decisions depends on the distribution of households.

\subsection{Model}
\subsubsection{Firms and technology}
Time is discrete and infinite, $t=0,1,\dots $. There is a continuum of firms that face idiosyncratic and aggregate risk featuring stochastic volatility in the spirit of \cite{bloometal_2018}.
At the beginning of each period, there is a continuum of firms of mass 1. We assume that only a fraction $\Gamma$ of firms survive to the end of the period and produce the consumption good. A fraction $(1-\Gamma)$ of firms exits the economy, and is replaced by an equal measure of start-ups that, conditional on surviving, start to produce in the next period.\footnote{Firm exit is modeled as in \cite{azinoviccolekubler_2023}. We assume that the capital of exiting firms is destroyed in the process.} 
Each of the firms $i\in[0,\Gamma]$ operates a decreasing returns to scale production technology
\begin{align}
y^i_{t} = A_t z^i_{t}(k_t^i)^{\alpha}(l_t^i)^{1-\alpha-\zeta},
\end{align}
where $A_t$ denotes stochastic aggregate productivity, which affects all firms in the economy, $z^i_t$ denotes the realization of firm specific productivity. The parameter $\alpha$ determines the elasticity of the output with respect to capital, and the parameter $\zeta$ controls the degree of returns to scale.\footnote{The capital share is $\alpha/(1-\zeta)$.}
Aggregate productivity follows an AR(1) process
\begin{align}
\log(A_t) = \rho^A \log(A_{t-1}) + \sigma^A_{t-1} \epsilon_t^A,
\end{align}
where $\epsilon_t^A\sim \mathcal{N}(0, 1)$. The volatility of innovations to aggregate TFP, $\sigma^A_t = \sigma(U_t)\in \{\sigma^A_L, \sigma^A_H\}$, takes two values depending on the uncertainty regime $U_t\in\{L, H\}$. The uncertainty regime follows a Markov chain as in \cite{bloometal_2018}. We denote the transition probabilities between the uncertainty regimes with $\pi^U_{L, H}$, $\pi^U_{L, L} = 1 - \pi^U_{L, H}$, $\pi^U_{H, L}$, and $\pi^U_{H,H} = 1 - \pi^U_{H, L}$, respectively. The firm-level idiosyncratic productivity follows a three-state Markov process $z_t \in \{z_1, z_2, z_3\}$, with $z_2 = (z_1 + z_3)/2$. 
The transition probabilities vary with the uncertainty regime: $\Pi^z_t\in\{\Pi^z_L,\Pi^z_H\}$.
We are interested in the effect of a pure uncertainty shock, \emph{i.e.} in an increase in the variance of future idiosyncratic productivity without changing the current productivity or the expected future productivity level $\E{z_{t+1}^i|z_t}$.
To do so, assume that the transition matrices in the low and high uncertainty regimes are given by
\begin{align}
\Pi^z_L = \begin{bmatrix}
\pi^z_{11} & \pi^z_{12} & \pi^z_{13} \\
\pi^z_{21} & \pi^z_{22} & \pi^z_{23} \\
\pi^z_{31} & \pi^z_{32} & \pi^z_{33} 
\end{bmatrix}, 
\Pi^z_H = \begin{bmatrix}
\pi^z_{11} + u & \pi^z_{12} - 2u& \pi^z_{13}+u\\
\pi^z_{21} + u & \pi^z_{22} - 2u& \pi^z_{23}+u \\
\pi^z_{31} + u & \pi^z_{32} - 2u& \pi^z_{33}+u 
\end{bmatrix},
\end{align}
where $0\leq u<\min\{\frac{1}{2}\pi^z_{12}, \frac{1}{2}\pi^z_{22}, \frac{1}{2}\pi^z_{13}\}$ parameterizes the increase in the idiosyncratic uncertainty. This formulation allows us to keep the idiosyncratic productivity level $z_t^i$, as well as the expected future productivity $\E{z_{t+1}^i|z_t}$ constant, when the economy transitions into the high volatility regime.\footnote{To the best of our knowledge, maintaining these two features would not easily be possible using, for example, a \cite{rouwenhorst_1995} or \cite{tauchen1986statistical} discretization methods for an AR(1) process.}
Firms accumulate capital subject to capital adjustment costs of the form
\begin{align}
\psi(k', k) = \frac{\xi(k', k)}{2}k\left(\frac{k'}{k}-1\right)^2,
\end{align}
where
\begin{align}
\xi(k', k) = 
\begin{cases}
\xi^{\text{up}} & \text{for }k'>k \\
\xi^{\text{down}} & \text{for }k'\leq k.
\end{cases}
\end{align}
We assume that $\xi^{\text{down}} > \xi^{\text{up}}$, so reducing the capital stock is associated with larger adjustment costs than increasing the capital stock by the same percentage.
Although the adjustment costs parameter changes discontinuously at $k' = k$, $\psi(k', k)$ is not only continuous but also differentiable. At $k'=k$ we have $\psi(k', k) = \frac{\partial \psi(k', k) }{\partial k'} = \frac{\partial \psi(k', k) }{\partial k} = 0$.\footnote{$\xi(k', k)$ is not continuous at $k' = k$, but the quadratic term $(k'/k- 1)^2$ ensures that $\psi(k', k)$ remains differentiable.
Additionally, we smooth out the jump in the adjustment cost parameter such that $
\xi(k', k) = w^{\text{up}}\xi^{\text{up}} + (1-w^{\text{up}})\xi^{\text{down}}$, where $w^{\text{up}} = \text{sigmoid}\left(s \frac{k' - k}{k}\right)$, and where $s>0$ denotes a smoothing parameter. For large $s$, $w^{\text{up}}$ converges to a step function, \emph{i.e.} $w^{\text{up}}\approx 1$ for $s \frac{k' - k}{k}>>0$ and $w^{\text{up}}\approx 0$ for $s \frac{k' - k}{k}<<0$.}

Each firm sets its investment and dividend payout policies to maximize its objective function: the present discounted value of its dividends, subject to a budget constraint, and a non-negative dividends constraint. The firm problem is characterized by the following Bellman equation:
\begin{align}
v_t(z_t^j, k_t^j) &= \max_{k_{t+1}^j\geq 0} d^j_t + \Gamma\,\Eind{t}{\Lambda_{t+1,t}\, v_{t+1}(z_{t+1}^j, k_{t+1}^j)} \\
\text{subject to:}&\nonumber\\
d_t^j &= y_t^j - w_t l_t^j - i_t^j - \psi(k^j_{t+1}, k^j_t)\\
d_t^j &>\underline{d}.
\end{align}
$\Gamma$ denotes the mass of surviving firms, which is equal to the (idiosyncratically independent) survival probability of an individual firm between $t$ and $t+1$.
The time index of the value function stands for the dependence of the firm problem on the aggregate state of the economy. $\Lambda_{t+1, t}$ denotes the aggregate stochastic discount factor given by the distribution of shareholders.
The policy functions of the firms are characterized by a set of KKT conditions. 
\begin{align}
\left(1 + \frac{\partial}{\partial k_{t+1}^i}\psi(k_{t+1}^i, k_t^i)\right) (1 + \lambda^i_t) &= \Gamma \E{\Lambda_{t+1, t}\left(1 + r_{t+1}^{i, k} -  \frac{\partial}{\partial k_{t+1}^i}\psi(k_{t+2}^i, k_{t+1}^i)  \right)(1 + \lambda_{t+1}^i)} \nonumber \\
\Leftrightarrow \lambda_t^i &= \frac{\Gamma \E{\Lambda_{t+1, t}\left(1 + r_{t+1}^{i, k} -  \frac{\partial}{\partial k_{t+1}^i}\psi(k_{t+2}^i, k_{t+1}^i)  \right)(1 + \lambda_{t+1}^i)}}{\left(1 + \frac{\partial}{\partial k_{t+1}^i}\psi(k_{t+1}^i, k_t^i)\right)} - 1\label{eq:firm_foc_lambd}\\
0 &\leq (d_t^i - \underbar{d}) \label{eq:firm_foc_constr}\\
0 &\leq \lambda_t^i\\
0 &= (d_t^i- \underbar{d}) \lambda_t^i, \label{eq:firm_foc_compl}
\end{align}
where $\lambda^i_t$ denotes the Lagrange multiplier on the non-negative dividend constraint and where $r^{i, k}_t$ denotes the marginal product of capital installed in the firm
\begin{align}
r_t^{i, k} = A_t z^i_t \alpha (k_t^i)^{\alpha - 1} (l^i_t)^{1-\alpha-\zeta} - \delta. 
\end{align}
Using the Fischer-Burmeister function, we rewrite the KKT conditions of the firm problem as a single equation $\psi^{FB}(\lambda_t^i, d_t^i - \underbar{d} ) = 0$, where $\lambda_t^i$ is given in equation \eqref{eq:firm_foc_lambd}.

\paragraph{Aggregation of firms}
Firms hire efficiency units of labor on a competitive spot market. There is a single equilibrium wage $w_t$ per efficiency unit of labor, which is paid by all the firms. The labor demand of an individual firm is given by a static first-order condition
\begin{align}
w_t &= (1-\alpha - \zeta) \frac{y_t^i}{l_t^i} = (1-\alpha - \zeta) A_t z_t^i(k_t^i)^{\alpha}(l_t^i)^{-\alpha-\zeta} \\
\Leftrightarrow l_t^i &= \left(A_t z_t^i \frac{1-\alpha - \zeta}{w_t}(k_t^i)^\alpha\right)^{\frac{1}{\alpha + \zeta}}.\label{eq:firms_labor}
\end{align}
Let $\mu^F_t(z^i, k^i)$ denote the distribution of surviving firms over idiosyncratic productivity and capital. The overall mass of surviving firms is given by $\Gamma = \int_{i}\mu^F_t(z^i, k^i)di $.
Since we assume that all households supply their labor inelastically, the aggregate labor supply is constant at $L= 1$. This allows us to solve for the equilibrium wage as a function of the productivity level $A_t$, and the cross-sectional distribution of firms. We obtain
\begin{align}
w_t &= A_t(1 - \alpha - \zeta)  \left( \int \left(z^i (k^i)^{\alpha}\right)^{\frac{1}{\alpha+\zeta}}\mu^F_t(z^i, k^i) di \right)^{\alpha + \zeta}\label{eq:firms_wage}.
\end{align}
Given the expression for the wage \eqref{eq:firms_wage}, the labor demand is given in a closed form, as described in equation \eqref{eq:firms_labor}. Note that, because of decreasing returns to scale, the full firm distribution matters for the determination of the aggregate wage.
Similarly, aggregate dividends are given by
\begin{align}
D_t = \int d_t(z^i, k^i) \mu_t(z^i, k^i)di.
\end{align}

\paragraph{Startups}
In each period, households trade shares in the aggregate index of existing firms. Owning a share entitles a household to receive a fraction of the aggregate dividend stream. The firms that exist in period $t$ include a measure $\Gamma$ of surviving firms that produce in period $t$, as well as a measure $(1 - \Gamma)$ of period $t$ startups, which start producing only in period $t+1$.
Holding $\theta_t^i$ shares, which have been purchased in period $t-1$, entitles households to a payout of $\theta_t^i(\Gamma p_t + D_t)$. This is because period $t$ startups can only be traded once they exist. The stocks traded in period $t-1$ therefore only make up fraction $\Gamma$ of the stocks traded in period $t$.

A measure $(1-\Gamma)$ of startups replaces exiting firms every period. The average value of period $t$ startups is $(1-\Gamma) p_t$. We set the initial capital of startups to be equal to the capital of the firms they are replacing.
The required investment is provided uniformly by all the households who therefore also own the startups. The initial investment is not subject to adjustment costs. The size of startup investment per household is given by
\begin{align}
I^{\text{su}}_t &= (1-\Gamma)\int k^i_{t+1}(z_t^i, k_t^i) \frac{\mu(z_t^i, k_t^i)}{\Gamma}\mathrm{d}i.
\end{align}
The aggregate rents from startup creation are given by 
\begin{align}
\Pi^{\text{su}}_t= (1 - \Gamma)p_t - I^{\text{su}}_t. 
\end{align}
The rents from startup creation are equal to the difference between the average capital stock in the economy and the price of an equity share.
We assume that the profits from startup creation are equally distributed to all households.

\subsubsection{Households}
There is a unit-mass continuum of infinitely lived households. Households have time separable expected utility preferences over stochastic consumption streams
\begin{align}
U\left(\{c_t\}_{t=0}^{\infty}\right)=\E{\sum_{t=0}^{\infty}\beta^tu(c_t)}\text{, where }u(c)=\frac{c^{1-\gamma}}{1-\gamma},
\end{align}
and where $\beta$ denotes the patience parameter, and $\gamma$ denotes the coefficient of relative risk aversion. Households' labor endowment is stochastic and follows an AR(1) process
\begin{align}
\log(e_t^i) = \rho^e \log(e_{t-1}^i)+\sigma^e\epsilon_t^{e, i}\text{, where } \epsilon_t^{e,i}\sim \mathcal{N}(0, 1).
\end{align}
We discretize the idiosyncratic labor endowment process into a two-state Markov process using the \cite{rouwenhorst_1995} algorithm. Households supply their efficient units of labor inelastically at equilibrium wage. While households can save by purchasing equity shares $\theta_t^i$, their borrowing is subject to a short-sale limit  $\theta_{t+1}^i \geq \underline{\theta}$. The budget constraint of household $i$ reads as
\begin{align}
c_t^i = e_t^i w_t + \theta_t^i(D_t + \Gamma p_t) -\theta_{t+1}^ip_t+\Pi^{\text{SU}}_t.
\end{align}
The solution to the household savings problem is characterized by the following set of KKT conditions 
\begin{align}
p_t u'(c^i_t) &= \beta \E{u'(c^i_{t+1})(D_{t+1} + \Gamma p_{t+1})} + \lambda_t^i \\
0&\leq \lambda^i_t \\
0&\leq (\theta_{t+1}^i - \underline{\theta}) \\
0&=(\theta_{t+1}^i - \underline{\theta})\lambda^i_t.
\end{align}
Households are exposed to idiosyncratic risk because of their labor endowments and to aggregate risk via the wage, dividends, the stock price, and the profits from startup creation. Because of that, aggregate shocks lead to fluctuations in the wealth distribution. We denote the household distribution by $\mu^H_t(e^i, \theta^i)$, where $e$ denotes idiosyncratic labor endowment, and $\theta$ share holdings.
\subsubsection{Asset markets}
Households trade claims on the aggregate dividend stream of \emph{existing} firms. The aggregate asset demand of households is given by
\begin{align}
\Theta_{t+1}^H = \int \theta^i_{t+1}(e^i, \theta^i)\mu^H(e^i, \theta^i) di.
\end{align}
Market clearing requires that $\Theta^H_{t+1}= 1$.
Firms take the stochastic discount factor as given. We assume that firms use the intertemporal marginal rate of substitution of shareholder households weighted by their corresponding share holdings as the stochastic discount factor.
For a specific realization of aggregate shocks in period $t+1$ and household $i$ with idiosyncratic labor endowment $e^i$ and shares $\theta^i$, we define its intertemporal marginal rate of substitution as
\begin{align}
\Lambda^i_{t+1, t}(e^i, \theta^i) = \frac{\beta \Eind{e^i_{t+1}}{u'(c^i_{t+1})}}{u'(c^i_t)},
\end{align}
where the expectations operator is taken over the realizations of idiosyncratic labor endowment. The aggregate stochastic discount factor is given by
\begin{align}
\Lambda_{t+1, t}=\frac{\int\Lambda^i_{t+1, t}(e^i, \theta^i) \theta_{t+1}(e^i, \theta^i) \mu^H_t(e^i, \theta^i) di}{\int \theta_{t+1}(e^i, \theta^i) \mu^H_t(e^i, \theta^i) di}.
\end{align}

\subsubsection{Equilibrium}
\paragraph{State space}
The aggregate state of the economy is given by
\begin{align}
\Xstagg_t=[\chi_t, A_t, \mu_t^F,\mu_t^H] \in \{0, 1\}\times \mathbb{R}_+ \times \mathbb{R}_{+}^{N_z \times N_k}\times \mathbb{R}_{+}^{N_e \times N_\theta} ,
\end{align}
where $\chi_t$ is an indicator denoting the uncertainty regime, $A_t$ denotes the aggregate productivity level, and $\mu_t^H$ and $\mu_t^F$ denote the distribution of households and firms. We represent the distributions using a finite histogram approximation \citep[see][]{young2010solving} with $N_z \times N_k$ and $N_e \times N_\theta$ histogram points for firms and households. $N_z$ and $N_e$ denote the number of histogram points along the exogenous idiosyncratic productivity dimension and $N_k$ and $N_\theta$ denote the number of histogram points for the capital and asset holdings respectively.

\paragraph{Functional rational expectations equilibrium}
The functional rational expectations equilibrium consists of firm policy functions, $\pi^F(\Xstagg_t, z^i_t, k^i_t) = k_{t+1}^i$, household policy function, $\pi^H(\Xstagg_t, e^i_t, \theta^i_t) = \theta_{t+1}^i$, as well as a price function $\pi^q(\Xstagg_t) = p_t$, consistent with optimizing households, optimizing firms, and market clearing.

\subsection{Parameterization}
One model period corresponds to one calendar year. We set patience to $\beta = 0.95$ and the coefficient of relative risk aversion to $\gamma = 2$. For the idiosyncratic labor process we choose $\rho^e = 0.871$ and $\sigma^e = 0.246$, as in \cite{auclertetal_2021}.\footnote{We convert the quarterly process in \cite{auclertetal_2021} into a yearly process.} We use the \cite{rouwenhorst_1995} method to discretize the labor process into a two state Markov chain with states $e_0=0.538, e_1=1.462$ and transition matrix
\begin{align}
\pi^e = \begin{bmatrix}
0.935 & 0.065\\
0.065 & 0.935
\end{bmatrix}
\end{align}
The short-sale limit is set to $\underline{\theta}=0$.

Following \cite{bloometal_2018}, we set the depreciation rate of capital to $\delta = 0.1$ and we choose $\alpha = 0.25$ and $\zeta=0.25$, consistent with a capital share of 1/3 in production. The persistence of the logarithm of aggregate productivity is set to $\rho^A = 0.8145$ and the standard deviations of innovations to productivity are set to $\sigma^A_L = 0.0124$ and $\sigma^A_H = 0.0199$, consistent with the quarterly process in \cite{bloometal_2018} and \cite{khan2008idiosyncratic}. We set the survival rate of firms to $\Gamma = 0.965$. There are $N_z = 3$ idiosyncratic productivity levels for firms corresponding to $z_0 = 0.5$, $z_1 = 1.0$, and $z_2 = 1.5$. The corresponding Markov transition matrices are given by 
\begin{align}
\Pi^z_L = \begin{bmatrix}
0.850 & 0.150 & 0.000 \\
0.075 & 0.850 & 0.075 \\
0.000 & 0.150 & 0.850
\end{bmatrix}\text{ and }\Pi^z_H = \begin{bmatrix}
0.925 & 0.000 & 0.075   \\
0.150 & 0.700 & 0.150\\
0.075 & 0.000 & 0.925 
\end{bmatrix}.
\end{align}
The transition matrix between the two uncertainty regimes is given by
\begin{align}
\Pi^U = 
\begin{bmatrix}
0.90 & 0.10 \\
0.21 & 0.79
\end{bmatrix},
\end{align}
implying a quarterly persistence of the high uncertainty regime of $0.943$ and a quarterly persistence of the low uncertainty regime of $0.974$, as in \cite{bloometal_2018}. The adjustment costs parameters are set to $\xi^{\text{up}} = 1.0$ and $\xi^{\text{down}} = 2.5$.\footnote{
The smoothing parameter in the adjustment cost function is set to $s=400$, such that $\xi(1.01k, k) = 0.993 \xi^\text{up} + 0.007\xi^\text{down}$ and $\xi(0.99k, k) = 0.007 \xi^\text{up} + 0.993\xi^\text{down}$.}
The parameter values are summarized in table \ref{tab:het_params}.

\subsection{Implementation}
As in the previous examples, we solve the model by training neural networks to approximate a core set of equilibrium functions, and then solve for the remaining equilibrium objects in closed form. Relative to the previous two applications, this model poses an additional challenge: individual policy functions of firms and households depend not only on aggregate quantities, but also on their corresponding idiosyncratic states. To solve for the firm and household policies, we train neural networks to approximate a mapping from truncated histories of aggregate shocks to \emph{policy functions}, that map the idiosyncratic state variables to equilibrium policies. 
To do so, we build on, and extend the operator learning approach of \cite{zhong_2023}. 

\subsubsection{Operator learning\label{sec:operatorlearning}}
Suppose that we want to obtain an approximation $\hat{f}(X, x)$ to the function $f(X, x)$, where $X\in\mathbb{R}^{N_1}$, $x\in\mathbb{R}^{N_2}$, and $f(X, x)\in \mathbb{R}$. 
Let $F(\mathbb{R}^{N_2}, \mathbb{R})$ denote the space of functions mapping $\mathbb{R}^{N_2}$ into $\mathbb{R}$.
The idea of operator learning is to decompose the mapping $f$ as $f(X, x)=g(X)(x)$. Here, the \emph{operator} $g : X \rightarrow g(X): \mathbb{R}^{N_1} \rightarrow F(\mathbb{R}^{N_2}, \mathbb{R})$, maps the vector $X \in \mathbb{R}^{N_1}$ to a \emph{function} $g(X)\in F(\mathbb{R}^{N_2}, \mathbb{R})$. The function $g(X)$ is then evaluated at $x$.

For example, let $X$ denote the aggregate state vector, let $x$ denote the idiosyncratic state vector, and let $f(X, x)$ denote a policy function which depends on aggregate and idiosyncratic state variables. In this example $g: \mathbb{R}^{N_1}\rightarrow F(\mathbb{R}^{N_2}, \mathbb{R})$ maps the aggregate state to a policy function. The policy function $g(X)$ in turn maps the idiosyncratic state $x$ to the choice variable $g(X)(x)$. Consider the case of $N_1 = N_2 = 1$ and suppose that, conditional on the aggregate state $X$, the policy function is a linear function of the idiosyncratic state $x$, that is, $f(X, x) = g(X)(x) = \alpha_X x$. Although the policy function is linear in the idiosyncratic state $x$, the slope coefficient $\alpha_X$ depends on the aggregate state $X$ in a potentially nonlinear way. We can split the problem of approximating $f(X, x)$ into two parts. First, approximating $g(X)$ by $\hat{g}(X): X \rightarrow \hat{\alpha}_X$.\footnote{A linear function from $\mathbb{R} \rightarrow \mathbb{R}$ is fully characterized by its slope coefficient. Therefore, we can essentially predict a function by predicting a single real number.} Second, evaluating $\hat{g}(X)(x) = \hat{\alpha}_X x$. As we show in this paper, a key advantage of approximating operators, as opposed to directly parameterizing policy functions, is that it is easy to ensure desirable properties of $\hat{g}(X)(x)$, such as monotonicity or concavity in $x$ for all $X$.

\paragraph{Shape-preserving operators}
For concreteness, let us consider the household consumption function from the model above: $\pi^c(\Xstagg_t, e_t^i, \theta_t^i) = c_t^i$. Following the operator learning approach, we decompose the problem of predicting the individual policy into two steps: First we predict a \emph{function} $\pi^{c\text{, ind}}_t$ for each aggregate state $\Xstagg_t$. Then we evaluate the resulting function $\pi^{c\text{, ind}}_t(e_t^i, \theta_t^i) = c_t^i$ for all idiosyncratic states of interest. 

Often, some crucial shape properties of the equilibrium functions $\pi^{c\text{, ind}}_t(e_t^i, \theta_t^i)$ are known ex-ante. In the context of the model we present in this section, or for a large class of other macroeconomic models, we know that the consumption function is strictly increasing and concave in the idiosyncratic asset holding $\theta^i$.\footnote{For example, this is the case in standard incomplete-market models in the spirit of \cite{imrohorouglu1989cost, bewley_1976, huggett1993risk, aiyagari_1994, krusell1998income}.} In this section, we show a simple way to ensure that all predicted policy functions fulfill these properties. Thus, we effectively restrict the search space to the space of economically more meaningful functions, increasing the robustness of our algorithm. What is more, a monotone consumption function allows us to apply the endogenous grid method. Using the endogenous grid method, we obtain targets to train the policy function with supervised learning, increasing the robustness of our approach. 

The first step is to choose a functional form to approximate the function $\pi^{c\text{, ind}}_t(e_t^i, \theta_t^i)$. We consider the case where $e_t^i$ takes two discrete values $\mathcal{E}^{\text{grid}} = [e_0, e_1]$, where $e_0 < e_1$, and where $\theta_t^i$ is continuous. Because $e^i_t$ is a discrete variable, we can split $\pi^{c\text{, ind}}$ into two univariate functions. Then, we approximate those functions using piecewise linear interpolation on a grid of interpolation nodes $\Theta^{\text{grid}}= [\theta_0, \theta_1, \dots, \theta_{N-1}]$, where $\theta_0 < \theta_1 < \dots < \theta_{N-1}$
\begin{align}
\mathcal{C}^{\text{grid}}_t=
\begin{bmatrix}
c_{t, 0, 0}, c_{t, 0, 1}, \dots, c_{t, 0, N-1} \\
c_{t, 1, 0}, c_{t, 1, 1}, \dots, c_{t, 1, N-1} 
\end{bmatrix} \in \mathbb{R}^{2\times N},
\end{align}
where $c_{t, i, j} = \pi^{c\text{, ind}}_t(e_i, \theta_j)$ for a given aggregate state $\Xstagg_t$, idiosyncratic states $e_i \in \mathcal{E}^{\text{grid}}$, and $\theta_j \in \Theta^{\text{grid}}$. For asset holdings $\theta$ between grid points, \emph{i.e.} $\theta_j < \theta < \theta_{j+1}$, we interpolate linearly between $c_{t, i, j}$ and $c_{t, i, j+1}$. Conditional on an aggregate state and fixed grids $\mathcal{E}^\text{grid}$ and  $\Theta^{\text{grid}}$, the approximate household consumption function is fully described by the $2\times N$ values of $\mathcal{C}^{\text{grid}}_t$, which in turn are parameterized by a neural network. Following our sequence space approach, this neural network takes the truncated sequence of aggregate shocks $\Xse_t$ as an input.

\emph{Monotonicity}: 
To ensure that the predicted consumption functions are increasing in idiosyncratic asset holdings $\theta$, we need to make sure that the prediction by the neural network always satisfies $c_{t, i, j} < c_{t, i, j+1}$ $\forall t, i, j$. To ensure that this condition holds, we follow a two-step procedure. First, instead of predicting $\mathcal{C}^{\text{grid}}_t$ directly, the neural network predicts the boundary consumption value\footnote{Consumption at the lowest gridpoint in the asset grid.} and consumption increments
\begin{align}
\nn^{c}(\Xse_t)\approx d\mathcal{C}^{\text{grid}}_t:=
\begin{bmatrix}
c_{t, 0, 0}, dc_{t, 0, 1}, \dots, dc_{t, 0, N-1} \\
c_{t, 1, 0}, dc_{t, 1, 1}, \dots, dc_{t, 1, N-1} 
\end{bmatrix} \in \mathbb{R}^{2\times N}_{>0},
\end{align}
where we ensure that all the $2\times N$ predicted entries of $d\mathcal{C}^{\text{grid}}_t$ are positive. This is easy to ensure, for example, by using a \emph{softplus} activation function in the output layer. In the second step we construct 
\begin{align}
\mathcal{C}^{\text{grid}}_t = \begin{bmatrix}
c_{t, 0, 0}, c_{t, 0, 1}, \dots, c_{t, 0, N-1} \\
c_{t, 1, 0}, c_{t, 1, 1}, \dots, c_{t, 1, N-1} 
\end{bmatrix},
\end{align}
where 
\begin{align}
c_{t, i, j} = c_{t, i, 0} + \sum_{k = 1}^{j}dc_{t, i, k}.\label{eq:het_consgrid}
\end{align}
Since the architecture of our neural network guarantees that all the predicted increments $dc_{t, i, k}$ are positive, this two-step procedure guarantees monotonicity of the predicted consumption at the interpolation nodes: $c_{t, i, j}<c_{t, i, j+1}$. Piecewise linear interpolation guarantees that monotonicity of the grid values translates into monotonicity of the resulting function. A large number of gridpoints, $N$, and, for example, a log-log spaced grid\footnote{\emph{Log-log} ensures high grid density for low asset holdings, where the short sale limit might bind.} for the asset holdings $\Theta^{\text{grid}}$ can ensure that the loss in accuracy coming from the linear interpolation is minimal.

\emph{Concavity}: In order to jointly guarantee monotonicity and concavity of the consumption function, one needs to ensure that the slope of the consumption function is decreasing in the individual asset holding, while it remains bounded from below by zero. Let $d\theta_{j} :=\theta_j - \theta_{j-1}$ define the distance between two grid points $\theta_j$ and $\theta_{j-1}$ on the asset grid $\Theta^{\text{grid}}$. To ensure concavity and monotonicity of consumption, we follow a three-step procedure. 
In a first step, we predict:
\begin{align}
\nn^{c}(\Xse_t)\approx dd\mathcal{C}^{\text{grid}}_t:=
\begin{bmatrix}
c_{t, 0, 0}, ddc_{t, 0, 1}, ddc_{t, 0, 2}\dots,ddc_{t, 0, N-1} \\
c_{t, 1, 0}, ddc_{t, 1, 1}, ddc_{t, 1, 2} \dots, ddc_{t, 1, N-1} 
\end{bmatrix},
\end{align}
where we ensure that $c_{t, i, 0}>0$ and $ddc_{t, i, j}<0$ using a suitable activation function. 
In the second step, we construct the matrix $d\mathcal{C}^{\text{grid}}_t$, using 
\begin{align}
dc_{t, i, j} = d\theta_{j} \times  exp\left(\sum_{k = 1}^j ddc_{t, i, k}\right).
\end{align}
Since all predictions $ddc_{t, i, k}<0$, our architecture enforces that $dc_{t, i, j} / d\theta_{j}$ is decreasing in $j$, ensuring concavity of the consumption function. The exponential function further guarantees that all $dc_{t, i, j} / d\theta_{j}>0$, making sure that the consumption function is also monotone.
In the third step, we use the cumulative sum to construct the implied consumption grid, as in equation \eqref{eq:het_consgrid}.

\emph{Borrowing constraints}:
With a slight modification of the procedure described above, we can additionally encode the borrowing constraint $\theta_{t+1}^i\geq\underline{\theta}=0$ directly into the network architecture. For a given asset price and given dividends, we can convert the distance between points in the asset grid, $d\theta_j$, to differences in cash-at-hand ($cah$) by multiplying the asset grid distance with the payout of the asset.\footnote{The income form other sources, such as labor income and startup rents, is constant across the asset grid, and can hence be omitted in this calculation.} In our model: $dcah_j = d\theta_j (D_t + \Gamma p_t)$. We know that the consumption function is increasing and concave in cash-at-hand. We can ensure monotonicity and concavity of the consumption function by making sure that the marginal propensity to consume, \emph{i.e.} $dc/dcah$, is positive and decreasing. To additionally ensure that the borrowing constraint is always satisfied, it is sufficient to ensure that the consumption share at the borrowing constraint and the marginal propensities to consume are bounded from above by 1. We can achieve this by following a three-step procedure, similar to the one outlined in the previous paragraph.
The neural network now predicts
\begin{align}
\nn^{c}(\Xse_t)\approx d\mathcal{MPC}^{\text{grid}}_t:=
\begin{bmatrix}
cs_{t, 0, 0}, dmpc_{t, 0, 1}, dmpc_{t, 0, 2}\dots, dmpc_{t, 0, N - 1} \\
cs_{t, 1, 0}, dmpc_{t, 1, 1}, dmpc_{t, 1, 2} \dots, dmpc_{t, 1, N - 1} 
\end{bmatrix},
\end{align}
where we use a sigmoid activation function to ensure that $0 < cs_{t, i,0}\leq 1$. Where $cs_{t, i, 0}$ denotes the consumption share out of cash-at-hand at the first grid point on the asset grid. We use a softplus activation and multiplication with $-1$ to ensure that all predicted $dmpc_{t, i, j}<0$.
In the second step we construct the consumption value at the first asset grid point $c_{t, i, 0} = cs_{t, i, 0} \times cah_{t,i,0}$. Since $cs_{t, i, 0}$ always lies between 0 and 1, the borrowing constraint is satisfied. We then obtain the marginal propensity to consume
\begin{align}
mpc_{t,i,j} = \underbrace{cs_{t, i, 0}}_{\in (0, 1)} \times \underbrace{e^{\sum_{k=1}^j dmpc_{t, i, k}}}_{\in (0, 1), \text{ decr in j}},
\end{align}
for $j>0$ and obtain
\begin{align}
\mathcal{MPC}_t^{\text{grid}} := \begin{bmatrix}
c_{t, 0, 0}, mpc_{t, 0, 1}, mpc_{t, 0, 2}, \dots, mpc_{t, 0, N - 1} \\
c_{t, 1, 0}, mpc_{t, 1, 1}, mpc_{t, 1, 2}, \dots, mpc_{t, 1, N - 1} 
\end{bmatrix}.
\end{align}
This parameterization ensures that all predicted $mpc_{t,i, j}$ are positive, bounded from above by 1, and decreasing in the index $j$. 
In the third step, we again compute the remaining consumption values for all gridpoints
\begin{align}
c_{t,i,j} = c_{t, i, 0} +\sum_{k = 1}^j mpc_{t, i, k} \times dcah_{t, i, k}.
\end{align}
As in the case of parameterizing $ddc$, the resulting consumption grid is increasing and concave.
Because the predicted marginal propensities to consume are bounded from above by 1, this construction additionally ensures that the borrowing constraint is never violated.\footnote{Given that we ensure that the consumption value at the first gridpoint is consistent with the borrowing constraint.} It also makes it easy for the neural network to predict the consumption for borrowing constraint households extremely precisely (by predicting the maximum mpc of 1). In models, where the marginal propensity to consume plays a central role, ensuring a precisely satisfied borrowing constraint can be particularly important for the implied model predictions.

\emph{Shape-preservation in other settings}:
It is worth noting that our approach of shape-preserving operator learning is not limited to the sequence-space approach or to discrete-time models. Shape-preserving operator learning can be analogously applied in state-space-based methods or in continuous-time models. Lastly, note that our architecture guarantees only the monotonicity and concavity of the predicted values at the interpolation nodes. With linear interpolation in one dimension, the monotonicity and concavity of the values at the interpolation nodes are sufficient for the monotonicity and (weak) concavity of the interpolating function. Shape-preserving interpolation of multivariate functions is a more complicated problem. Multivariate piecewise linear interpolation is guaranteed to preserve monotonicity; however, concavity of the interpolating function is generally not guaranteed for simple interpolation methods in higher dimensions. 

\subsubsection{Firm policies}
As part of computing an approximate recursive equilibrium, we need to solve for the firms' policy functions $\pi^F(\Xstagg_t, z^i_t, k_t^i) = k_{t+1}^{i}$ and $\pi^\lambda(\Xstagg_t, z^i_t, k_t^i) = \lambda_{t}^{i}$, such that these functions are consistent with the optimality conditions given in equations \eqref{eq:firm_foc_lambd} to \eqref{eq:firm_foc_compl}.
We approximate both functions using the operator learning approach. Employing the shape-preservation techniques described above, we make sure that the predicted KKT multiplier is non-negative and weakly decreasing in firm capital $k$. Similarly, we ensure that the firm savings function is increasing in $k$.

The firm-level productivity follows a three-state Markov chain. We set a grid of $N_k$ points over the firm capital holdings. For each aggregate shock, we obtain the following predictions
\begin{align}
\mathcal{K}^{\text{grid}}_t &= 
\begin{bmatrix}
k_{t+1, 0, 0} & k_{t+1, 0, 1} & \dots & k_{t+1, 0, N_k-1} \\
k_{t+1, 1, 0} & k_{t+1, 1, 1} & \dots & k_{t+1, 1, N_k-1} \\
k_{t+1, 2, 0} & k_{t+1, 2, 1} & \dots & k_{t+1, 2, N_k-1} \\
\end{bmatrix}\\
\mathcal{\lambda}^{\text{grid}}_t &= 
\begin{bmatrix}
\lambda_{t, 0, 0} & \lambda_{t, 0, 1} & \dots & \lambda_{t, 0, N_k-1} \\
\lambda_{t, 1, 0} & \lambda_{t, 1, 1} & \dots & \lambda_{t, 1, N_k-1} \\
\lambda_{t, 2, 0} & \lambda_{t, 2, 1} & \dots & \lambda_{t, 2, N_k-1} \\
\end{bmatrix},
\end{align}
corresponding to the gridded idiosyncratic firm policies for a given aggregate state at the $3 \times N_k$ interpolation nodes. To evaluate the policies for capital values between grid points, we rely on linear interpolation.

We use a shape-preserving operator network to ensure that the predicted values for next period's capital, $k_{t+1, i, j}$ are increasing in $j$ (\emph{i.e.} increasing in $k_{t,i,j}$). For the KKT multiplier, $\lambda_{t, i, j}$, we similarly ensure that all the predictions are non-negative and weakly decreasing in $j$. Following our sequence space approach, the input to the neural networks is given by a truncated sequence of aggregate shocks
\begin{align}
\Xse_t = [\underbrace{\epsilon^A_{t-T+1}, \epsilon^A_{t-T+2}, \dots, \epsilon^A_t}_{\text{shocks to aggregate TFP}}, \underbrace{\chi_{t-T+1}, \chi_{t-T+2}, \dots, \chi_t}_{\text{uncertainty regime}}].
\end{align}
In the first step, the neural networks predict
\begin{align}
\nn^{k}(\Xse_t) &= \begin{bmatrix}
k_{t+1, 0, 0} & dk_{t+1, 0, 1} & dk_{t+1, 0, 2} & \dots & dk_{t+1, 0, N_k -1} \\
k_{t+1, 1, 0} & dk_{t+1, 1, 1} & dk_{t+1, 1, 2} & \dots & dk_{t+1, 1, N_k -1} \\
k_{t+1, 2, 0} & dk_{t+1, 2, 1} & dk_{t+1, 2, 2} & \dots & dk_{t+1, 2, N_k -1}
\end{bmatrix} \\
\nn^{\lambda}(\Xse_t) &= \begin{bmatrix}
\log(\lambda_{t, 0, 0}) & d\lambda_{t, 0, 1} & d\lambda_{t, 0, 2} & \dots & & d\lambda_{t, 0, N_k -1} \\
\log(\lambda_{t, 1, 0}) & d\lambda_{t, 1, 1} & d\lambda_{t, 1, 2} & \dots & & d\lambda_{t, 1, N_k -1} \\
\log(\lambda_{t, 2, 0}) & d\lambda_{t, 2, 1} & d\lambda_{t, 2, 2} & \dots & & d\lambda_{t, 2, N_k -1},
\end{bmatrix}
\end{align}
where we use the softplus activation in the output layer to ensure that $k_{t+1,i, j}\geq 0$, $dk_{t+1,i,j}\geq 0$, and $-$softplus to ensure that $d\lambda_{t,i,j}<0$.
In the second step, we construct $\mathcal{K}^{\text{grid}}_{t+1}$ and $\mathcal{\lambda}^{\text{grid}}_t$ to ensure non-negativity and monotonicity of both functions:
\begin{align}
k_{t+1, i, j} &= k_{t+1, i, 0} + \sum_{h = 1}^j dk_{t+1, i, h} \\
\lambda_{t, i, j} &= exp\left(\log(\lambda_{t, i, 0}) + \sum_{h=1}^j d\lambda_{t, i, h}\right).
\end{align}

\paragraph{Simulation}
We represent the cross-sectional distribution of firms over idiosyncratic capital and productivity using the finite histogram method of \cite{young2010solving}. We base the firm histogram on the same grid as we use for interpolation nodes of firms' policy functions. Because of that, we can evaluate the histogram transition using the gridded capital policies predicted by the neural network without the need to resort to interpolation.  

\paragraph{Loss function}
To evaluate the performance of our neural network in producing policies consistent with firm optimality conditions, we need to define a loss function to map equilibrium conditions error into a scalar quantity. As in \cite{azinovic2022deep}, we choose the standard mean-squared function as our loss. Through the stochastic discount factor the firms' optimality conditions also depend on household policies and on the equity price. Because of that, the loss function also depends on the household network $\nn^c$, and on the price network $\nn^p$.

We use Gauss-Hermite quadrature with $N^{GH}$ quadrature nodes to approximate the expectation over realizations of shocks to aggregate productivity next period, for each of the two uncertainty regimes. Let $\mathcal{A}_t:= \left(\Xst_t, \{\Xst_{t+1, i}\}_{i=1}^{2 N_{GH}}\right.$, $\left.\Xse_t, \Xse_{t+1,i}\}_{i=1}^{2N_{GH}}\right)$, denote aggregate states and truncated histories of aggregate shocks for a given realization in period $t$, and at all $2N_{GH}$ integration nodes for the next period $t+1$. Given $\mathcal{A}_t$, the wage can be computed as a closed form function of the state-vector, as given in equation \eqref{eq:firms_wage}. The outputs of the firm network $\nn^k$, the outputs of the household network $\nn^c$, and the output of the price network $\nn^p$ allows us to construct the aggregate stochastic discount factor $\Lambda_{t+1, t}$, aggregate dividends $\left(D_t,  \{D_{t+1, i}\}_{i=1}^{2N_{GH}}\right)$, as well as the stock prices $\left(p_t, \{p_{t+1, i}\}_{i=1}^{2N_{GH}}\right)$. We then sample idiosyncratic states $(z^i_t, k_t^i)$,\footnote{We sample the idiosyncratic labor endowment states from the ergodic distribution of the corresponding Markov chain. For the idiosyncratic capital, we sample half of the points from the capital grid, and the other half from a uniform distribution.} and evaluate the firm policies on those states to obtain $k_{t+1}^i, \{k_{t+2, j}^{i}\}_{j=1}^{2N_{GH}}$, and $\{\lambda_{t+1, j}^{i}\}_{j=1}^{2N_{GH}}$. Then we recover the implied KKT multiplier, $\lambda_t^{i\text{, implied}}$, from the optimality condition \eqref{eq:firm_foc_lambd}: 
\begin{align}
\lambda_t^{i\text{, implied}} &= \frac{\Gamma \E{\Lambda_{t+1, t}\left(1 + r_{t+1}^{i, k} +  \frac{\partial}{\partial k_{t+1}^i}\psi(k_{t+2}^i, k_{t+1}^i)  \right)(1 + \lambda_{t+1}^i)}}{\left(1 + \frac{\partial}{\partial k_{t+1}^i}\psi(k_{t+1}^i, k_t^i)\right)} -1.
\end{align}
The difference between the predicted and the implied Lagrange multiplier summarizes the error in the firm Euler equation. Besides the Euler equation error, we also need to evaluate the error in the remaining KKT conditions.
\begin{align}
err^{\text{firm}}_1\left(\mathcal{A}_t, \nn^c, \nn^p, \nn^k,\nn^\lambda,z^i_t,k_t^i\right) &= \frac{\lambda_t^i - \lambda_t^{i\text{, implied}}}{1 + \lambda_t^{i\text{, implied}}} \label{eq:firmloss_lambda_error} \\
err^{\text{firm}}_2\left(\mathcal{A}_t, \nn^c, \nn^p, \nn^k,\nn^\lambda,z^i_t,k_t^i\right) &= \psi^{FB}\left(\frac{d_t^i - \underline{d}}{k_t^i}, \lambda_t^{i\text{, implied}}\right) \label{eq:firmloss_KKT_error}\text{, where }\\
\psi^{FB}(a, b) &= a - b - \sqrt{a^2 + b^2},
\end{align}
denotes the Fischer-Burmeister function \citep[see, \emph{e.g.}][]{jiang_1999}. The optimal firm policies are characterized by 
\begin{align}
err^{\text{firm}}_1\left(\mathcal{A}_t, \nn^c, \nn^p, \nn^k,\nn^\lambda,z^i_t,k_t^i\right) &= 0  \label{eq:firmloss_zeroerr1}\\
err^{\text{firm}}_2\left(\mathcal{A}_t, \nn^c, \nn^p, \nn^k,\nn^\lambda,z^i_t,k_t^i\right) &= 0,\label{eq:firmloss_zeroerr2}
\end{align}
for all $\mathcal{A}_t$ and all $z^i_t, k^i_t$.

\subsubsection{Household policies and stock price}
The two additional equilibrium functions, which we need to compute are the household policy function, and the equity price function. For households, we choose to parameterize the consumption function $\pi^C(\Xstagg_t, e_t^i, \theta_t^i) = c_t^i$. For the stock price we need to approximate $\pi^P(\Xstagg) = p_t$. We parameterize the equity price using a simple densely connected feed-forward neural network that maps the truncated history of aggregate shocks into a non-negative equity price\footnote{To make sure the network predicts non-negative price, we activate the output layer with a \emph{softplus} activation function.}
\begin{align}
\nn^p(\Xse_t) = p_t.
\end{align}
To parameterize household consumption function, we use our shape-preserving neural network operator architecture, to obtain a matrix $\mathcal{C}^{\text{grid}}_t$, that represents the consumption function on a discrete grid over the idiosyncratic assets $\theta_j$ and the labor endowment $e_i$:
\begin{align}
\mathcal{C}^{\text{grid}}_t = \begin{bmatrix}
c_{t, 0, 0}, c_{t, 0, 1}, \dots, c_{t, 0, N} \\
c_{t, 1, 0}, c_{t, 1, 1}, \dots, c_{t, 1, N} 
\end{bmatrix},
\end{align}
where $c_{t, i, j}$ denote the consumption in period $t$ for each idiosyncratic productivity level $e_i$ in the productivity grid $\mathcal{E}^{\text{grid}} = [e_0, e_1]$, and for each asset holding $\theta_j$ in the asset grid $\Theta^{\text{grid}}= [\theta_0, \theta_1, \dots, \theta_{N_\theta - 1}]$, with $\theta_0 < \theta_1 < \dots < \theta _{N_{\theta} - 1}$. We use our shape-preserving neural network architecture to ensure that the resulting consumption function satisfies three properties: first, it is increasing in cash-at-hand, second, it is concave in cash-at-hand, and third, it is consistent with the borrowing constraint $\theta_{t+1} \geq \underline{\theta} = 0.$ We impose these three restrictions by following the three-step procedure, outlined in section \ref{sec:operatorlearning}.
The neural network learns to predict
\begin{align}
\nn^{c}(\Xse_t)\approx d\mathcal{MPC}^{\text{grid}}_t:=
\begin{bmatrix}
cs_{t, 0, 0}, dmpc_{t, 0, 1}, dmpc_{t, 0, 2}\dots, dmpc_{t, 0, N} \\
cs_{t, 1, 0}, dmpc_{t, 1, 1}, dmpc_{t, 1, 2} \dots, dmpc_{t, 1, N} 
\end{bmatrix},
\end{align}
where $cs_{t, i, 0}$ denotes the consumption share out of cash-at-hand at the first grid point on the asset grid. As outlined in section \ref{sec:operatorlearning}, the neural network architecture ensures that all predicted $dmpc_{t, i, j}<0$, as well as $0 < cs_{t, i, 0} \leq 1$.
In a second step we then construct
\begin{align}
c_{t,i,0} &= cs_{t, i, 0} \times cah_{t,i,0}\\
mpc_{t,i,j} &= cs_{t, i, 0} \times e^{\sum_{k=1}^j dmpc_{t, i, k}},
\end{align}
for $j>0$ and obtain $\mathcal{MPC}_t^{\text{grid}}$. By construction, all predicted $mpc_{t,i, j}$ are positive, bounded from above by 1, and decrease in the index $j$. 
In the third step, we construct the consumption values for all the remaining grid points 
\begin{align}
c_{t,i,j} &= c_{t, i, 0} +\sum_{k = 1}^j mpc_{t, i, k} \times dcah_{t, i, k} \\ dcah_{t,i, j} &= d\theta_{t,i,j}(D_t + \Gamma p_t)
\end{align}
completing the construction of $\mathcal{C}^{\text{grid}}_t$. 

\paragraph{Loss function}
Instead of backpropagating through the complex computational graph defined by the forward-looking optimality condition, we use the method of endogenous gridpoints \citep[see][]{carroll_2006},\footnote{For the endogenous grid method in higher dimensions, see \cite{druedahl2017general}.} to derive the period $t$ consumption function implied by the current guess of equilibrium functions in period $t+1$. Exploiting speed and robustness of the endogenous gridpoint method, we are furthermore able to use a simple Newton-Raphson method to solve for market clearing equity price $p_t^{\text{solved}}$ together with implied consumption function $\mathcal{C}_t^{\text{grid, solved}}$. Using this procedure, we obtain \emph{training targets} for price and household networks, allowing us to train these networks on a simple supervised learning loss.

From $\mathcal{C}_t^{\text{grid, solved}}$ and $p_t^{\text{solved}}$, we can also obtain target values for the intermediate predictions, \emph{i.e.} $\mathcal{MPC}_t^{\text{grid, solved}}$.
The supervised learning loss terms for household and price blocks are summarized in the following equations:
\begin{align}
\text{err}^{\text{hh}}_1\left(\mathcal{A}_t, \nn^c, \nn^p, \nn^k\right) &= \left(\mathcal{C}_t^{\text{grid}} - \mathcal{C}_t^{\text{grid, solved}}\right)\oslash \mathcal{C}_t^{\text{grid, solved}}\label{eq:hhloss_zeroerr1_def}\\
\text{err}^{\text{hh}}_2\left(\mathcal{A}_t, \nn^c, \nn^p, \nn^k\right) &= \left(\mathcal{MPC}_t^{\text{grid}} - \mathcal{MPC}_t^{\text{grid, solved}}\right)\oslash \mathcal{MPC}_t^{\text{grid, solved}}\label{eq:hhloss_zeroerr2_def}\\
\text{err}^{\text{p}}_1\left(\mathcal{A}_t, \nn^c, \nn^p, \nn^k\right) &= \frac{p_t - p_t^{\text{solved}}}{p_t^{\text{solved}}}\label{eq:priceloss_zeroerr2_def},
\end{align}
where $\oslash$ denotes element-wise division, and $p_t$, $\mathcal{C}_t^{\text{grid}}$, and $\mathcal{MPC}_t^{\text{grid}}$ denote predictions by the neural networks.

\subsubsection{Loss function}
Our overall loss function to train the neural networks is given by
\begin{align}
\mathcal{L}(\bm{\rho}, \mathcal{A}_t) &= w^\text{firm}_1 \left(\frac{1}{N_{\text{sample}}}\sum_{i = 1}^{N_{\text{sample}}} \text{err}^{\text{firm}}_1\left(\mathcal{A}_t, \nn^c, \nn^p, \nn^k,\nn^\lambda,z^i_t,k_t^i\right) ^ 2 \right) \nonumber \\
&+ w^\text{firm}_2 \left(\frac{1}{N_{\text{sample}}}\sum_{i = 1}^{N_{\text{sample}}} \text{err}^{\text{firm}}_2\left(\mathcal{A}_t, \nn^c, \nn^p, \nn^k,\nn^\lambda,z^i_t,k_t^i\right) ^ 2 \right) \nonumber \\
&+ w_1^{\text{hh}} \left( \frac{1}{2N_{\theta}} \sum_{i=0}^1\sum_{j=0}^{N_\theta - 1} \left[\text{err}^{\text{hh}}_1\left(\mathcal{A}_t, \nn^c, \nn^p, \nn^k\right)\right]_{i, j}^2\right) \nonumber \\
&+ w_2^{\text{hh}} \left( \frac{1}{2N_{\theta}} \sum_{i=0}^1\sum_{j=0}^{N_\theta - 1} \left[\text{err}^{\text{hh}}_2\left(\mathcal{A}_t, \nn^c, \nn^p, \nn^k\right)\right]_{i, j}^2\right) \nonumber \\
&+ w^p \left( \text{err}^{\text{p}}_1\left(\mathcal{A}_t, \nn^c, \nn^p, \nn^k\right) \right) ^ 2,\label{eq:het_fullloss}
\end{align}
where $w^\text{firm}_1$, $w^\text{firm}_2$, $w_1^{\text{hh}}$, $w_2^{\text{hh}}$, and $w^p$ denote the weights of the different components of the loss function and where, with slight abuse of notation, $\bm{\rho}$ collects the trainable parameters of all four neural networks.

\subsubsection{Forward simulation}
Given the aggregate state-history pair, $\mathcal{A}_t$, we obtain $\mathcal{A}_{t+1}$ by drawing new random shocks for productivity and for the uncertainty regime according to the corresponding probability distributions, and by evolving the endogenous state according to the current guess of equilibrium policy functions. The evolution of the distribution of firms is computed using policy functions obtained by evaluation of the firm network $\nn^k$. To make sure that the evolution of the distribution of households satisfies market clearing condition, we use the market clearing policies $\mathcal{C}^{\text{grid, solved}}_t$ and $p_t^{\text{solved}}$ computed using a Newton-Raphson root-finding algorithm.\footnote{Note that the consumption policies, together with the stock price, pin down the asset policies via the households' budget constraint.}

\paragraph{Computing the target values} In order to compute the target values for household consumption, and for the stock price, $\mathcal{C}^{\text{grid, solved}}_t$ and $p_t^{\text{solved}}$, that are consistent with household optimality and market clearing, we rely on the endogenous gridpoint method (EGM). Specifically, given a guess for current stock price $p_t^{\text{guess}}$, and current guess for equilibrium dynamics of the economy,\footnote{\emph{i.e.} we evaluate the evolution of the firm distribution using the firm policy functions encoded in the firm network, and then we evaluate the evolution of the household distribution using household network as an input to solve for market clearing price and policy.} we use the EGM to derive the period $t$ consumption function implied by expectations of $t+1$ marginal utility of consumption and asset payouts. We denote this consumption function $c_t^{i\text{, guess}}$ and corresponding savings function as $\theta_{t+1}^{i\text{, guess}}$. Next, we evaluate the excess demand function implied by household policies generated by the current price guess $ED(p_t^{\text{guess}}, \mathcal{A}_t, \nn^c, \nn^p, \nn^k) = \Theta^{\text{guess}}_{t+1} - 1$. Then we update the price guess using a Newton-Raphson algorithm
\begin{align}
p_t^{\text{new}} = p_t^{\text{old}} - \frac{ED(p_t^{\text{old}}, \mathcal{A}_t, \nn^c, \nn^p, \nn^k)}{\frac{\partial}{\partial p_t^{\text{old}}}ED(p_t^{\text{old}}, \mathcal{A}_t, \nn^c, \nn^p, \nn^k)}.
\end{align}
Our shape-preserving neural network architecture ensures that the excess demand function is well behaved in the price guess. Therefore, a small number of Newton-Raphson steps is sufficient to compute the market clearing price.\footnote{For GPU efficiency reasons, we use a predetermined number of Newton-Raphson steps, so that the same number of steps is used for all the states. The derivative of the excess demand function with respect to the guessed price can be computed efficiently using automatic differentiation.} Furthermore, the prediction of the neural network $\nn^p$ provides us with an excellent starting guess for the market clearing price, once the initial training stage is completed.\footnote{Hence, the algorithm might start with the initial number of Newton-Raphson iterations set to $10$, which can be reduced to $3$ once the algorithm converges to sufficiently low level of price prediction errors.} When the market clearing price $p_t^{\text{solved}}$ is found, we save the price and the corresponding household policies $\mathcal{C}^{\text{grid, solved}}_t$ as target for supervised training.

\subsection{Training}
\subsubsection{Outline of the step-wise training procedure}
We again follow a step-wise training procedure. Our step-wise procedure ensures that our algorithm learns reasonable firm policies, and hence generates sensible wages and dividends before it starts solving the household block.

\emph{Step 1: Firm side only}: In the first step we only train the firm side of the model. To do so, we set the stochastic discount factor of the firm to be equal to the patience parameter $\Lambda_{t+1, t} = \beta$, and set the weights in the loss function to $w_1^{\text{firm}} = w_2^{\text{firm}} = 1$ and $w_1^{\text{hh}} = w_2^{\text{hh}} = w^p = 0$. Imposing an exogenous stochastic discount factor and setting weights on household and price error to zero allows us to isolate and solve the firm problem, so we can then start solving the household problem with a good guess for firm policies already at hand. Furthermore, we start the training procedure with lower values for the adjustment cost parameters $\xi^{\text{up}} = 0.1$ and $\xi^{\text{down}} = 0.25$. The policies for the simplified parameterization can be learned quickly by the neural network.

\emph{Step 2, Pre-training price and household policies}
In the second step, we ensure that the neural networks parameterizing household policy and the equity price function also start with a good initial guess. To facilitate initial convergence, we again start the training procedure in a simplified economy: we introduce an artificial parameter $\tau$. $\tau$ modifies the payout of equity to $p_{t+1}\Gamma(1 - \tau) + D_{t+1}$. With $\tau = 0$, we recover the original economy. Setting $\tau = 1$ makes equity effectively a claim on $t+1$ dividends, reducing it into a short-lived asset. Given $\tau = 1$, we use $p_t^{\text{pre-train}} = D_t$ as a pre-training target for the price network $\nn^p$. For the household policy function, we use $\theta_{t+1}^{\text{i}} = \max(0, 0.7 \theta_t^i - 0.1)$ for $e^i_t = e_0$ and $\theta_{t+1}^{\text{i}} = 0.7 \theta_t^i + 0.6$ for $e^i_t = e_1$ as pre-training targets. This provides us with a good starting guess for the price function (in the short-lived asset calibration, $\tau = 1$), and the household policy functions.

\emph{Step 3, Training the firm and household side together:}
Retaining the simplified parameterization of the model, we now train all price and policy functions on the full equilibrium loss function, given in equation \eqref{eq:het_fullloss} with weights set to $w_1^{\text{firm}} = w_2^{\text{firm}} = w_1^{\text{hh}} = w_2^{\text{hh}} = w^p = 1$.

\emph{Step 4, Step-wise model transformation to full parameterization:} We then gradually change the parameters to the desired parameterization of the full model. We linearly increase the adjustment cost parameters to the final values of $\xi^{\text{up}} = 1.0$ and $\xi^{\text{down}} = 2.5$. At the same time, we decrease $\tau$ from $\tau = 1$ to $\tau = 0$ following a quadratic schedule. Additionally, we gradually change the stochastic discount factor used by the firms from $\beta$ to $\Lambda_{t+1, t}$. At each step, firms use a weighted average between $\beta$ and $\Lambda_{t+1, t}$ as a stochastic discount factor. We start with a full weight of $1$ on $\beta$. Then we gradually decrease it to $0$ while we increase the weight on $\Lambda_{t+1, t}$ from $0$ to $1$.

\emph{Step 5, Training with the final model parameters:} We train the neural networks on the full loss function with the final model parameterization, until we reach a sufficiently low level of remaining errors in equilibrium conditions.

\subsubsection{Hyperparameters}
We summarize the hyperparameters that we used to solve the model in table \ref{tab:het_hp}. We choose all four neural networks, $\nn^k$, $\nn^\lambda$, $\nn^c$, and $\nn^p$, to be densely connected feed-forward neural networks with three hidden layers and gelu activation functions. We truncate the history of shocks after 300 periods. The input layers consist of 600 neurons, corresponding to the truncated history of uncertainty regimes and the innovations to productivity. Each hidden layer consists of $1024$ gelu activated neurons.

\subsection{Accuracy}
In this section, we demonstrate the accuracy of our solution. Since a closed form or a high-quality numerical reference solution is not available, we investigate the errors in the equilibrium conditions, implied by the price and policy functions encoded in the learned parameters of our neural networks. We report the accuracy measures in economically interpretable units. In addition, we provide a comprehensive set of statistics on the distribution of errors across the state space.

\subsubsection{Firm policies}
\begin{figure}
    \centering
    \includegraphics[width=0.45\linewidth]{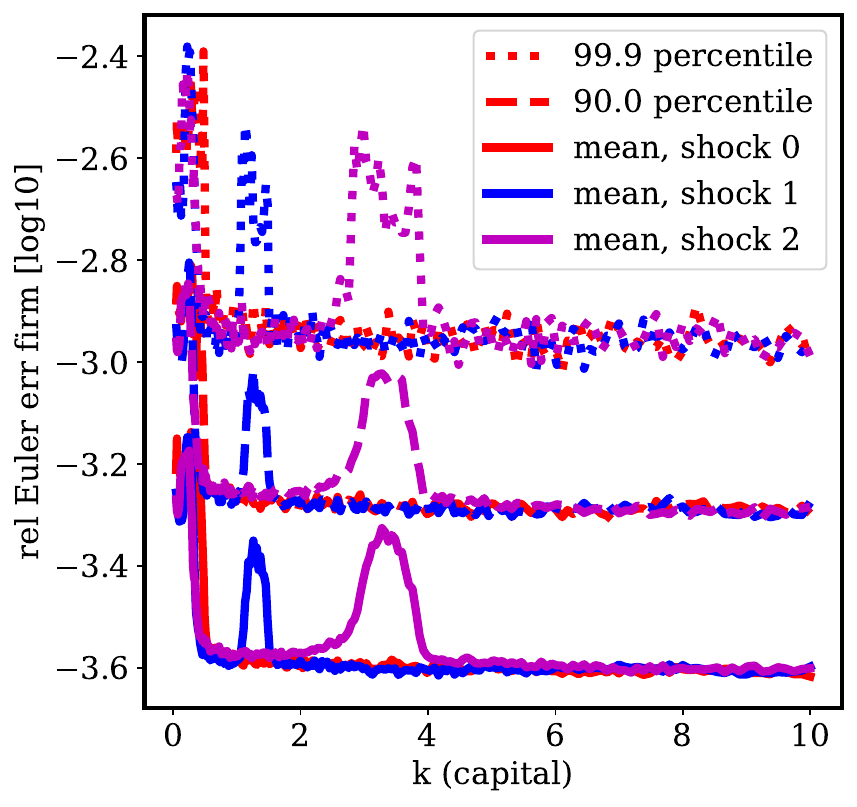}
    \includegraphics[width=0.45\linewidth]{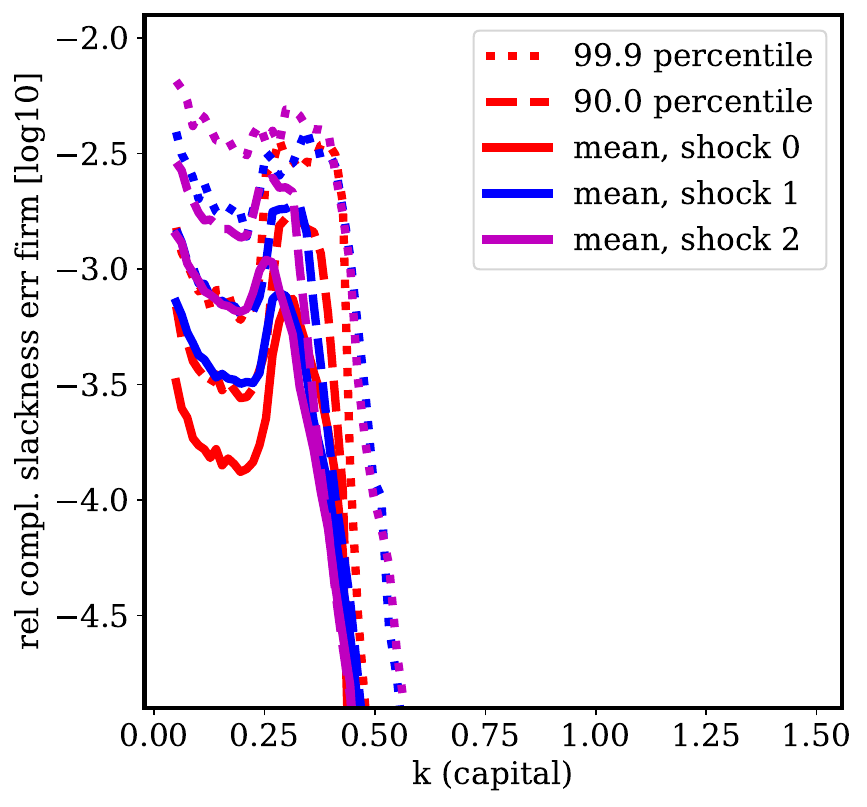}
    \caption{Remaining errors in the equilibrium conditions for the firm problem. The left panel shows the errors in the firms' Euler equations and the right panel shows the errors in the remaining KKT conditions.\label{fig:het_accuracy_firms}}
\end{figure}

The optimal firm policies are characterized by the set of firms' KKT conditions, given in equations \eqref{eq:firm_foc_lambd} to \eqref{eq:firm_foc_compl}.
Rearranging equation \eqref{eq:firm_foc_lambd} we obtain the following:
\begin{align}
0 &= \frac{\Gamma \E{\Lambda_{t+1, t}\left(1 + r_{t+1}^{i, k} +  \frac{\partial}{\partial k_{t+1}^i}\psi(k_{t+2}^i, k_{t+1}^i)  \right)(1 + \lambda_{t+1}^i)}}{\left(1 + \frac{\partial}{\partial k_{t+1}^i}\psi(k_{t+1}^i, k_t^i)\right)\left(1 + \lambda_t^i\right)} - 1,\label{eq:firm_foc_mbmc}
\end{align}
where the numerator in the first term denotes the expected marginal benefit of additional capital in the next period, and the denominator gives the marginal cost in the current period. The errors in equation \eqref{eq:firm_foc_mbmc} are therefore interpretable as relative marginal cost errors.\footnote{An error of 0.01, for example, would indicate that the marginal benefit is 1\% higher than the marginal cost.}
The left panel of figure \ref{fig:het_accuracy_firms} shows statistics of the distribution of errors in equation \eqref{eq:firm_foc_mbmc} for different idiosyncratic productivity shocks, idiosyncratic levels of capital, and 1024 aggregate states drawn from the ergodic distribution of the economy. 
The mean marginal cost error remains below 0.07\%, the 90th percentile below 0.16\%, and the 99.9th percentile below 0.4\% for all levels of firm capital and the three productivity levels. Even though the errors are very low everywhere, they are slightly larger in the areas of the state space where the firm policies feature a kink due to the constraint on dividends and due to asymmetric adjustment costs.\footnote{For a fixed idiosyncratic level of capital and a fixed idiosyncratic productivity level, the mean and the percentiles are computed across 1024 aggregate states drawn from the ergodic distribution of the economy.}

The right panel of figure \ref{fig:het_accuracy_firms} shows statistics of the distribution of errors in the remaining KKT condition of firms
\begin{align}
0 &= \psi^{\text{FB}}\left(\frac{\lambda_t^{i}}{1 + \lambda_t^{i}}, \frac{d^i_t - \underline{d}}{k_t^i}\right),\label{eq:firm_check_compl}
\end{align}
where $\psi^{\text{FB}}(x, y)$ denotes the Fischer-Burmeister function. The errors in equation \eqref{eq:firm_check_compl} summarize the violations of all the remaining KKT conditions associated with the firm problem.\footnote{Because the Fischer-Burmeister function satisfies $\psi^{\text{FB}}(x, y) = 0 \Leftrightarrow x\geq0,y\geq0,xy=0$.} Except for very low levels of capital, the error is virtually zero. Throughout, the mean error remains below 0.2\% and even the 99.9th percentile of errors remains below 0.8\%.
We conclude that firm policies are computed with high accuracy across the aggregate and idiosyncratic state space.

\subsubsection{Household policies}
As described above, the endogenous gridpoints method combined with a Newton-Raphson loop allows us to jointly solve for the market clearing equity price and the corresponding consumption policies of households. Hence, we can assess the accuracy of the consumption policies learned by the neural network by comparing the neural network predictions for period $t$ to the period $t$ consumption policies implied by household optimality and market clearing prices.\footnote{Where we evaluate the expectations over $t+1$ objects using policies and prices encoded by neural networks.}
\begin{figure}
    \centering
    \includegraphics[width=0.45\linewidth]{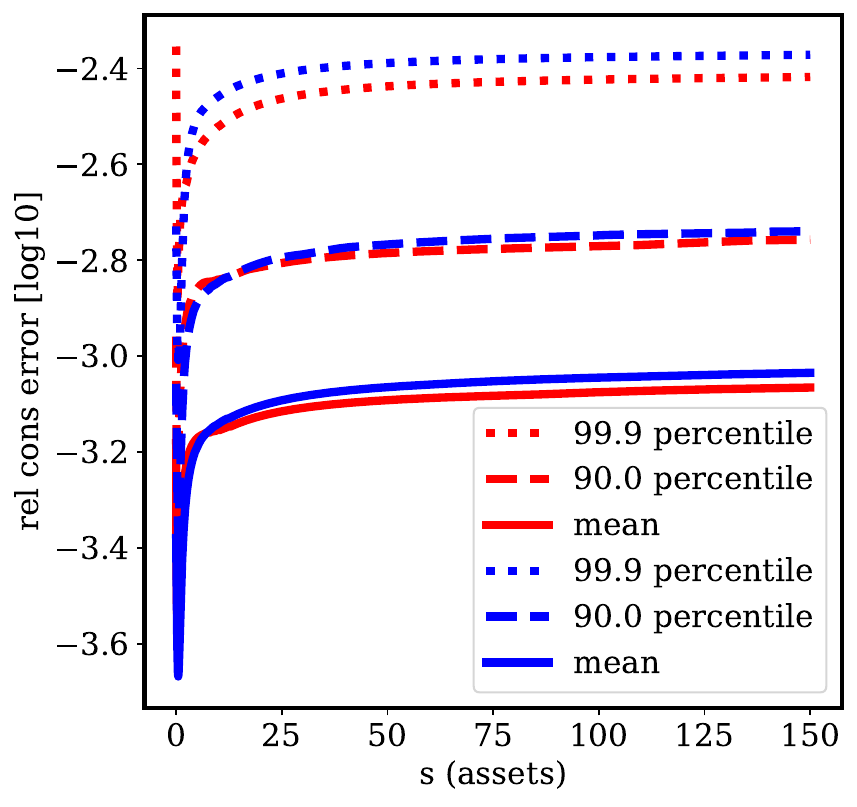}
    \caption{Remaining errors in the equilibrium conditions for the households' optimality conditions, expressed in units of relative consumption errors.\label{fig:het_accuracy_cons}}
\end{figure}
Figure \ref{fig:het_accuracy_cons} shows the accuracy of the consumption function learned by the neural network across the aggregate and idiosyncratic state space. For all the idiosyncratic states, the mean error remains below 0.08\%, the 90th percentile below 0.17\%, and the 99.9th percentile below 0.4\%.\footnote{For a fixed idiosyncratic asset holding and a fixed idiosyncratic productivity level, the mean and the percentiles are computed across 1024 aggregate states drawn from the ergodic distribution of the economy.}  We conclude that the household policies are approximated to a high accuracy.

\subsubsection{Stock price}
\begin{figure}
    \centering
    \includegraphics[width=0.45\linewidth]{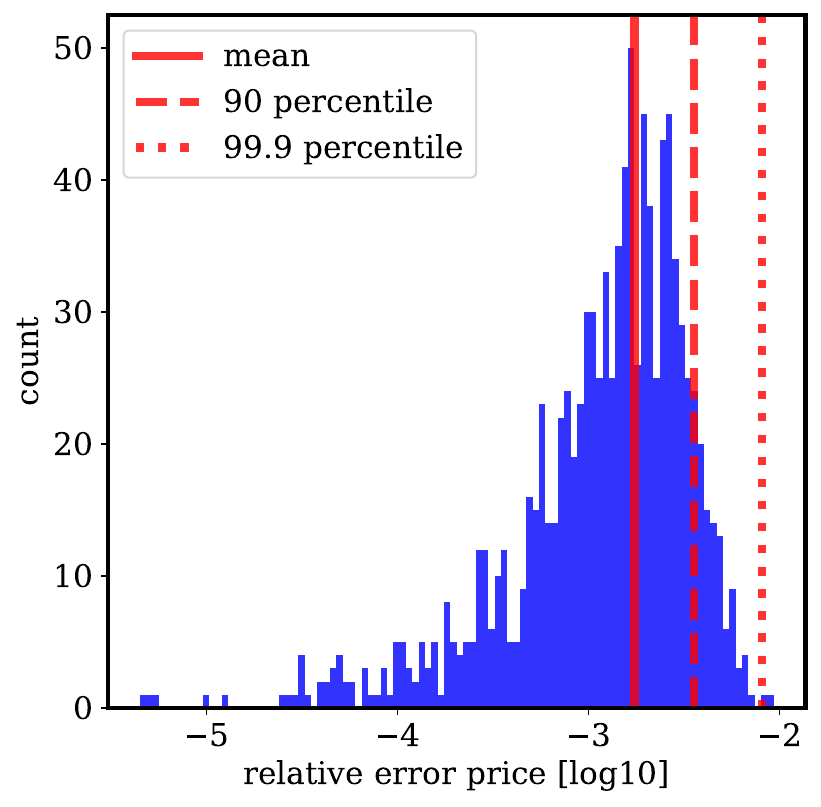}
    \includegraphics[width=0.45\linewidth]{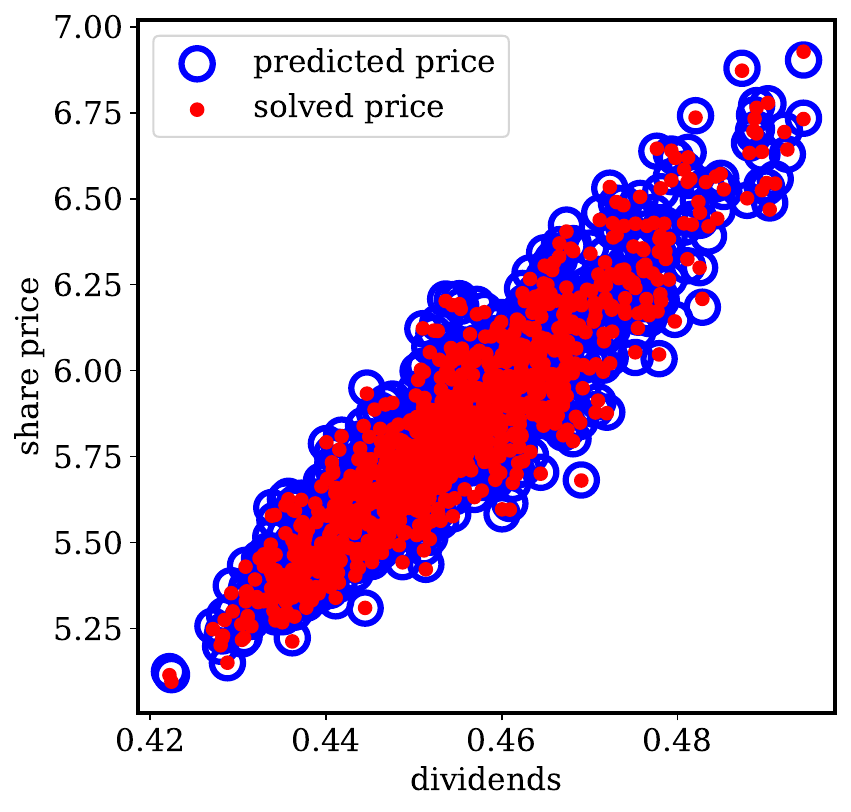}
    \caption{Prediction errors for the price network. The left panel shows the distribution of relative prediction errors across 1024 aggregate states drawn from the ergodic distribution of the economy. The right panel shows the network predictions (blue circles) and the solved market clearing prices (red dots). \label{fig:het_accuracy_price}}
\end{figure}

The left panel of figure \ref{fig:het_accuracy_price} shows the distribution of discrepancies between the equity price predicted by the neural network and the market clearing equity price obtained using the EGM-Newton-Raphson algorithm. The mean error of the price prediction is 0.17\%, the 90th percentile 0.35\% and the 99.9th percentile is 0.80\%. The right panel of figure \ref{fig:het_accuracy_price} shows the price predictions by the neural network together with the computed market clearing prices in a scatter plot against aggregate dividends.
Although the accuracy of the price function is slightly lower than the accuracy we achieved for the policy functions of firms and households, it remains high. The resulting errors in asset demand are less than the price errors and less than 0.1\% over the entire simulated ergodic set.

\subsection{Inspecting the learned equilibrium policies}
\subsubsection{Firms}
\begin{figure}
    \centering
    \includegraphics[width=0.32\linewidth]{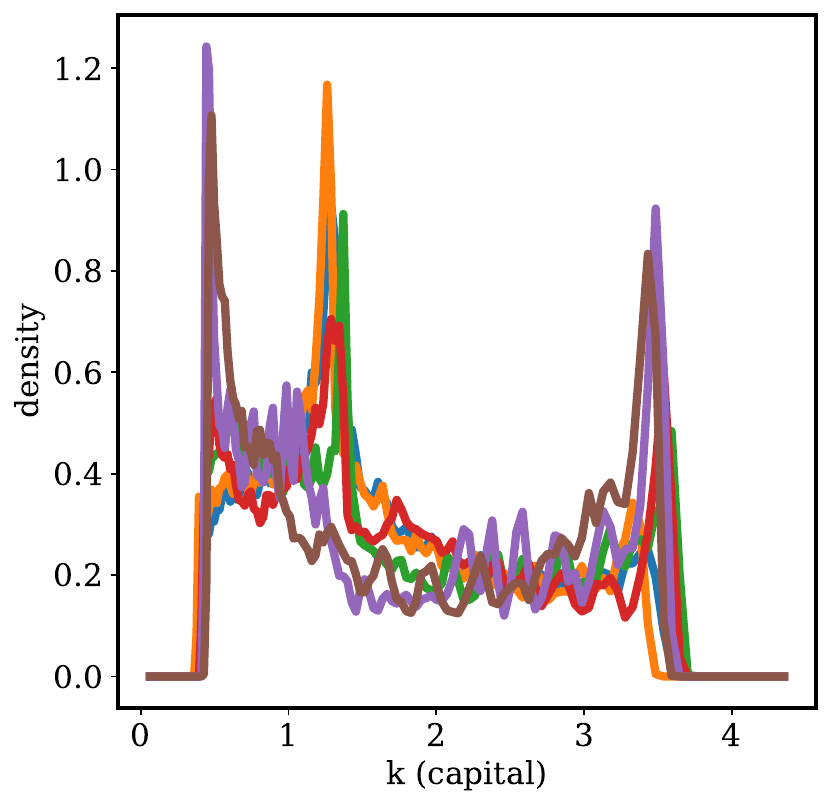}
    \includegraphics[width=0.32\linewidth]{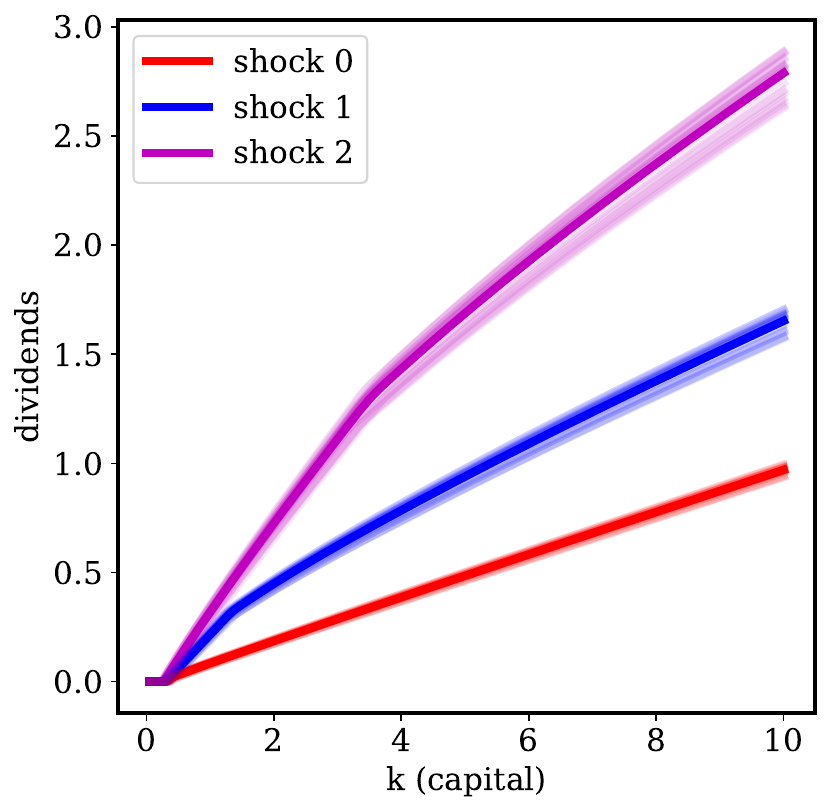}
    \includegraphics[width=0.32\linewidth]{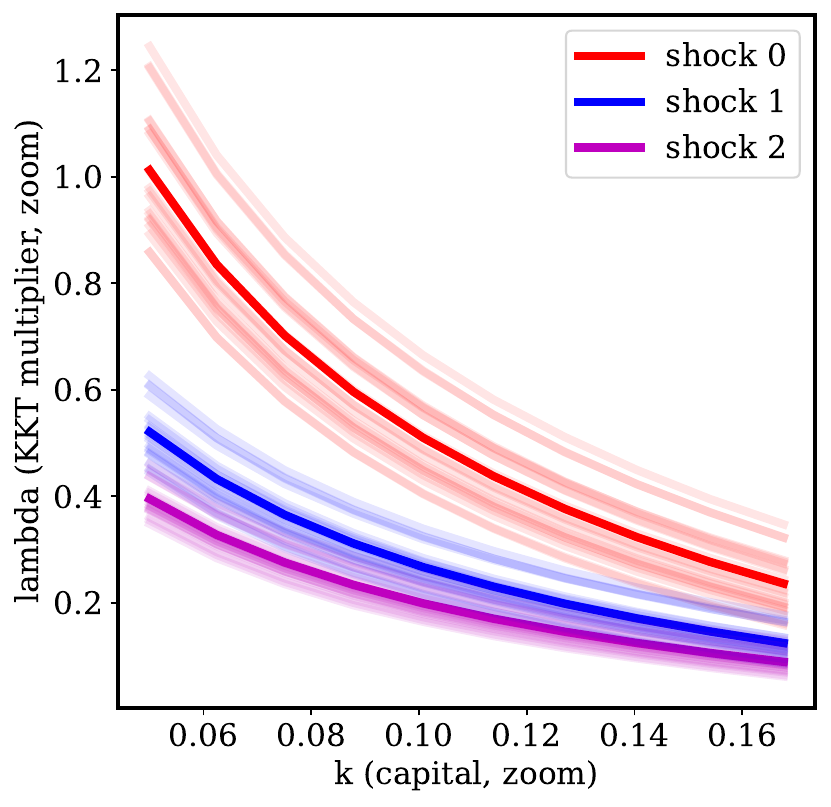}
    \caption{Left panel: density of firms across the capital grid for different simulated aggregate states. Middle panel: dividends policy for firms over the capital grid (horizontal axis) and across different productivity levels (different colors). The shaded lines show the policies for different aggregate states, and the solid lines show the average taken across the sample of aggregate states. Right panel: firms' KKT multiplier associated with the non-negative dividend constraint over the different capital and productivity levels, as well as across different aggregate states.\label{fig:het_policy_firms}}
\end{figure}

We turn to inspecting the equilibrium policies.
The left panel of figure \ref{fig:het_policy_firms} shows the density of firms across the capital grid for different simulated aggregate states.
We can see that the distribution of firms across capital varies substantially. Due to the decreasing returns to scale production function, together with capital adjustment costs, this variation directly affects aggregate wages and dividends (see, \emph{e.g.}, equation \eqref{eq:firms_wage}), such that the households' problem depends crucially on the firm distribution.

The middle panel in figure \ref{fig:het_policy_firms} shows the dividend policies of firms for different capital levels (horizontal axis) and different idiosyncratic productivity levels (represented by different colors). The shaded lines show the policies for several randomly selected aggregate states, and the solid lines show the mean taken over aggregate states. Each of the policy functions shows two kinks: at the first kink, the non-negativity constraint on dividends ceases to bind. At the second kink, firms switch from adjusting capital upward to adjusting it downward. The exact location of the kinks depends on the firm's productivity level and the aggregate state of the economy.

Similarly, the right panel shows the policy functions for the KKT multiplier on the dividend constraint. Perhaps surprisingly, the multiplier for very low values of capital is larger for the low productivity values than for the high productivity values. This is the case because, while the firms are constrained for all three productivity levels, the high productivity firms are able to choose a higher capital level for the next period.
As guaranteed by the monotonicity preserving neural network architecture, the shaded lines illustrate that the KKT multiplier is weakly decreasing and bounded from below by zero.
\subsubsection{Households}
\begin{figure}[H]
    \centering
    \includegraphics[width=0.32\linewidth]{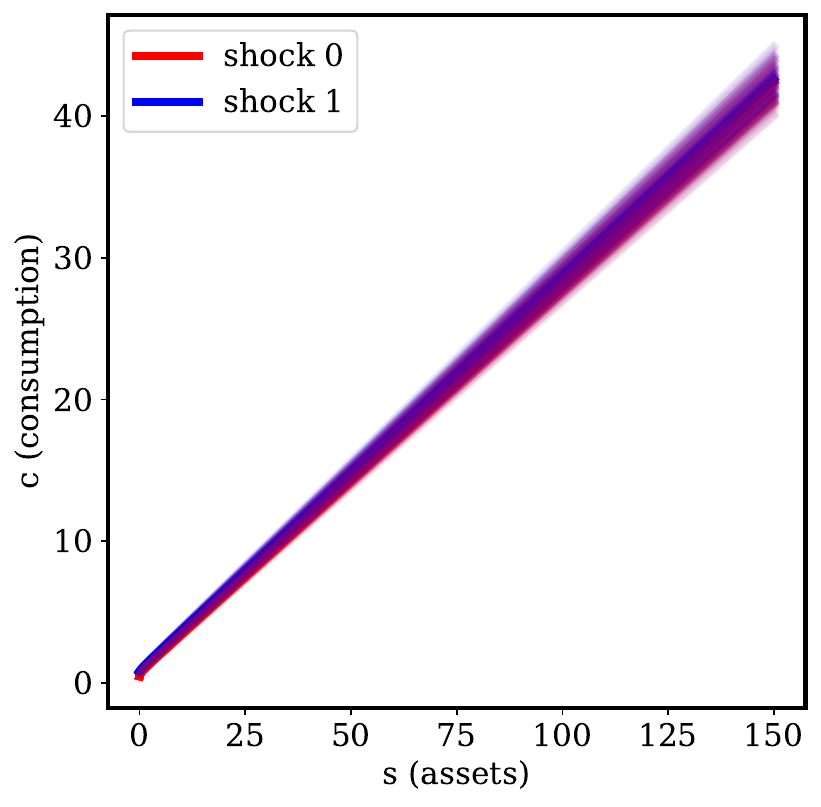}
    \includegraphics[width=0.32\linewidth]{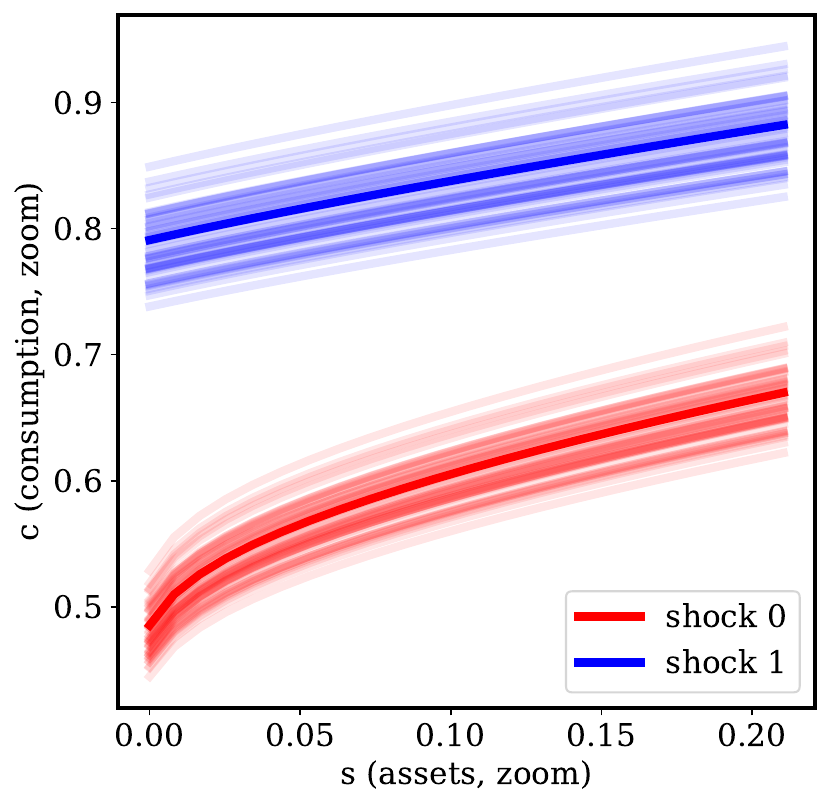}
    \includegraphics[width=0.32\linewidth]{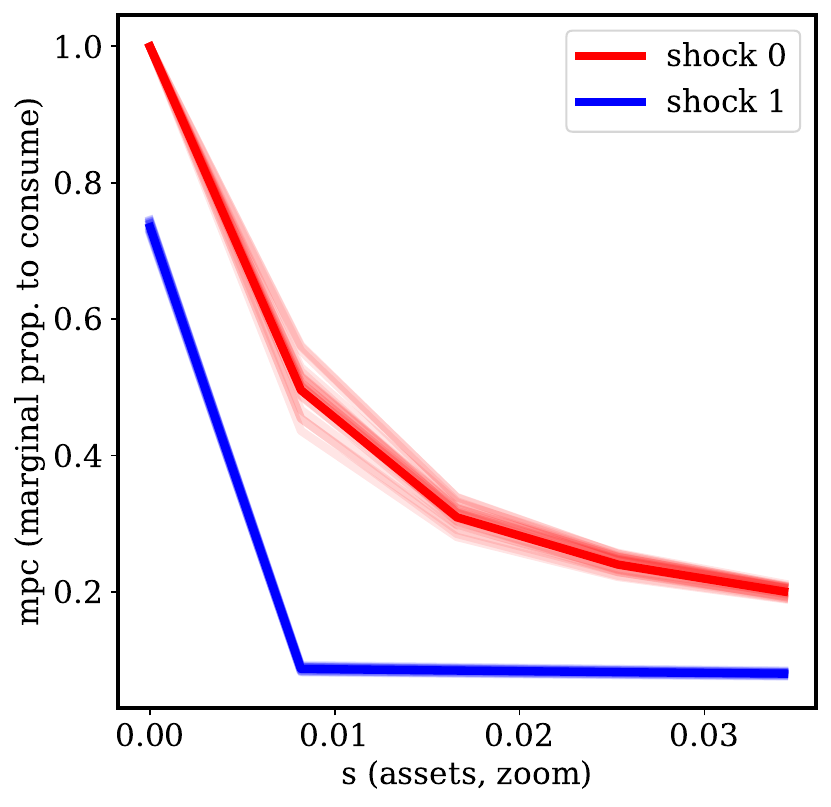}
    \caption{Left and middle panel: consumption function of households for their idiosyncratic productivity level (in different colors) and for different idiosyncratic asset holdings (horizontal axis). The shaded lines show the consumption policies for several different aggregate states, and the solid lines show the mean across aggregate states. Right panel: marginal propensity to consume for different productivity and wealth levels, as well as for several aggregate states. For the asset level $s=0$, the plot shows the consumption share out of cash at hand. \label{fig:het_policy_households}}
\end{figure}

Figure \ref{fig:het_policy_households} shows the household policies predicted by the neural network. Each of the shaded lines corresponds to a different aggregate state, and the solid lines correspond to the mean, averaged over aggregate states, for a specific idiosyncratic productivity and wealth level.
As figure \ref{fig:het_policy_households} shows, consumption varies substantially between different aggregate states. We can also see that, for each of the aggregate states, consumption is increasing and concave in the wealth of households. Our shape-preserving neural network architecture ensures that this economically important feature of the consumption functions is guaranteed. We achieve this using the three-step procedure described above. The households' marginal propensities to consume are constructed from the predictions of the neural network, such that the marginal propensity to consume, for a given aggregate state and idiosyncratic productivity level, is always decreasing in cash-at-hand, as well as bounded between 1 and 0 (see the right panel in figure \ref{fig:het_policy_households}). This ensures on the one hand the monotonicity and concavity of the consumption function, and on the other hand facilitates the precise prediction of the marginal propensity to consume (and hence consumption) of borrowing constraint households.

\section{Application to heterogeneous households with discrete and continuous choices\label{sec:retirement_model}}
In this section, we illustrate that our method is also applicable to economies where individual agents face a mix of discrete and continuous choices. The combination of discrete and continuous choices poses an additional challenge because it often generates convex-concave value functions.\footnote{See, for example,  \cite{fella2014generalized, iskhakov2017endogenous, druedahl2017general}, and \cite{druedahl2021guide}.} As a result, the Euler equations characterizing the continuous choices generally have multiple solutions in the areas of the state space that feature a locally convex value function. Hence, Euler equations alone are not \emph{sufficient} for characterizing agents' behavior. Additionally, the presence of discrete choices can lead to discontinuous policy functions, which can therefore only be approximated with flexible function approximators. Nevertheless, the sequence-space approach, coupled with shape-preserving operator learning allows us to accurately solve a general equilibrium life-cycle model featuring aggregate shocks, uninsurable idiosyncratic risk, and a discrete early retirement decision.\footnote{See \cite{bardoczy2020spousal} for an application of Sequence-Space Jacobians, i.e. the seminal perturbation method developed in \cite{auclertetal_2021}, to models with discrete and continuous choices.}

\subsection{Model}
The economy is inhabited by finitely lived households, who face an \emph{optional} early retirement decision. During their lifetime, households' labor endowment is subject to idiosyncratic risk. Households can self-insure by investing in claims to physical capital subject to a no short-sale constraint. The only source of aggregate risk is a three-state Markov process that determines the depreciation rate of capital.

\subsubsection{Technology}
A representative firm rents capital $K_t$ and efficient units of labor $L_t$ on competitive spot markets and pays a rental rate $r_t^K$ for capital and wage $w_t$ per efficiency unit of labor. The firm produces output $Y_t$ using a Cobb-Douglas production technology
\begin{align}
Y_t &= K_t^\alpha L_t^{1-\alpha} \\
r_t^K &= \alpha K_t^{\alpha - 1} L_t^{1-\alpha} \label{eq:dc_rtk}\\
w_t &= (1-\alpha) K_t^\alpha L_t^{-\alpha} \label{eq:dc_w}.
\end{align}

\subsubsection{Aggregate risk}
The exogenous part of the aggregate state follows a three-state Markov process $z_t \in \mathcal{Z}:=\{z_0,z_1,z_2\}$ with transition matrix $\Pi^z \in \mathbb{R}^{3\times 3}$.
The state $z_t$ determines the depreciation rate of capital
\begin{align}
\delta_t = \delta(z_t).
\end{align}
The return on capital carried from $t$ to $t{+}1$ equals $R_{t+1} = 1-\delta_{t+1}+r_{t+1}^K$.
We interpret $z_2$ as a disaster state with substantially higher depreciation rate. States $z_0$ and $z_1$ are ``normal'' regimes with the same current depreciation rate but different transition probabilities, so they differ in \emph{expected} future depreciation even when $\delta_t$ is identical. 

\subsubsection{Demographics}
Households live for $H=10$ periods. Age is indexed by $h \in \mathcal{H} := \{0,1,\dots,H-1\}$.
Households work for the first six periods,
\begin{align}
\mathcal{H}^{\text{work}} := \{0,1,2,3,4,5\},
\end{align}
make an \emph{optional} retirement decision at age
\begin{align}
h\in \mathcal{H}^{\text{optional}} := \{6\},
\end{align}
and are all retired for the last three periods of their life
\begin{align}
h \in \mathcal{H}^{\text{retired}} := \{7,8,9\}.
\end{align}
We assume a stationary population with equal mass in each cohort. Households die with certainty at the end of the final period of their life. 
Bequests left by the dying cohort become the initial assets of newborns in the next period.

\subsubsection{Idiosyncratic labor productivity}
Each household draws an idiosyncratic productivity state $e_t \in \mathcal{E}^{\text{grid}} = \{e_0,\dots,e_{N_e-1}\}$ that follows a time-homogeneous Markov chain with transition matrix $\Pi^e$.
Efficiency units of labor supplied by a working household of age $h$ are $n^h e_t$, where $n^h$ follows a deterministic life-cycle profile.
Idiosyncratic productivity is dynastic: newborns inherit the family productivity state (on top of the remaining wealth at death), which continues to evolve according to $\Pi^e$ across generations.

\subsubsection{Preferences, optional retirement, and bequests}
Households have CRRA utility over consumption,
\begin{align}
u(c) = \frac{c^{1-\gamma}-1}{1-\gamma}.
\end{align}
At the optional retirement age $h=6$, households choose whether to work ($\ell=1$) or retire early ($\ell=0$).
Working in the optional period carries a deterministic utility cost $\epsilon>0$.
To obtain non-degenerate choice probabilities, we add i.i.d.\ Type-I extreme value taste shocks $\varepsilon_i$ to the discrete alternatives $i\in\{\text{work},\text{ret}\}$.
Let $\sigma>0$ denote the scale parameter.
Finally, in the last period of life ($h=9$) households have a warm-glow bequest motive over the assets passed on to newborns in the next period.
Let $\psi^b>0$ denote the bequest strength, $\nu>0$ its curvature, and $\bar b>0$ a shifter:
\begin{align}
b(k_{t+1}) = \psi^b \frac{\left(k_{t+1} + \bar b\right)^{1-\nu}-1}{1-\nu}.
\end{align}

\subsubsection{Asset markets and non-market income}
Capital is the only asset. Households face a borrowing constraint
\begin{align}
k_{t+1} \geq \underline{k}. \label{eq:dc_bc}
\end{align}
Retired households (including early retirees) receive a constant income $y^{\text{home}}$ from home production. 

\subsubsection{Household problem}
Let the individual labor choice be $\ell_t\in\{0,1\}$ (to avoid confusion with aggregate labor $L_t$).
Outside the optional retirement age, $\ell_t$ is exogenously fixed:
\begin{align}
\ell_t = 1 \ \text{for } h\in\mathcal{H}^{\text{work}}, 
\qquad
\ell_t \in \{0,1\}\ \text{for } h=6,
\qquad
\ell_t = 0 \ \text{for } h\in\mathcal{H}^{\text{retired}}.
\end{align}
The household budget constraint is
\begin{align}
c_t = R_t k_t + y_t(h,e_t,\ell_t) - k_{t+1}, \label{eq:dc_budget}
\end{align}
where income $y_t(h, e_t, l)$ is  given by
\begin{align}
y_t(h,e,\ell) :=
\begin{cases}
w_t\, n^h e, & \text{if }\ell=1,\\
y^{\text{home}}, & \text{if }\ell=0.
\end{cases}
\end{align}
It is helpful to define the \emph{savings-choice-specific} continuation value
\begin{align}
W_t(h, e_t, k_{t+1}) =
\begin{cases}
\mathbb{E}_t\!\left[V_{t+1}(h+1,e_{t+1},k_{t+1})\right], & \text{for } h<9,\\[3pt]
b(k_{t+1}), & \text{for } h=9,
\end{cases}
\end{align}
where $\mathbb{E}_t[\cdot]$ is taken with respect to the idiosyncratic transition $\Pi^e$ and the aggregate regime transition $\Pi^z$, conditional on current states and where $V(\cdot)$ denotes the value function. The savings-choice-specific continuation values, $W_t(h, e_t, k_{t+1})$, are the equilibrium objects, which we approximate with neural networks.

\paragraph{Continuous savings problem (all ages except $h=6$)}
For $h\in  \left\{\mathcal{H}^{\text{work}} \cup  \mathcal{H}^{\text{retired}}\right\} $, the Bellman equation is
\begin{align}
V_t(h,e_t,k_t)
=
\max_{k_{t+1}\ge \underline{k}}
\left\{
u(c_t) + \beta W_t(h, e_t, k_{t+1})
\right\},
\end{align}
subject to the borrowing constraint \eqref{eq:dc_bc} and the budget constraint \eqref{eq:dc_budget}.

\paragraph{Optional retirement at $h=6$}
At age $h=6$ we work with discrete-choice-specific value functions excluding taste shocks.
Let $\tilde V_t^{\text{work}}(e_t,k_t)$ and $\tilde V_t^{\text{ret}}(e_t,k_t)$ solve
\begin{align}
\tilde V_t^{\text{work}}(e_t,k_t)
&= \max_{k_{t+1}\ge \underline{k}}
\left\{
u(c_t^{\text{work}}) - \epsilon + \beta W_t(6, e_t, k_{t+1})
\right\}, \\
\tilde V_t^{\text{ret}}(e_t,k_t)
&= \max_{k_{t+1}\ge \underline{k}}
\left\{
u(c_t^{\text{ret}}) + \beta W_t(6, e_t, k_{t+1})
\right\},
\end{align}
where $c_t^{\text{work}}$ uses $\ell=1$ (labor income $w_t n^6 e_t$) and $c_t^{\text{ret}}$ uses $\ell=0$ (home income $y^{\text{home}}$) in the budget constraint \eqref{eq:dc_budget}.
Since both choices lead to retirement from age $h=7$ onward, the conditional continuation term $W_t(6,e_t,k_{t+1})$ is common across the two alternatives.
With i.i.d.\ Type-I extreme value taste shocks of scale $\sigma$, the ex-ante value is
\begin{align}
V_t(6,e_t,k_t)
= \sigma \log\!\Big(
\exp(\tilde V_t^{\text{work}}(e_t,k_t)/\sigma)
+\exp(\tilde V_t^{\text{ret}}(e_t,k_t)/\sigma)
\Big),
\end{align}
and the conditional probability of working is
\begin{align}
p_t^{\text{work}}(e_t,k_t)
= \frac{\exp\!\big(\tilde V_t^{\text{work}}(e_t,k_t)/\sigma\big)}
{\exp\!\big(\tilde V_t^{\text{work}}(e_t,k_t)/\sigma\big)+\exp\!\big(\tilde V_t^{\text{ret}}(e_t,k_t)/\sigma\big)}.
\end{align}
In the limit $\sigma\to 0$, this expression converges to the deterministic discrete choice.

\subsubsection{Equilibrium}
\paragraph{State space}
The aggregate state is
\begin{equation}
\Xstagg_t = [z_t, \mu_t(h, e, k)],
\end{equation}
where $z_t$ is the aggregate depreciation regime and $\mu_t(h, e, k)$ is the cross-sectional distribution of households over age, idiosyncratic productivity and assets.

\paragraph{Aggregation and prices}
The aggregate capital supply is given by
\begin{align}
K_t &= \sum_{h\in \mathcal{H}}\sum_{e \in \mathcal{E}^{\text{grid}}}\int_k k_t(h, e,k) \ \mu_t(h,e,k) \mathrm{d}k.
\end{align}
Efficient units of labor supplied by an individual of age $h$ are $\ell_t(h,e,k)\,n^h e$, where
\begin{align}
\ell_t(h,e,k) =
\begin{cases}
1, & \text{for } h \in \mathcal{H}^{\text{work}},\\
p_t^{\text{work}}(e,k), & \text{for } h=6,\\
0, & \text{for } h \in \mathcal{H}^{\text{retired}}.
\end{cases}
\end{align}
The aggregate labor supply is given by
\begin{align}
L_t &= \sum_{h\in \mathcal{H}}\sum_{e \in \mathcal{E}^{\text{grid}}}\int_k \ell_t(h, e,k) n^h\, e  \ \mu_t(h,e,k) \mathrm{d}k.
\end{align}
It is useful to summarize the optional cohort's labor supply by the fraction of its \emph{maximum} possible efficiency units supplied:
\begin{align}
L_t^{6,\max} &:= \sum_{e \in \mathcal{E}^{\text{grid}}}\int_k n^6\, e \ \mu_t(6,e,k) \mathrm{d}k,\\
L_t^{6} &:= \sum_{e \in \mathcal{E}^{\text{grid}}}\int_k  p_t^{\text{work}}(e, k)\, n^6\, e \ \mu_t(6,e,k) \mathrm{d}k,\\
\lambda_t &:= \frac{L_t^{6}}{L_t^{6,\max}} \in [0,1].
\end{align}
By construction, $\lambda_t=1$ if no one retires early at $h=6$, and $\lambda_t=0$ if everyone retires early.
Using $\lambda_t$, aggregate labor can be written as
\begin{align}
L_t
=
\sum_{h\in \mathcal{H}^{\text{work}}}\sum_{e \in \mathcal{E}^{\text{grid}}}\int_k  n^h\, e  \ \mu_t(h,e,k) \mathrm{d}k.
\;+\;
\lambda_t\, L_t^{6,\max}.
\end{align}
Given $(K_t,L_t)$, prices can be computed in closed form using the firm's first-order conditions \eqref{eq:dc_rtk} and \eqref{eq:dc_w}.
Market clearing conditions consist of:
(i) capital-market clearing, where the representative firm rents the aggregate household capital stock $K_t$;
(ii) labor-market clearing, where firm labor demand equals aggregate labor supply $L_t$; and (iii) goods-market clearing. Given that (i) and (ii) hold, (iii) is implied by Walras' law.

\paragraph{Functional rational expectations equilibrium}
A functional rational expectations equilibrium consists of (i) cohort- and productivity-specific savings and consumption functions for all ages, (ii) choice-specific savings and value functions at the optional retirement age $h=6$ that imply a logit working probability $p_t^{\text{work}}$, and (iii) an induced law of motion for $\mu_t$ such that households' choices are optimal given prices implied by $(K_t,L_t)$ and the evolution of aggregate regimes, and markets clear.

\subsection{Parameterization}
One model period corresponds to six calendar years.
We set $\gamma=2$ and $\beta = 0.95^6 \approx 0.735$.
Idiosyncratic productivity takes three values $e\in\mathcal{E}^{\text{grid}}=\{0.5,1.0,1.5\}$ with transition matrix
\begin{align}
\Pi^e = \begin{bmatrix}
0.70 & 0.20 & 0.10 \\
0.15 & 0.70 & 0.15 \\
0.10 & 0.20 & 0.70
\end{bmatrix}.
\end{align}
We impose $\underline{k}=0$ and approximate the asset space with an evenly spaced grid of $N_k=300$ points on $[0,5.0]$.
We set $\alpha=\frac{1}{3}$.
Depreciation is regime-dependent with
\begin{align}
\delta(z_0)=\delta(z_1)=\delta_n,\qquad \delta(z_2)=\delta_d,
\end{align}
where $\delta_n = 1-(1-0.05)^6 \approx 0.265$ and $\delta_d = 1.5\, \delta_n \approx 0.397$.
The Markov transition matrix for $z_t$ is
\begin{align}
\Pi^z = \begin{bmatrix}
0.90 & 0.07 & 0.03 \\
0.70 & 0.20 & 0.10 \\
0.50 & 0.25 & 0.25
\end{bmatrix}.
\end{align}
The life-cycle component $n^h$ is set to
\begin{align}
(n^0,\dots,n^6) = (0.53,\,0.83,\,1.04,\,1.17,\,1.21,\,1.17,\,1.04),
\end{align}
and $n^h=0$ for $h\in\{7,8,9\}$.
The deterministic utility cost of working in the optional period is set to $\epsilon=1.68$ and the logit smoothing parameter to $\sigma=0.08$.
Retirees receive $y^{\text{home}}=0.1$ each period. For bequests, we set $\psi^b=0.1$, $\nu=2$, and $\bar b = 0.1$.
The parameters are summarized in table \ref{tab:params_retirement}.

\subsection{Implementation}
We rely on the monotone operator framework introduced in section \ref{sec:operatorlearning}. In contrast to the previous applications, we chose to approximate the choice-specific continuation values.
Given prices and the predicted continuation values, savings policies and the model-implied work probabilities are computed via generalized EGM (with an upper-envelope step) and the logit choice rule.

\subsubsection{State representation}
The endogenous state is the cross-sectional distribution $\mu_t$ over age $h\in\mathcal{H}$, idiosyncratic productivity $e\in\mathcal{E}^{\text{grid}}$, and assets $k\in\mathcal{K}^{\text{grid}}$.
We represent $\mu_t$ as a histogram on the tensor grid
\[
\mathcal{G}^{\text{grid}} := \mathcal{H}\times \mathcal{E}^{\text{grid}}\times \mathcal{K}^{\text{grid}},
\]
with $H=10$, $N_e=3$, and $N_k=200$ grid points. We write $\mu_t(h,e_i,k_j)$ for the mass in cell $(h,e_i,k_j)$.
Following the sequence-space approach, the neural operator takes a truncated history of regime realizations as input
\begin{align*}
\Xse_t=(z_t,z_{t-1},\dots,z_{t-T+1}).
\end{align*}

\subsubsection{Household policies with operator learning}
Let $\nn$ denote the neural operator with parameters $\brho $.
Given a history $\Xse_t$, the operator outputs two objects: (i) grids of continuation values for all cohorts except the terminal cohort, and (ii) a scalar guess for optional-cohort labor supply.

\paragraph{Continuation values}
We predict the savings-choice-specific continuation value $W_t(h,e,k')$ on the fixed asset grid for $h\in\{0,\dots,8\}$:
\begin{align}
\nn^{W}(\Xse_t)
=
\hat{\mathcal{W}}_t
=
\{\hat W_t(h,e_i,k_j)\;|\; h\le 8,\ e_i\in\mathcal{E}^{\text{grid}},\ k_j\in\mathcal{K}^{\text{grid}}\}.
\end{align}
For the terminal cohort, the continuation value is given by the bequest function, $W_t(9,e,k')=b(k')$.

To enforce monotonicity in assets, we parameterize each $\hat W_t(h,e,\cdot)$ by a boundary value and nonnegative increments.
Specifically, the network outputs $\hat{W}_t(h,e,k_0)$ and raw increments $\tilde\Delta_t(h,e,k_j)$ for $j=1,\dots,N_k - 1$, sets $\Delta_t(h,e,k_j)=\text{softplus}(\tilde\Delta_t(h,e,k_j))$, and constructs
\[
\hat W_t(h,e,k_j)=\hat W_t(h,e,k_0)+\sum_{m=1}^j \Delta_t(h,e,k_m),
\]
which guarantees $\hat W_t(h,e,k)$ is increasing in k.

\paragraph{Optional-cohort labor-supply}
We also output a scalar
\begin{align}
\nn^{\lambda}(\Xse_t)=\hat\lambda_t\in(0,1),
\end{align}
interpreted as the fraction of the optional cohort's \emph{maximum} possible efficiency units supplied (so $\lambda_t=1$ corresponds to no early retirement and $\lambda_t=0$ to universal early retirement).
The scalar prediction $\hat\lambda_t$\footnote{In our implementation we enforce $\hat\lambda_t\in(0,1)$ by applying a sigmoid activation in the output layer of $\nn^{\lambda}$.
} is used only to initialize Newton's method for labor-market clearing (Step~0 in Section~\ref{sec:dc_forward_sim}); the realized $\lambda_t$ is computed from the household discrete-choice problem given equilibrium prices.

\paragraph{From continuation values to policies}
Given $(K_t,z_t)$, a candidate labor input $L$ determines prices $(w_t,r_t^K)$.
Using these prices and the predicted continuation values $\hat{\mathcal{W}}_t$, we compute savings and consumption policies for all cohorts via generalized EGM.
For worker cohorts we apply an upper-envelope step to handle potential nonconcavities.
At $h=6$, we solve the two conditional saving problems (work/retire), compute the corresponding choice-specific value functions, and obtain the model-implied work probabilities $p_t^{\text{work}}(e,k)$ from the logit formula.

\subsubsection{Loss function}
Let the network outputs be the continuation-value grids $\hat{\mathcal{W}}_t$ and the scalar $\hat\lambda_t$.
Given supervised learning targets $(\mathcal{W}_t^{\text{tar}},\lambda_t^{\text{tar}})$ constructed by the equilibrium routine described in Section~\ref{sec:dc_training}, we minimize mean squared \emph{relative} errors on economically interpretable components: (i) the fraction of optional-cohort labor-supply, (ii) boundary values of the continuation-value, and (iii) slope of the continuation-value on the asset grid.

For a small $\varepsilon_{\text{loss}}>0$,\footnote{Since the slope of the continuation value can become close to zero during training, we use $\varepsilon_{\text{loss}}>0$ in order to define approximately relative errors.} define
\begin{align}
\text{err}_\lambda
&=
\frac{\lambda_t^{\text{tar}}-\hat\lambda_t}{\varepsilon_{\text{loss}}+\lambda_t^{\text{tar}}},\\
\text{err}_{W,0}(h,e)
&=
\frac{W_t^{\text{tar}}(h,e,k_1)-\hat W_t(h,e,k_1)}{\varepsilon_{\text{loss}}+\big|W_t^{\text{tar}}(h,e,k_1)\big|},\\
\text{err}_{W,\Delta}(h,e,j)
&=
\frac{\Delta W_t^{\text{tar}}(h,e,k_j)-\Delta \hat W_t(h,e,k_j)}{\varepsilon_{\text{loss}}+\big|\Delta W_t^{\text{tar}}(h,e,k_j)\big|},
\end{align}
where $\Delta W(h,e,k_j)=W(h,e,k_j)-W(h,e,k_{j-1})$ and $\Delta \hat W_t(h,e,k_j):=\hat W_t(h,e,k_j)-\hat W_t(h,e,k_{j-1})$.
The loss is the mean of $\text{err}_\lambda^2$, $\text{err}_{W,0}^2$, and $\text{err}_{W,\Delta}^2$ over grid indices and sampled aggregate state-history pairs:
\begin{align}
\mathcal{L}
&=
\frac{1}{B}\sum_{b=1}^B \text{err}_{\lambda,b}^2
+
\frac{1}{B}\sum_{b=1}^B \frac{1}{|\mathcal{I}_{W,0}|}\sum_{(h,e)\in\mathcal{I}_{W,0}} \text{err}_{W,0,b}(h,e)^2
+
\frac{1}{B}\sum_{b=1}^B \frac{1}{|\mathcal{I}_{W,\Delta}|}\sum_{(h,e,j)\in\mathcal{I}_{W,\Delta}} \text{err}_{W,\Delta,b}(h,e,j)^2,
\end{align}
where $\mathcal{I}_{W,0}=\{(h,e):h\le 8, e\in\mathcal{E}^{\text{grid}}\}$ and $\mathcal{I}_{W,\Delta}=\{(h,e,j):h\le 8, e\in\mathcal{E}^{\text{grid}}, j=2,\ldots,N_k\}$.
In the baseline implementation we set $\varepsilon_{\text{loss}}=10^{-2}$.

\subsubsection{Forward simulation}
\label{sec:dc_forward_sim}
This subsection describes the forward simulation step that maps the current aggregate state $(z_t,\mu_t)$ and regime history $\Xse_t$ into a successor state-history pair used for training, and (as part of the same routine) constructs the objects needed to form continuation-value targets.

\paragraph{Step 0: equilibrium policies and labor market clearing}
For each state-history pair $(z_t,\mu_t,\Xse_t)$ we first solve for the within-period general equilibrium objects and policy functions.
Given the current histogram $\mu_t$, aggregate capital is predetermined.
Aggregate labor $L_t$ is unknown because (i) wages and the rental rate follow from $(K_t,L_t,z_t)$ and (ii) the optional-cohort participation probability depends on wages and thus on $L_t$.
We therefore solve for a scalar fixed point $L$ satisfying:
\[
ED(L) = L - \mathcal{L}^{hh}(L)=0,
\]
where $\mathcal{L}^{hh}(L)$ is the aggregate labor implied by household policies when prices are computed for aggregates $(K_t,L,z_t)$.
In our implementation we take a small, fixed number of Newton steps and obtain $ED'(L)$ by automatic differentiation.
A convenient initial guess for $L$ is constructed as $L^{\text{guess}} = L^{\text{work}} + \hat\lambda_t\,L^{6,\max}$, where $L^{\text{work}}$ are efficiency units supplied by the non-optional working cohorts, $L^{6,\max}$ is the optional cohort's maximum possible efficiency units (i.e.\ if nobody retires early), and $\hat\lambda_t\in(0,1)$ is the network-predicted fraction of that maximum.

Each evaluation of $ED(L)$ calls the generalized EGM to solve the household problems at all cohorts given (i) prices implied by $(K_t,L,z_t)$ and (ii) the network-predicted continuation objects for the relevant cohorts.
This yields: (a) savings policies for worker cohorts $\{g_t^h(e,k)\}_{h\in\mathcal{H}^{\text{work}}}$, (b) the two conditional savings policies at the optional cohort, $g_t^{6,\text{work}}(e,k)$ and $g_t^{6,\text{ret}}(e,k)$, together with the logit work probabilities $p_t^{\text{work}}(e,k)$, and (c) savings policies for retired cohorts including the terminal bequest policy.
The converged Newton iterate provides the equilibrium $L_t$ and the policy bundle that is used in the distribution transition below.

\paragraph{Step 1: simulating the distribution forward}
Let $\mathcal{K}^{\text{grid}}=\{k_1<\dots<k_{N_k}\}$ and $\mathcal{E}^{\text{grid}}=\{e_1,\dots,e_{N_e}\}$.
Write $\mu_t(h,i,j)$ for the histogram mass at age $h$, productivity $e_i$, and assets $k_j$.
For a cohort-specific policy $g(e_i,k_j)$, the implied next assets $k' = g(e_i,k_j)$ typically lie off-grid.
Following the non-stochastic simulation method introduced by \cite{young2010solving}, we project mass to the fixed grid by splitting it linearly between the two bracketing grid points.
Let $j^-(i,j)$ and $j^+(i,j)$ denote indices such that $k_{j^-}\le k'\le k_{j^+}$ and define weights
\[
\omega^-(k')=\frac{k_{j^+}-k'}{k_{j^+}-k_{j^-}},
\qquad
\omega^+(k')=1-\omega^-(k').
\]
Idiosyncratic productivity transitions are applied exactly using the transition matrix $\Pi^e$.
Define the corresponding transport operator $\mathcal{T}$ acting on a mass array $m(i,j)$:
\begin{align}
\big[\mathcal{T}(g,m)\big](i',j')
=
\sum_{i=1}^{N_e}\sum_{j=1}^{N_k}
m(i,j)\,\Pi^e_{i i'}\,
\Big(
\omega^-(g(i,j))\,\mathbf{1}\{j'=j^-(i,j)\}
+
\omega^+(g(i,j))\,\mathbf{1}\{j'=j^+(i,j)\}
\Big).
\label{eq:dc_transport}
\end{align}
This step is deterministic and conserves mass by construction.

\paragraph{Step 2: cohort aging, optional retirement split, and newborns}
Using \eqref{eq:dc_transport}, the histogram update is given by
\begin{align}
\mu_{t+1}(0,\cdot,\cdot)
&=
\mathcal{T}\!\left(g_t^{9},\,\mu_t(9,\cdot,\cdot)\right),
\label{eq:dc_forward_newborn}
\\
\mu_{t+1}(h+1,\cdot,\cdot)
&=
\mathcal{T}\!\left(g_t^{h},\,\mu_t(h,\cdot,\cdot)\right),
\qquad h\in\{0,1,2,3,4,5\},
\label{eq:dc_forward_work}
\\
\mu_{t+1}(7,\cdot,\cdot)
&=
\mathcal{T}\!\left(g_t^{6,\text{work}},\,p_t^{\text{work}}\odot \mu_t(6,\cdot,\cdot)\right)
+
\mathcal{T}\!\left(g_t^{6,\text{ret}},\,(1-p_t^{\text{work}})\odot \mu_t(6,\cdot,\cdot)\right),
\label{eq:dc_forward_optional}
\\
\mu_{t+1}(h+1,\cdot,\cdot)
&=
\mathcal{T}\!\left(g_t^{h},\,\mu_t(h,\cdot,\cdot)\right),
\qquad h\in\{7,8\}.
\label{eq:dc_forward_ret}
\end{align}
Equation \eqref{eq:dc_forward_optional} is the only place where the discrete choice enters the forward step: the optional cohort's mass is split according to the EGM-implied logit probability $p_t^{\text{work}}(e,k)$, and each part follows its corresponding conditional savings policy.
Equation \eqref{eq:dc_forward_newborn} maps the assets of the terminal cohort to the newborn distribution via the bequest policy; since \eqref{eq:dc_transport} applies $\Pi^e$, productivity transitions are applied uniformly in all cohort transitions, including between generations.

\paragraph{Step 3: selection of the next state-history pair}
The distribution update \eqref{eq:dc_forward_newborn}--\eqref{eq:dc_forward_ret} does not depend on the realized $z_{t+1}$; only the regime component changes.
Hence we form candidate successor states for all regimes $z'\in\{z_0,z_1,z_2\}$ as
\[
\Xstagg_{t+1}(z') = (z',\mu_{t+1}),
\qquad
\Xse_{t+1}(z') = (z', z_t, z_{t-1}, \dots, z_{t-T+2}).
\]
To evolve the simulated training state-history pairs, we draw a realized regime $z_{t+1}\sim \Pi^z(z_t,\cdot)$ and select the corresponding $(\Xstagg_{t+1}(z_{t+1}),\Xse_{t+1}(z_{t+1}))$.

\subsection{Training\label{sec:dc_training}}

\subsubsection{Outline of the training procedure}
We train the neural network with supervised learning on simulated data.
Each training point consists of (i) an aggregate regime $z_t$, (ii) the cross-sectional distribution $\mu_t$ on the tensor grid $(h,e,k)$, and (iii) a truncated regime history
$\Xse_t=(z_t,z_{t-1},\dots,z_{t-T+1})$ that serves as the network input.
Given $\Xse_t$, the network outputs
(a) continuation-value arrays for all cohorts except the terminal cohort, parameterized in a monotone way over the asset grid (boundary plus positive increments), and
(b) a scalar prediction $\hat\lambda_t\in(0,1)$ for the optional cohort's effective participation rate.
The scalar $\hat\lambda_t$ is used only to initialize the labor-market-clearing routine.
The training procedure is organized in three stages.

\paragraph{Stage 1: deterministic steady state}
We first compute a deterministic steady state with constant depreciation and no regime switching.
We iterate on a guess for $(K,L)$, compute prices, solve the life-cycle problem via EGM (including the optional retirement decision), and simulate the stationary distribution.
The resulting steady-state policy, value objects and stationary distribution are stored and used to initialize (i) the simulated training data and (ii) the pretraining stage below.

\paragraph{Stage 2: stochastic pretraining with steady-state continuation}
We next introduce aggregate risk, but when solving the household problems we hold the \emph{next-period} continuation objects fixed at their steady-state values.
Operationally, the EGM step uses steady-state continuation values on the right-hand side of the Euler equation and in the discrete-choice log-sum, while prices and the current distribution are allowed to vary with the simulated history.
This pretraining stage stabilizes optimization and produces a diversified cloud of $(z_t,\mu_t,\Xse_t)$ before turning on full recursion.

\paragraph{Stage 3: full general equilibrium}
Finally, we train in the full model in which continuation objects depend on the operator itself.
To stabilize the training, we use a target network, whose parameters are updated more slowly.
Let $\brho$ denote the online parameters and $\bar\brho$ the target parameters, updated by
\[
\bar\brho \leftarrow \tau\,\brho + (1-\tau)\,\bar\brho.
\]
At each iteration we take one supervised-learning step on a batch of size $B$.
For each state-history pair $(z_t,\mu_t,\Xse_t)$, one iteration consists of:

\begin{enumerate}
\item \emph{Online prediction at $t$.} Evaluate the online network to obtain predictions
$\nn(\Xse_t)$, including the continuation-value arrays and the scalar participation guess $\hat\lambda_t$.

\item \emph{Target prediction at candidate $t{+}1$ histories.}
Construct the candidate next histories $\{\Xse_{t+1}(z')\}_{z'\in\mathcal{Z}}$ by prepending each possible next regime $z'$ and shifting the history vector.
Evaluate the \emph{target} network on these candidate histories to obtain $\nnt(\Xse_{t+1}(z'))$ for all $z'\in\mathcal{Z}$.

\item \emph{Current equilibrium with market clearing.}
Given $(z_t,\mu_t)$ and the online predictions $\nn(\Xse_t)$, we clear the labor market by solving
\[
ED(L_t) = L_t-\mathcal{L}^{hh}(L_t)=0
\]
with a small, fixed number of Newton--Raphson iterations.
Here $\mathcal{L}^{hh}(L)$ denotes aggregate labor implied by household policies under prices computed at $(K_t,L,z_t)$.
Each evaluation of $ED(L)$ solves the household problems for all cohorts via EGM using the online network's continuation objects.
The derivative $ED'(L)$ used in Newton's method is obtained by automatic differentiation.

This step delivers equilibrium prices and the full policy bundle at $t$, including (i) the worker savings policies, (ii) the optional-cohort work probabilities and the two conditional savings rules (work or retire), and (iii) the retirement and terminal bequest policies, as well as the \emph{realized} optional-cohort participation rate $\lambda_t$.

\item \emph{Forward simulation of the distribution.}
Using the current equilibrium policy bundle from step (3), we update the histogram deterministically to obtain $\mu_{t+1}$, following \cite{young2010solving}.
This produces the aggregate states in the next period $(z',\mu_{t+1})$ for all $z'\in\mathcal{Z}$.

\item \emph{Next-period equilibria for target construction.}
For each possible next regime $z'$, we solve the $t{+}1$ equilibrium as in step (3), but using the \emph{target-network} predictions $\nnt(\Xse_{t+1}(z'))$ to evaluate next-period continuation objects in the EGM step.
We then form continuation-value targets at $t$ by taking expectations over $z_{t+1}$ and $e_{t+1}$ using the Markov transition matrices.

\item \emph{Loss and parameter updates}
We define supervised targets as (i) the continuation-value arrays constructed in step (5) and (ii) the realized optional-cohort effective participation rate $\lambda_t$ from step (3).
The loss is the mean squared relative error between online predictions and the constructed training targets; for continuation values we compute errors on the boundary and on one-step increments (consistent with the monotone parameterization), and for participation we compute a scalar relative error.
Gradients are taken only through the loss function, taking the target values as fixed.
We update the online parameters $\brho$ using the Adam optimizer and update the target parameters $\bar\brho$ by the EMA rule above.

\item \emph{Evolve the training state-history pairs.}
To generate the next on-policy batch, we draw $z_{t+1}\sim\Pi^z(z_t,\cdot)$ and select the corresponding $(z_{t+1},\mu_{t+1},\Xse_{t+1}(z_{t+1}))$ as the next state history pair.
\end{enumerate}

\subsubsection{Hyperparameters}
$\nn$ is a feed-forward network with GELU activations (three hidden layers of width $1024$) and input history length $T=100$.
We train with Adam using a learning rate of $10^{-6}$ and an EMA parameter $\tau=10^{-4}$.
Each iteration uses a $B=1024$ state-history pairs and clears the labor market with $5$ Newton steps.
Relative-error losses use a stabilizer parameter $\varepsilon_{\text{loss}}$ in the denominator to avoid numerical issues when targets are near zero. At the beginning of training we set $\varepsilon_{\text{loss}} = 1.0$. During the training, we reduce $\varepsilon_{\text{loss}}$ first to $0.1$, and finally to $10^{-2}$.

\subsection{Accuracy}

To assess the accuracy of our solution, we calculate the distribution of prediction errors of our continuation value network. The continuation values are shown in figure \ref{fig:ret_model_contvalues} and error statistics are shown in figure \ref{fig:ret_model_errors}. Since the continuation values cross zero, relative errors are not meaningful. Because of that, we report absolute errors across idiosyncratic and aggregate states.

\subsection{Inspecting the equilibrium}
\begin{figure}
    \centering
    \includegraphics[width=0.32\linewidth]{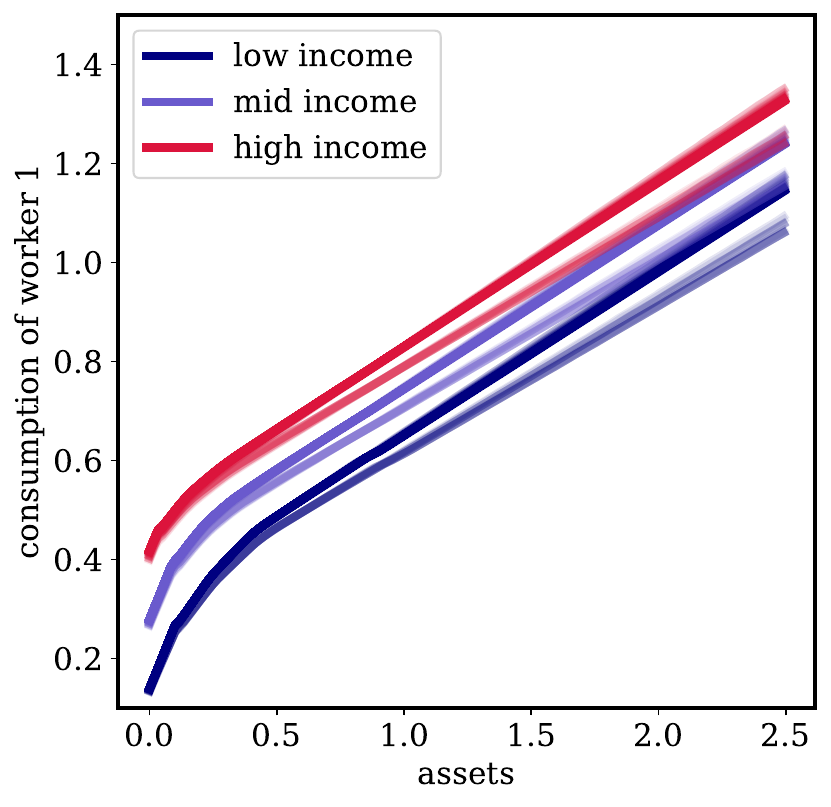}
    \includegraphics[width=0.32\linewidth]{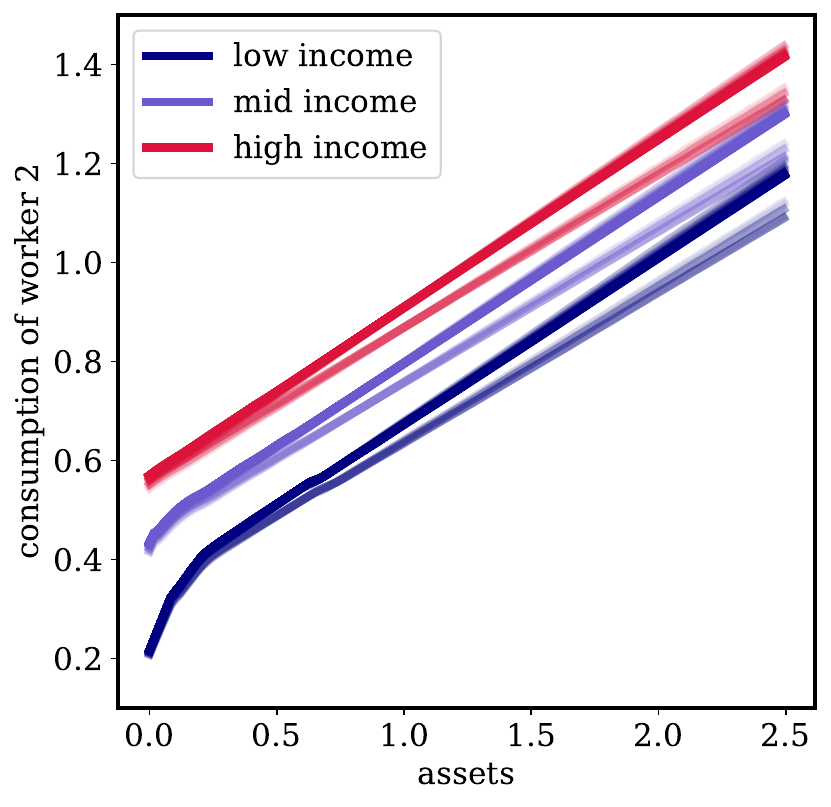}
    \includegraphics[width=0.32\linewidth]{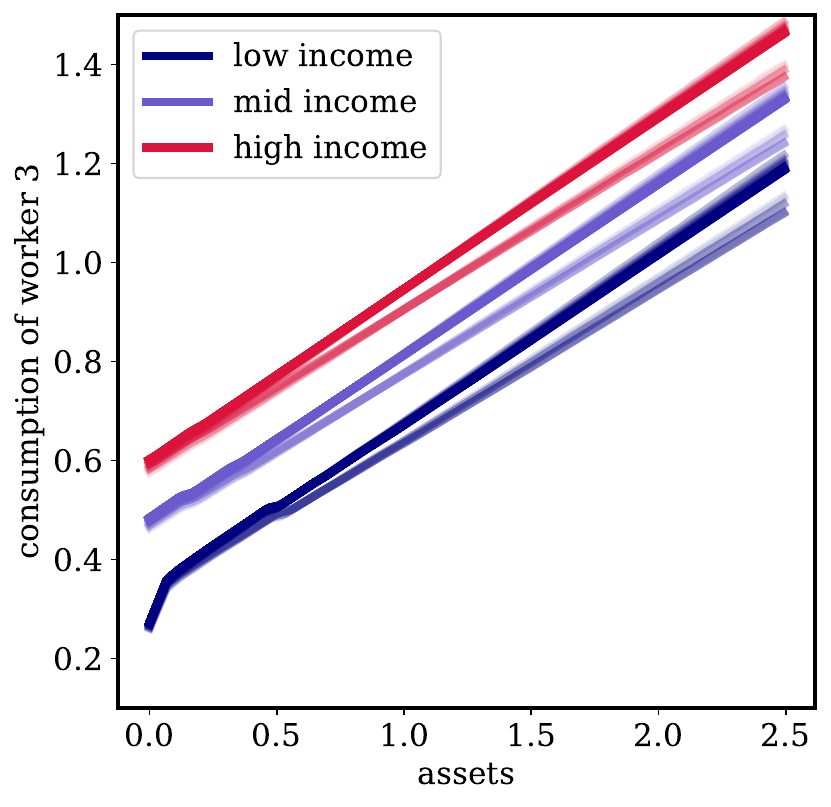}\\
    \includegraphics[width=0.32\linewidth]{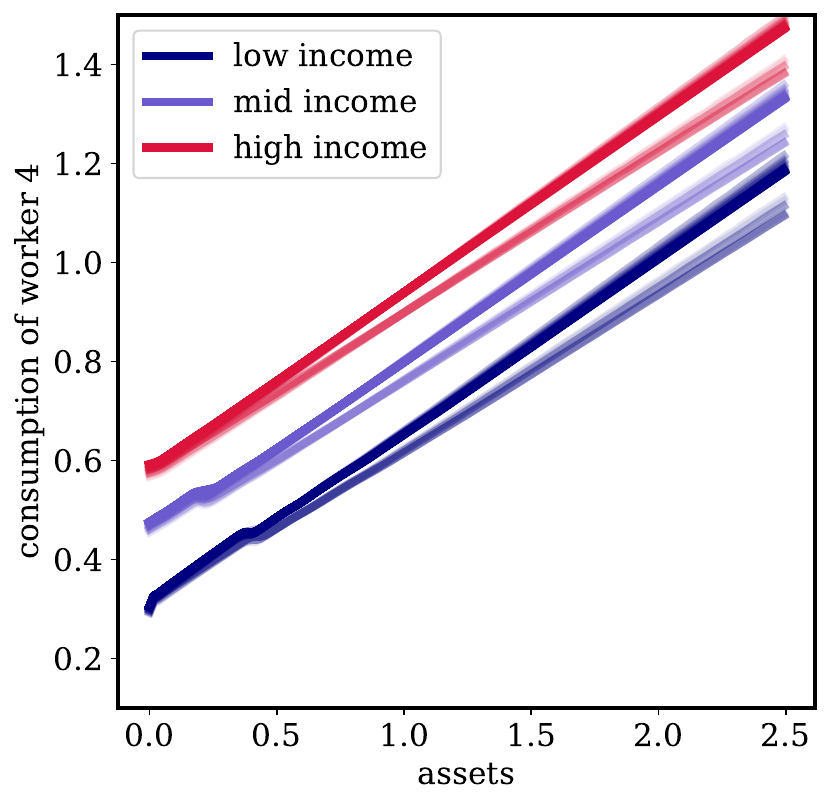}
    \includegraphics[width=0.32\linewidth]{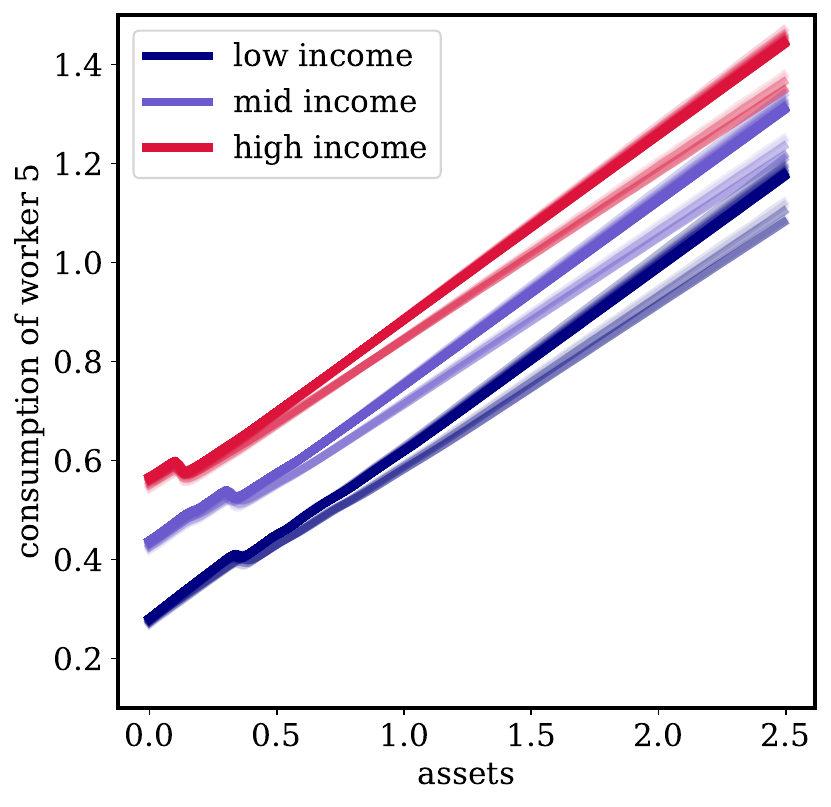}
    \includegraphics[width=0.32\linewidth]{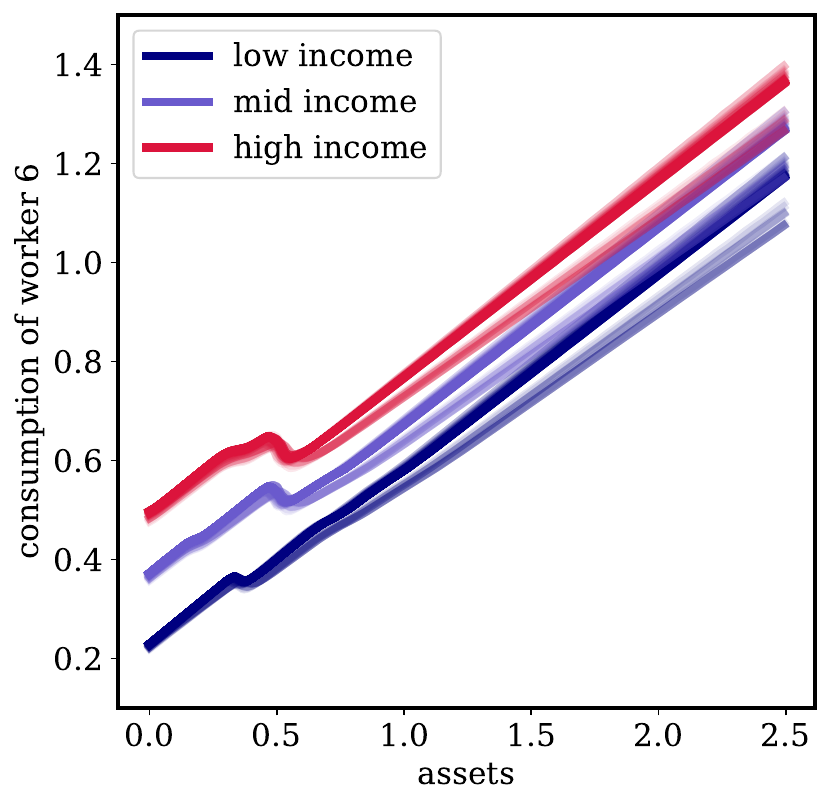}
    \caption{Consumption functions of working-age cohorts. The shaded lines show the policies for different aggregate states, and the solid lines show the average taken across the sample of aggregate states.\label{fig:ret_model_worker_consumption}}
\end{figure}

\begin{figure}
    \centering
\includegraphics[width=0.24\linewidth]{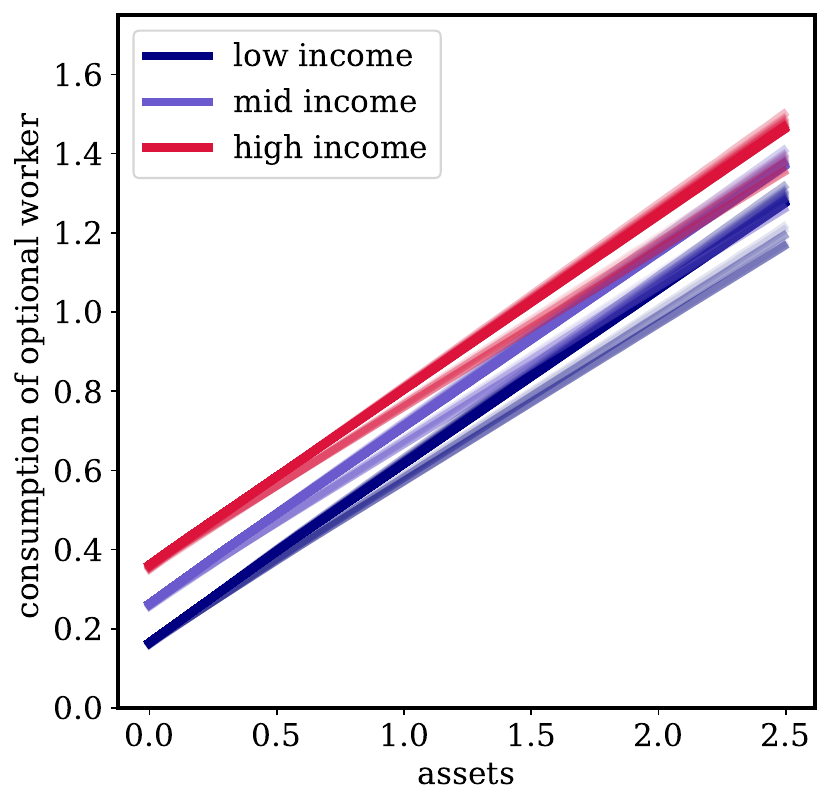}
\includegraphics[width=0.24\linewidth]{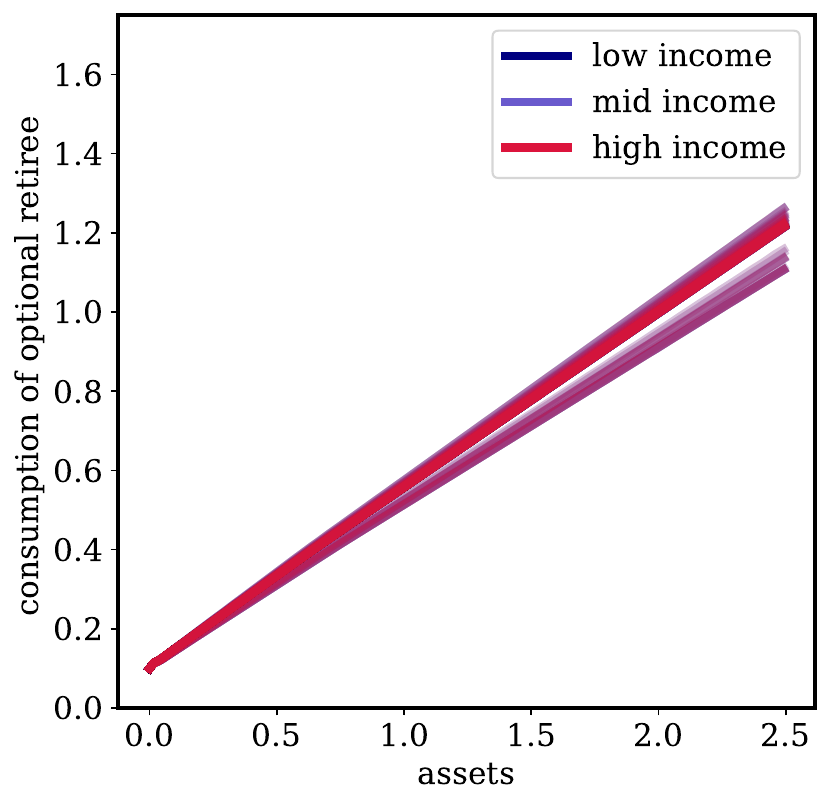}
\includegraphics[width=0.24\linewidth]{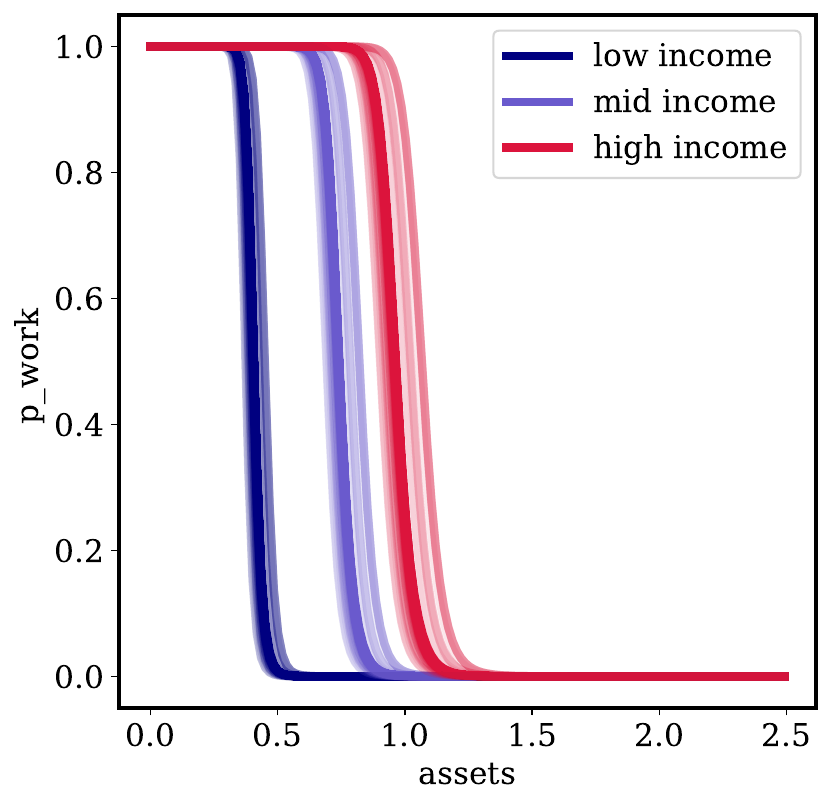}
\includegraphics[width=0.24\linewidth]{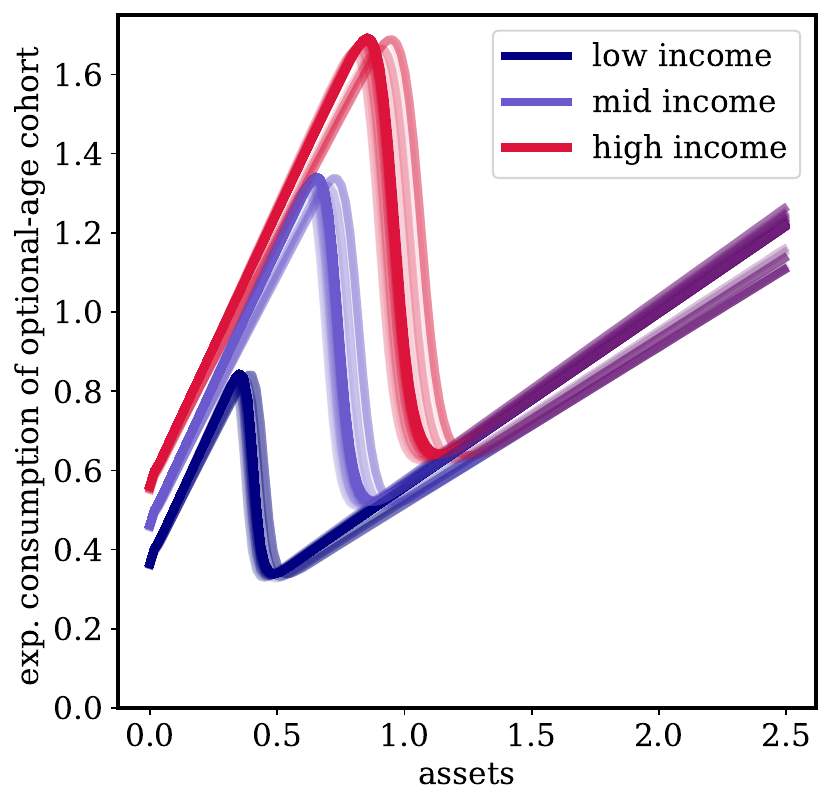}
    \caption{First and second panel: consumption function of the optional retirement cohort conditional on the retirement decision. Third and fourth panel: probability to work in the optional period together with the expected consumption conditional on assets, idiosyncratic productivity and the aggregate history, where the expectation is taken over the taste shock. The shaded lines show different aggregate histories and the solid shows the mean across aggregate histories. \label{fig:ret_model_opt_consumption}}
\end{figure}

In figure \ref{fig:ret_model_worker_consumption} we plot consumption functions for working-age cohorts. Figure \ref{fig:ret_model_opt_consumption} shows conditional consumption functions and work probabilities of the cohort facing optional retirement, and figure \ref{fig:ret_model_retiree_consumption} plots the consumption function of full retirees.
The bottom right panel in figure \ref{fig:ret_model_worker_consumption} shows the consumption functions for working-age households in the period before the optional early-retirement period. The discrete retirement choice leads to discontinuities in the consumption function of pre-retirement cohorts. The discontinuities propagate backward and slowly dissipate as uncertainty concavifies the continuation values of younger cohorts.
The two left panels in figure \ref{fig:ret_model_opt_consumption} show that the consumptions conditional on the retirement decision of the optional cohort are smooth. As shown in the right two panels, the discrete retirement choice, however, renders the expected consumption strongly non-linear and non-monotone. This is the source of the non-concavities in the continuation values and strongly non-linear policies in the prior periods.
Figure \ref{fig:ret_model_retiree_consumption} shows the consumption function of retired households.
Since retired households do not earn wage, their consumption does not depend on their idiosyncratic productivity state.

\section{Conclusion}
We introduce a new algorithm for computing global solutions of dynamic stochastic general equilibrium models. Exploiting the ergodic property of a large class of dynamic economies, we rely on the truncated history of aggregate shocks as an approximately sufficient statistic for the aggregate state. Based on this approximation, we use deep neural networks to parameterize the mapping from truncated aggregate shock histories to equilibrium objects of interest. Finally, we train our neural networks to satisfy all the equilibrium conditions along simulated paths of the economy.

We apply our method to three challenging economies, each constructed to feature a distinct difficulty that commonly arises in dynamic equilibrium models. First, we solve an overlapping generations model with 72 age groups, portfolio choice, and multiple sources of aggregate risk. The overlapping generations model showcases that our sequence space approach is applicable even when there is a clear dependence of current equilibrium objects on long histories of shocks.

Second, we consider an economy where a continuum of heterogeneous households trade in a long-lived financial asset, which constitutes a claim to a share in the dividends paid by a continuum of heterogeneous firms, who operate a decreasing returns to scale production technology and face asymmetric adjustment costs on capital. Additionally, the economy is subject to stochastic fluctuations in aggregate productivity and in the level of aggregate and idiosyncratic uncertainty. To solve models that combine aggregate and idiosyncratic risk, we introduce shape-preserving operator learning: we train deep neural networks to predict idiosyncratic policy \emph{functions} as a function of the truncated history of aggregate shocks, such that we can guarantee the predicted policy functions to be monotone or concave in key idiosyncratic state variables, such as household wealth. These guarantees allow us to use the method of endogenous gridpoints and a simple Newton-Raphson algorithm to obtain high-quality training targets for supervised learning of household policies and the market clearing stock price.

As a final example, we solve for an equilibrium in a life-cycle economy with aggregate and idiosyncratic risk where households have an early retirement option. Solving this model presents three key challenges. First, the early retirement option introduces non-convexities into the decision problems of working-age households. Because of that, Euler equations are not sufficient to pin-down consumption and savings behavior of working-age households. Second, the non-convexities make the policy functions discontinuous: expectations of future retirement introduce jumps in the savings function of workers, moreover the location of those jumps depends on the aggregate state. Third, this economy features an especially high-dimensional state space: there is a non-degenerate cross-sectional wealth distribution within each age cohort. We show that our method remains applicable and computationally tractable even in all three environments. 

\clearpage
\FloatBarrier
\bibliography{Bibliographie}{}
\bibliographystyle{apalike}

\FloatBarrier
\clearpage
\renewcommand{\appendixpagename}{Appendix}

\begin{appendices}
\renewcommand{\thefigure}{A\arabic{figure}}
\setcounter{figure}{0}
\renewcommand{\thetable}{A\arabic{table}}
\setcounter{table}{0}

\section{Additional Tables}
\begin{table}[H]
\small
\centering
\begin{tabular}{lllllll}
\toprule
Parameter & $\gamma$ & $\delta$ & $\beta$ & $\alpha$ & $\rho^A$ & $\sigma^A$ \\
\midrule 
Meaning &  \makecell[l]{relative \\ risk aversion} & \makecell[l]{depreciation \\ of capital} & patience & \makecell[l]{capital share \\ in production} & \makecell[l]{persistence of \\ log TFP} & \makecell[l]{std dev of \\ innovations log TFP}\\ 
Value & 2 & 0.1 & 0.95 & $\frac{1}{3}$ & 0.8 & 0.03\\
\bottomrule
\end{tabular}
\caption{\label{tab:RA_calib}Parameter values chosen for the \protect\cite{brock_1972} model.
}
\end{table}
\begin{table}[H]
\centering
\begin{tabular}{lllllll}
\toprule
Parameter & $N^{\text{quad}}$ & $T$ & $N^{\text{hidden 1}}$ & $N^{\text{hidden 2}}$ & $N^{\text{hidden 3}}$ & $N^{\text{output}}$ \\
\midrule 
Meaning &  \makecell[l]{Quad. \\ nodes} & \makecell[l]{Length of \\ hist. of shocks} & \makecell[l]{\# nodes \\
layer 1 \\ (activation)} & \makecell[l]{\# nodes \\ layer 2 \\ (activation)} & \makecell[l]{\# nodes \\ layer 3 \\ (activation)} & \makecell[l]{\# nodes \\ output layer \\ (activation)}\\ 
\midrule
Value & 8 & 100 & \makecell[l]{128 \\ (gelu)} & \makecell[l]{128 \\ (gelu)} & \makecell[l]{128 \\ (gelu)} & \makecell[l]{1 \\ (sigmoid)} \\
\bottomrule
\toprule
Parameter & $N^{\text{data}}$ & $N^{\text{mb}}$ & $N^{\text{episodes}}$ & Optimizer & $\alpha^{\text{learn}}$\\
\midrule 
Meaning &  \makecell[l]{States  per\\ episode} & \makecell[l]{mini-batch \\ size} & \makecell[l]{\# episodes} & Optimizer & \makecell[l]{learning \\ rate} \\ 
\midrule
Value & 4096 & 256 & 40{,}000 & Adam & $10^{-5}$ \\
\bottomrule
\end{tabular}
\caption{\label{tab:RA_hp}Hyperparameter values chosen for the neural network to solve the \protect\cite{brock_1972} model.
}
\end{table}
\begin{table}[H]
\centering
\begin{tabular}{lll}
\toprule
Parameter &  & Value \\
\midrule
$H$ & Number of cohorts & 72\\
$\gamma$ & Relative risk aversion & 2 \\
$\beta$ & Patience & 0.96 \\
$B$ & Bond supply & 0.75 \\
$\underline{b}$ & Borrowing constr. bond & 0 \\
$\underline{k}$ & Borrowing constr. capital & 0  \\
$\xi^\text{adj}$ & Adj. cost on capital & 0.1\\
$\rho$ & Persistence of log TFP & 0.85 \\
$\sigma^A$ & Std. dev. innovations to log TFP & 0.03\\
$\delta$ & Depreciation of capital in normal times & 0.1 \\
$\rho^\delta$ & Persistence of depr. in disaster & 0\\
$\sigma^\delta$ & Std. dev. of innovations to depreciation & 0.2\\
$\mu^\delta$ & Mean depreciation during disasters & -1.10 \\
$\pi^{n \rightarrow n}$ & Prob. to remain in normal times & 0.94 \\
$\pi^{d \rightarrow d}$ & Prob. to remain in the disaster regime & 2/3\\
\bottomrule
\end{tabular}
\caption{\label{tab:OLG_params}Parameter values for the OLG model in section \protect\ref{sec:OLG}.
}
\end{table}
\begin{table}[H]
\centering
\begin{tabular}{lll}
\toprule
Parameter &  & Value \\
\midrule
$N^{\text{quad}}$ & Number quadrature nodes TFP / depreciation & 4 / 4\\
$T$ & Length of history of shocks & 144 \\
$N^{\text{hidden 1}}$ & \# neurons in the first hidden layer (activation) & 720 (gelu) \\
$N^{\text{hidden 2}}$ & \# neurons in the second hidden layer (activation) & 720 (gelu) \\
$N^{\text{hidden 3}}$ & \# neurons in the third hidden layer (activation) & 720 (gelu) \\
$N^{\text{output}}$ & \# neurons in the output layer (activation) & 214 (None) \\
$N^{\text{data}}$ & States per episode & 8192  \\
$N^{\text{mb}}$ & mini-batch size & 64\\
$N^{\text{episodes}}$, step 1 & Training episodes (capital only) & 512 \\
$N^{\text{episodes}}$, step 2 & Training episodes (pretrain bond price) & 1536\\
$N^{\text{bond steps}}$, step 3 & Number of incremental increases in bond supply & $4$\\
$N^{\text{episodes}}$, step 3 & Training episodes (for each intermediate bond supply) & 16384\\
$N^{\text{episodes}}$, step 4 & Training episodes (final model) & $32768$\\
Optimizer & Optimizer & Adam\\
$\lr$ & Learning rate &  $10^{-5}$ \\
\bottomrule
\end{tabular}
\caption{\label{tab:OLG_hp}Hyper-parameter values for the OLG model in section \protect\ref{sec:OLG}.
}
\end{table}
\begin{table}[H]
\centering
\begin{tabular}{lll}
\toprule
Parameter &  & Value \\
\midrule
$\gamma$ & Relative risk aversion & 2 \\
$\beta$ & Patience & 0.95 \\
$\underline{\theta}$ & Borrowing constraint & 0 \\
$\rho^e$ & Persistence id. income process & 0.871  \\
$\sigma^e$ & Std. dev. id. income process & 0.246\\
$\rho^A$ & Persistence of aggregate TFP & 0.8145 \\
$\sigma^A_L$, $\sigma^A_H$ & Std. dev. innovations to TFP & 1.24\%, 1.99\%\\
$\delta$ & Depreciation of capital & 0.1 \\
$\alpha$ & Capital share in production & 0.25\\
$\zeta$ & Returns to scale & 0.25\\
$\Gamma$ & Survival rate of firms & 0.965\\
$z$ & Idiosyncratic firm productivity & [0.5, 1.0, 1.5] \\
$\Pi_z^L$, $\Pi_z^H$ & Transition matrices for id. firm prod. & See text \\
$\Pi^U_{L, L}$, $\Pi^U_{H, H}$ & Persistence of uncertainty regimes & 0.90, 0.79\\
$\xi^{\text{up}}$, $\xi^{\text{down}}$, $s$ & Adjustment costs firms & 1.0, 2.5, 400\\
\bottomrule
\end{tabular}
\caption{\label{tab:het_params}Parameter values for the economy with heterogeneous firms and heterogeneous households.
}
\end{table}
\begin{table}[H]
\centering
\begin{tabular}{lll}
\toprule
Parameter &  & Value \\
\midrule
$T$ & Length of history of shocks (per shock) & 300 \\
$N^{\text{input}}$ & \# nodes input layer & 600\\
$N^{\text{hidden 1}}$ & \# neurons in the first hidden layer (activation) & 1024 (gelu) \\
$N^{\text{hidden 2}}$ & \# neurons in the second hidden layer (activation) & 1024 (gelu) \\
$N^{\text{hidden 3}}$ & \# neurons in the third hidden layer (activation) & 1024 (gelu) \\
$N^{\text{output}}$ & \# neurons in the output layer & \makecell{$\nn^k$: 600 \\$\nn^\lambda$: 600 \\ $\nn^c$: 200 \\ $\nn^p : 1$} \\
$N^k$ & \# grid points for capital grid (log spaced) & 200 \\
$N^\theta$ & \# grid points for asset grid (log spaced) & 200 \\
$N^{\text{quad}}$ & \# quadrature nodes TFP  & 5\\
$N^{\text{data}}$ & States per episode & 131072  \\
$N^{\text{mb}}$ & mini-batch size & 128\\
$N^{\text{episodes}}$, step 1 & Training episodes & 100 \\
$N^{\text{episodes}}$, step 2 & Training episodes & 100\\
$N^{\text{episodes}}$, step 3 & Training episodes & 100\\
$N^{\text{episodes}}$, step 4 & Training episodes & 500\\
$N^{\text{episodes}}$, step 5 & Training episodes & $16300$\\
Optimizer & Optimizer & Adam\\
$\lr$ & Learning rate & $10^{-6}$ \\
\bottomrule
\end{tabular}
\caption{\label{tab:het_hp}Hyperparameter values for the heterogeneous firms and households model.
}
\end{table}
\begin{table}[H]
\centering
\begin{tabular}{lll}
\toprule
Parameter & Meaning & Value \\
\midrule
$H$ & Number of cohorts & $10$ \\
$\gamma$ & Relative risk aversion & $2$ \\
$\beta$ & Patience (per 6-year period) & $0.95^6 \approx 0.735$ \\
$\alpha$ & Capital share in production & $1/3$ \\
$\underline{k}$ & Borrowing constraint & $0$ \\
$N_k$ & Asset grid points & $300$ \\
$k\in[\underline{k},\bar k]$ & Asset grid bounds & $[0,5]$ \\
$\mathcal{E}^{\text{grid}}$ & Idiosyncratic productivity states & $\{0.5,1.0,1.5\}$ \\
$\Pi^e$ & Transition matrix for $e$ & See text \\
$\delta_n$ & Depreciation (normal) & $1-(1-0.05)^6 \approx 0.265$ \\
$\delta_d$ & Depreciation (disaster) & $1.5\delta_n \approx 0.397$ \\
$\Pi^z$ & Transition matrix for $z$ & See text \\
$\epsilon$ & Utility cost of work at $h=6$ & $1.68$ \\
$\sigma$ & Taste-shock scale (logit smoothing) & $0.08$ \\
$y^{\text{home}}$ & Income when retired & $0.1$ \\
$\psi^b$ & Bequest motive strength & $0.1$ \\
$\nu$ & Bequest curvature & $2$ \\
$\bar b$ & Bequest shifter & $0.1$ \\
\bottomrule
\end{tabular}
\caption{Parameter values for the discrete-continuous choice OLG model in section \protect\ref{sec:retirement_model}.}
\label{tab:params_retirement}
\end{table}

\FloatBarrier
\section{Additional Figures}
\begin{figure}[H]
    \centering
    \includegraphics[width=0.5\linewidth]{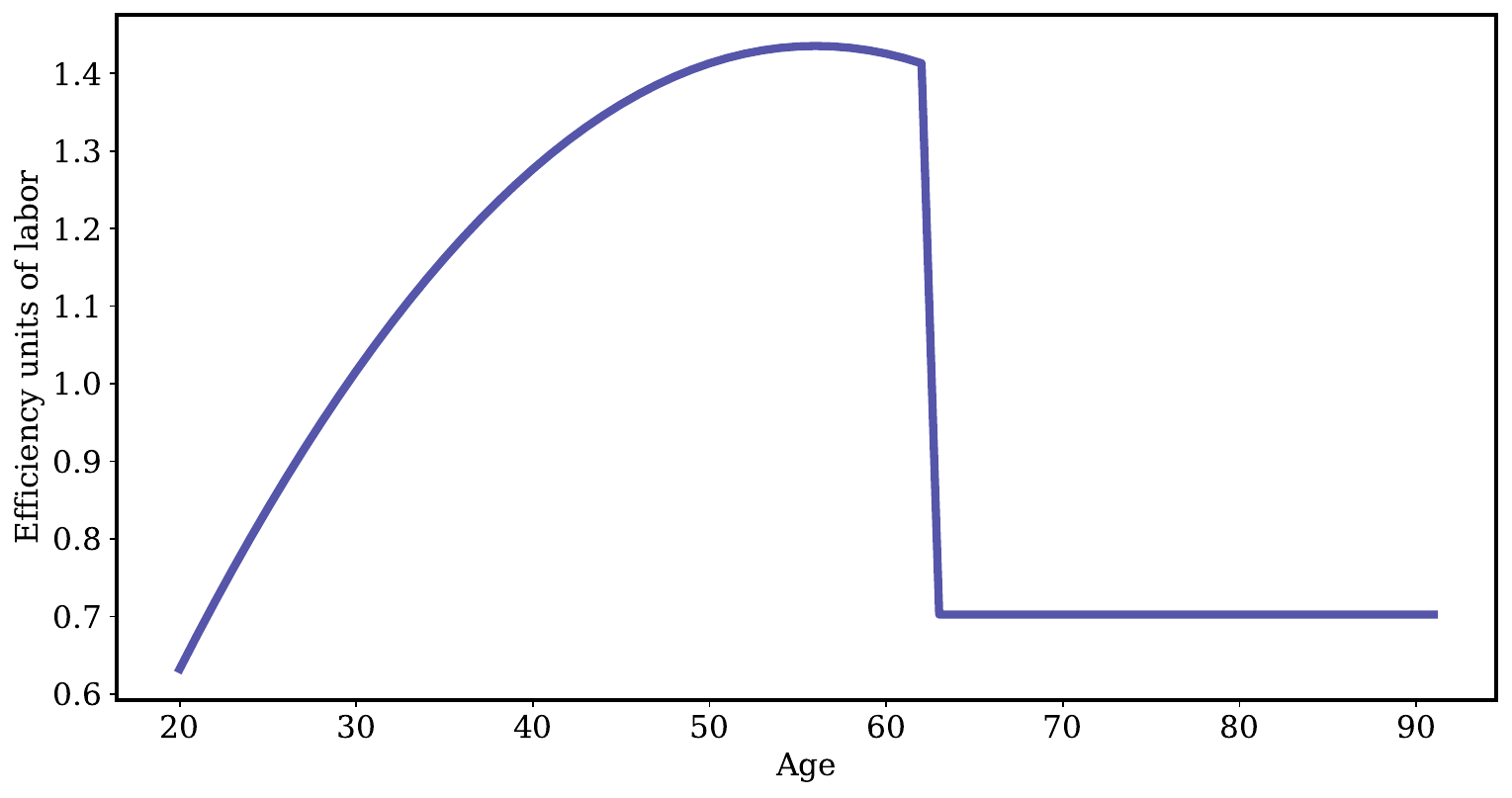}
    \caption{Efficiency units of labor over the life-cycle for the OLG model in section \protect\ref{sec:OLG}.\label{fig:OLG_laborendowment}}
\end{figure}
\begin{figure}
    \centering
    \includegraphics[width=0.32\linewidth]{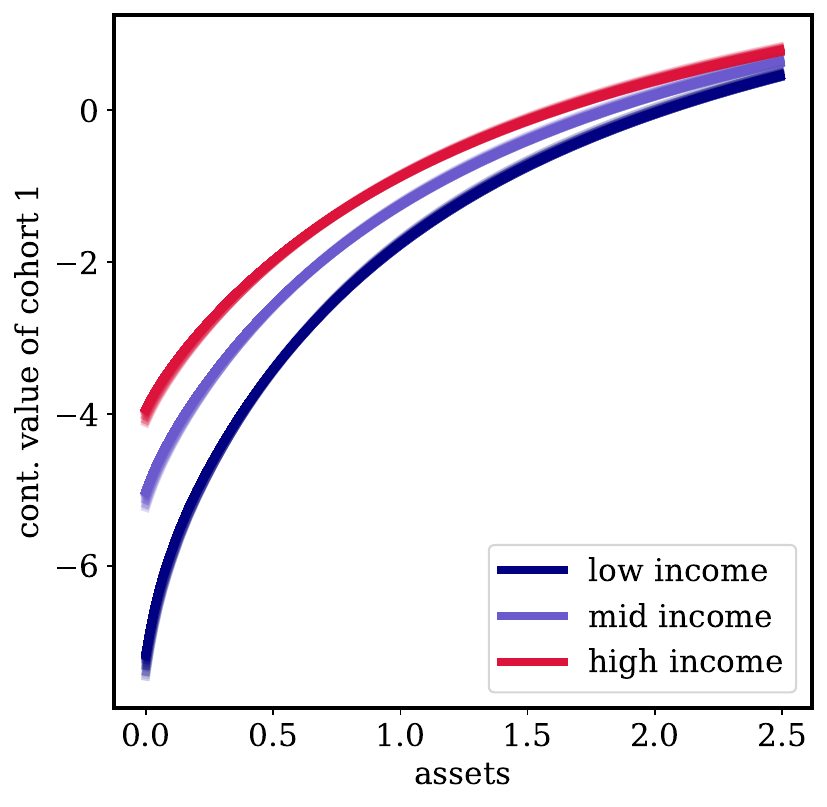}
    \includegraphics[width=0.32\linewidth]{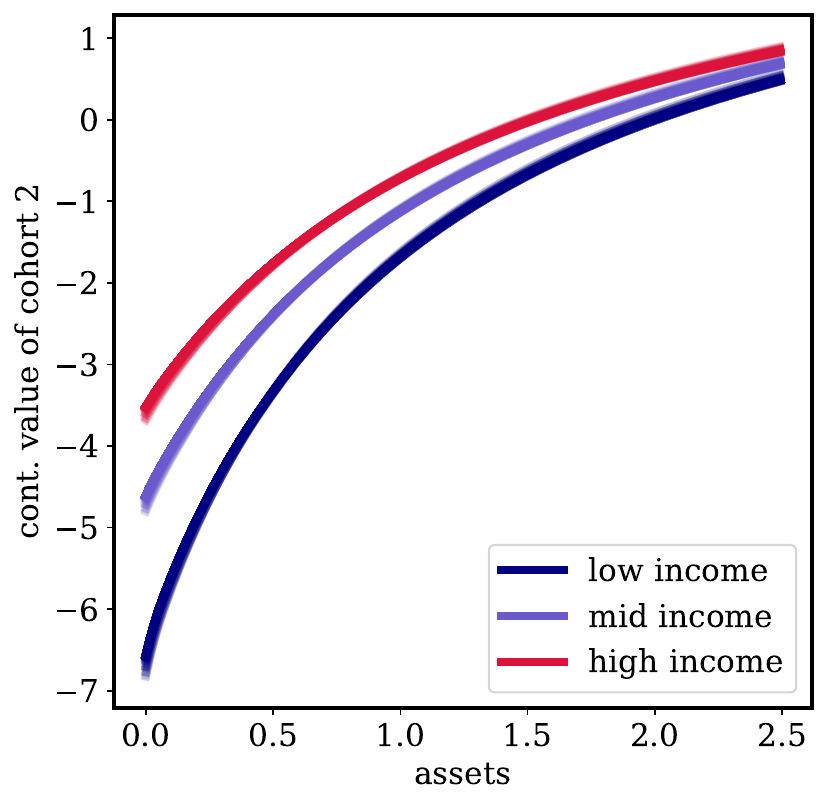}
    \includegraphics[width=0.32\linewidth]{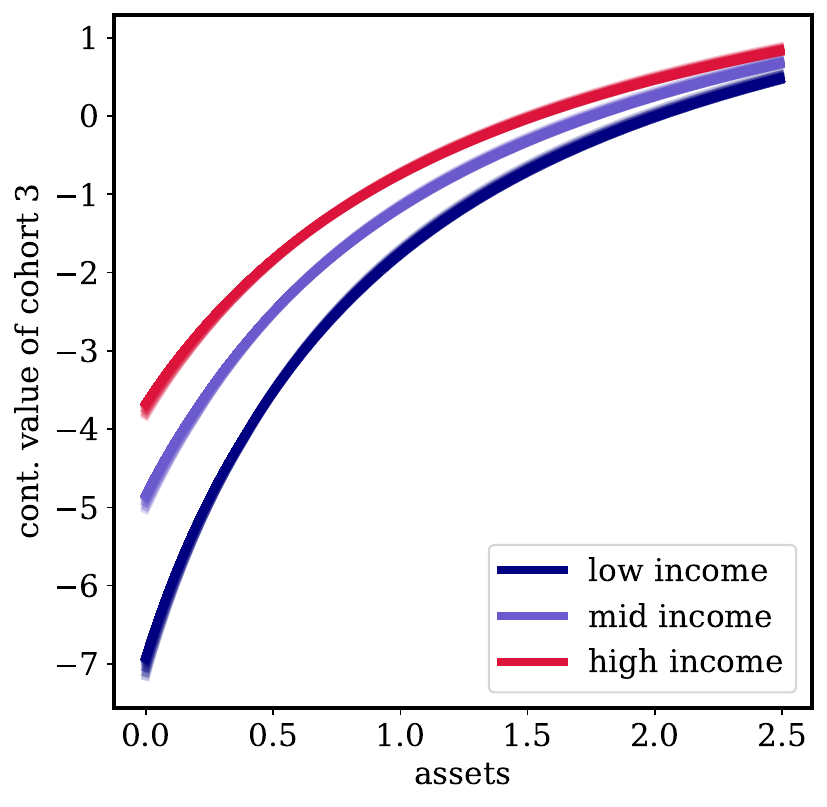}\\
    \includegraphics[width=0.32\linewidth]{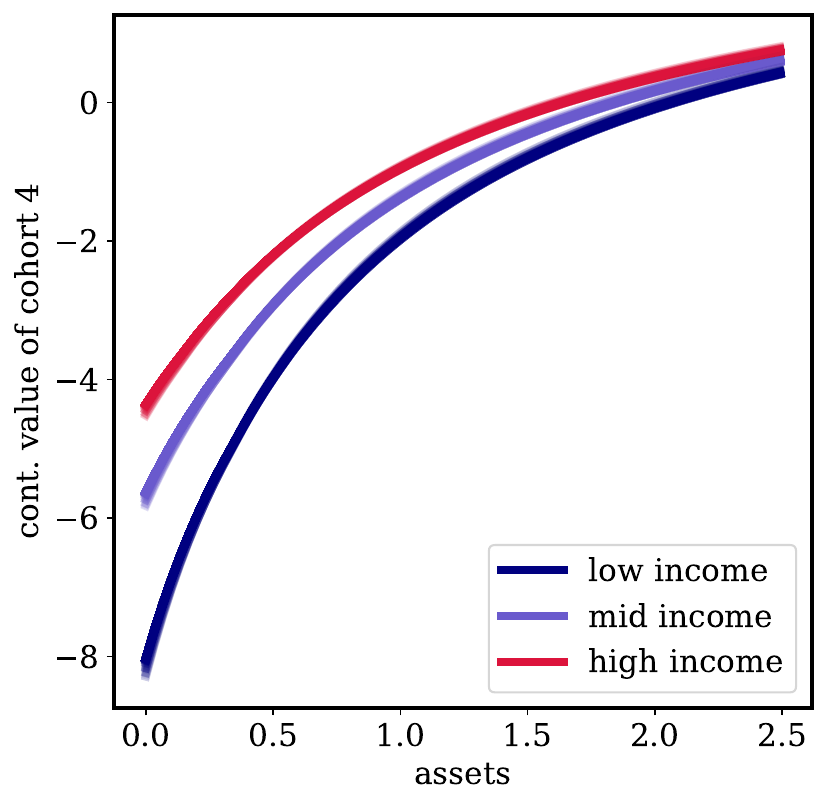}
    \includegraphics[width=0.32\linewidth]{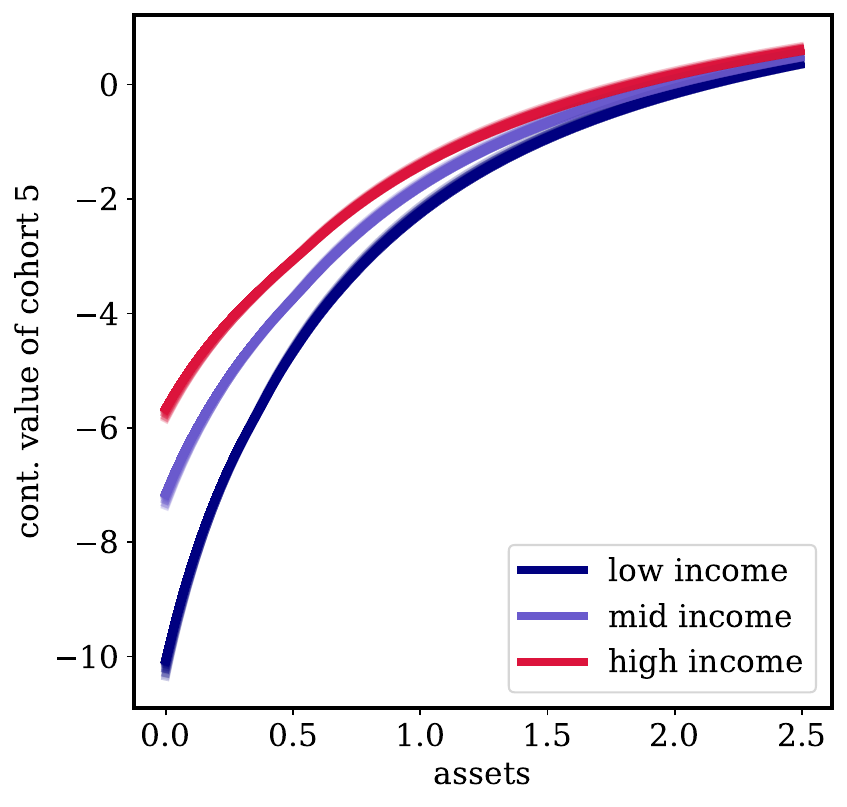}
    \includegraphics[width=0.32\linewidth]{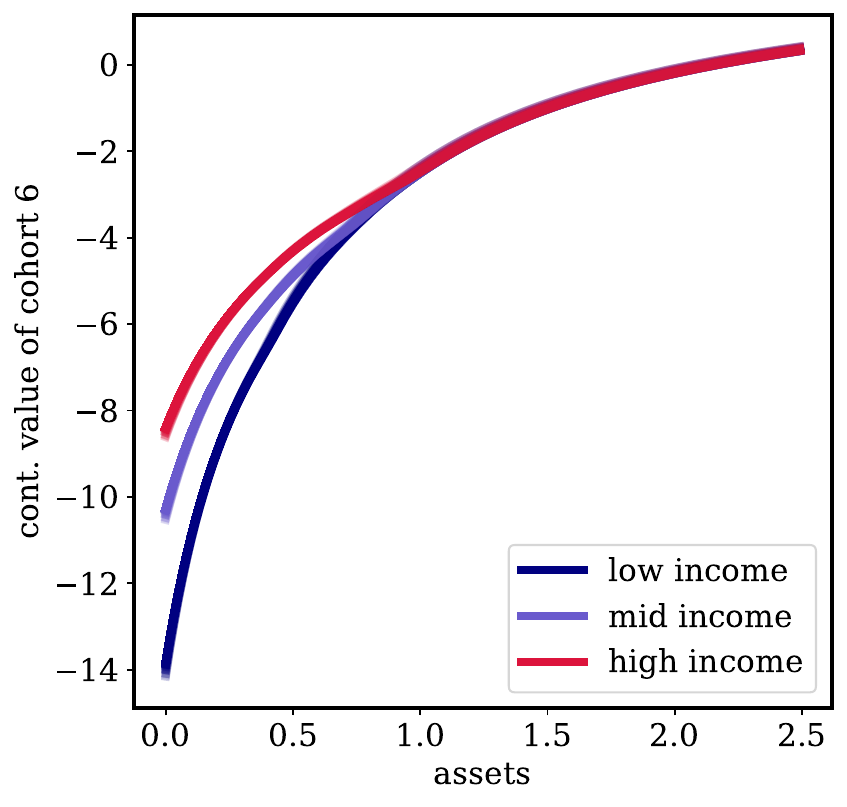}\\
    \includegraphics[width=0.32\linewidth]{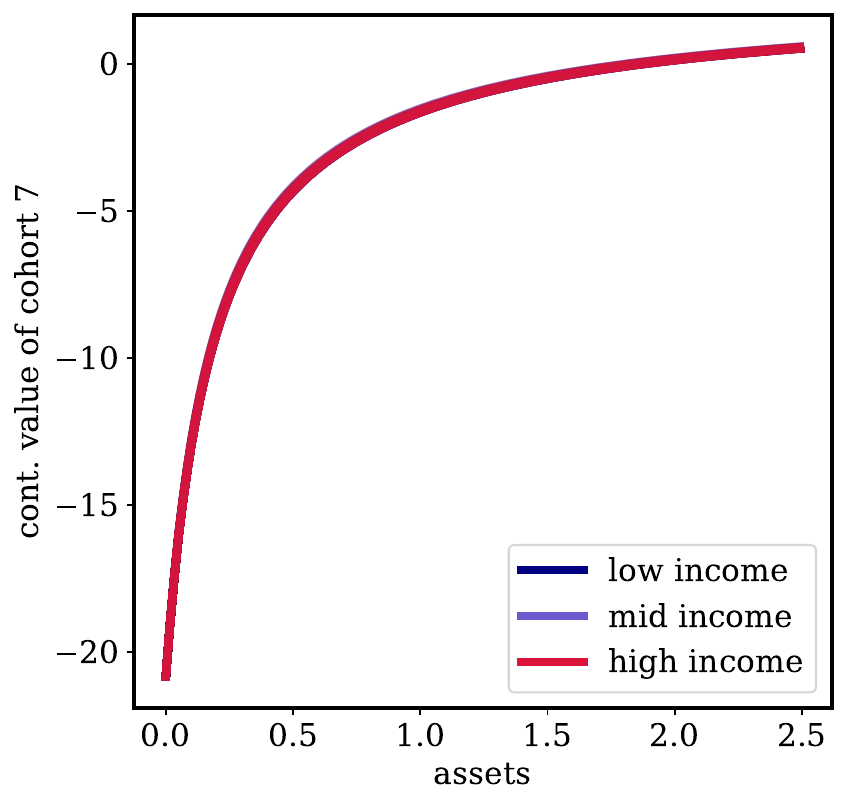}
    \includegraphics[width=0.32\linewidth]{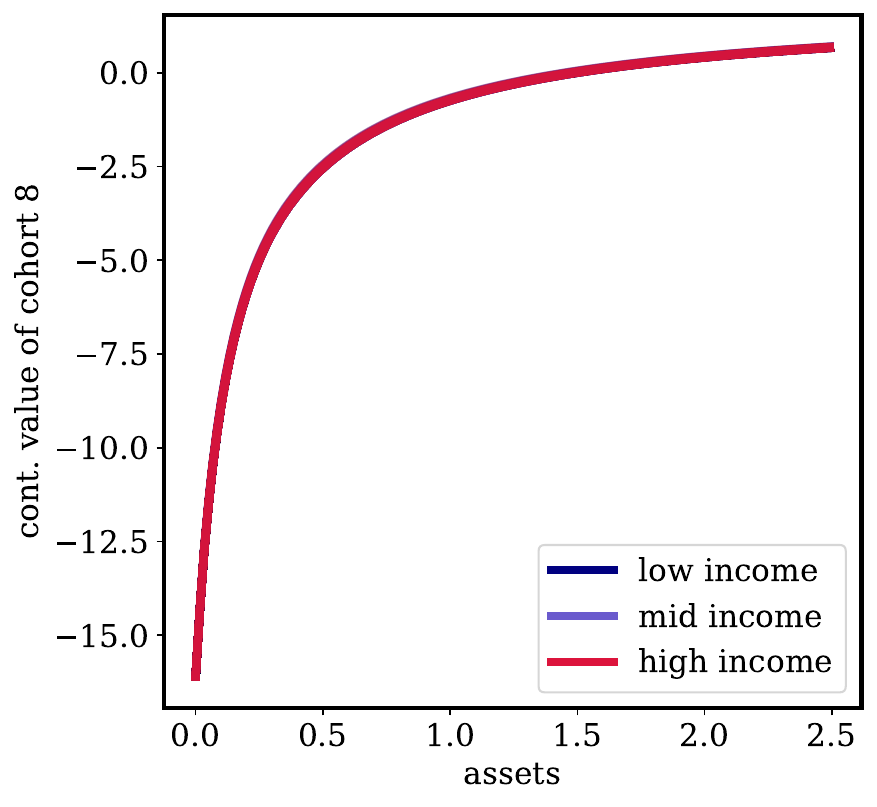}
    \includegraphics[width=0.32\linewidth]{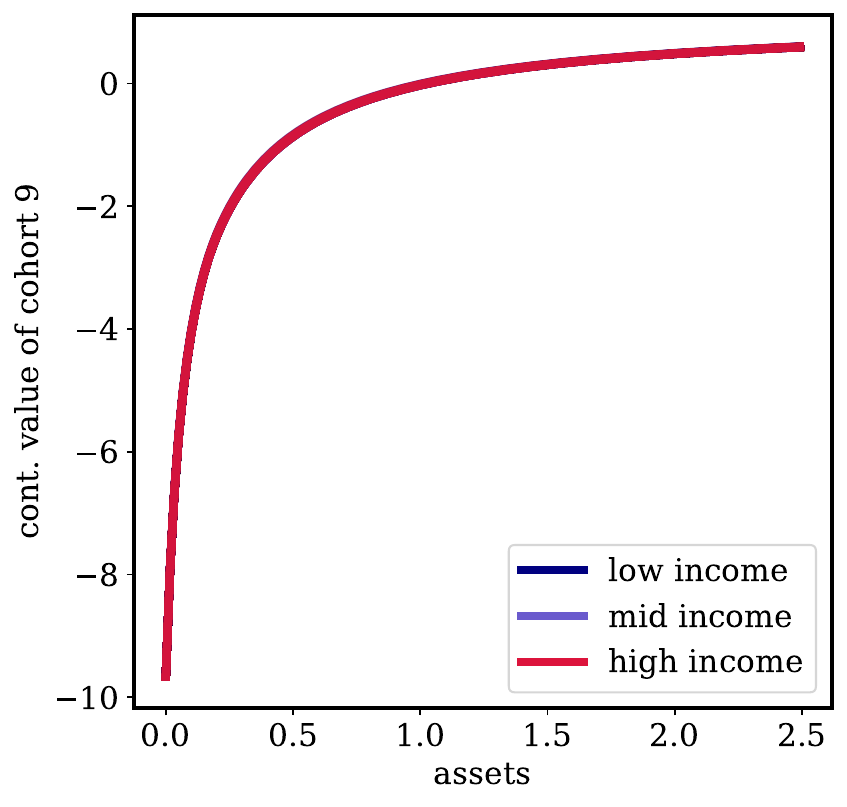}
    \caption{Continuation values by age and asset holdings for different idiosyncratic productivity and several aggregate states. 
    \label{fig:ret_model_contvalues}}
\end{figure}
\begin{figure}
    \centering
    \includegraphics[width=0.32\linewidth]{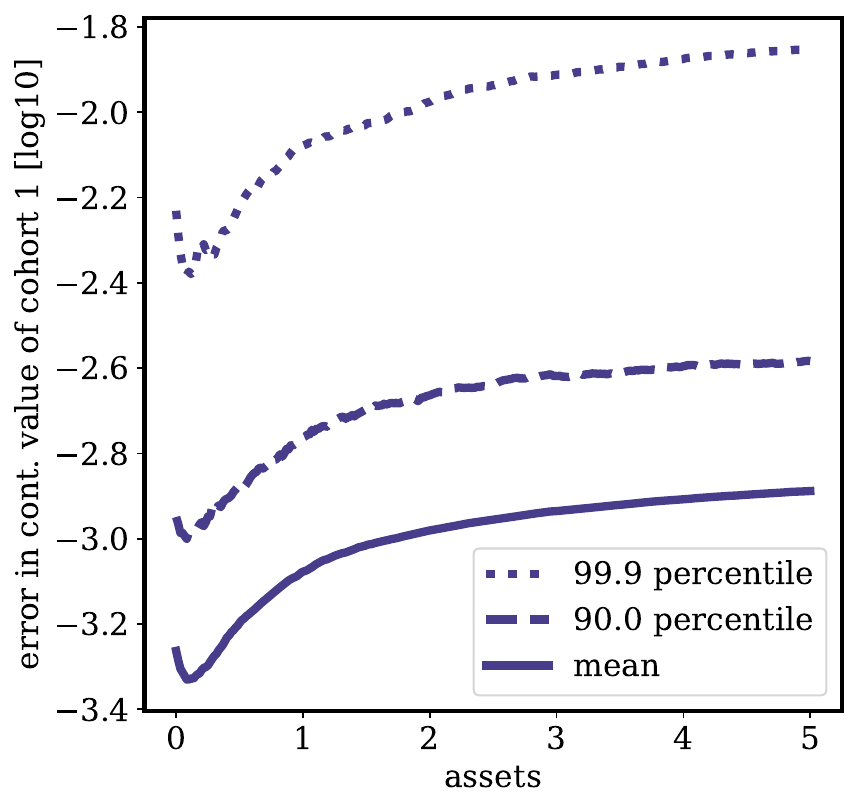}
    \includegraphics[width=0.32\linewidth]{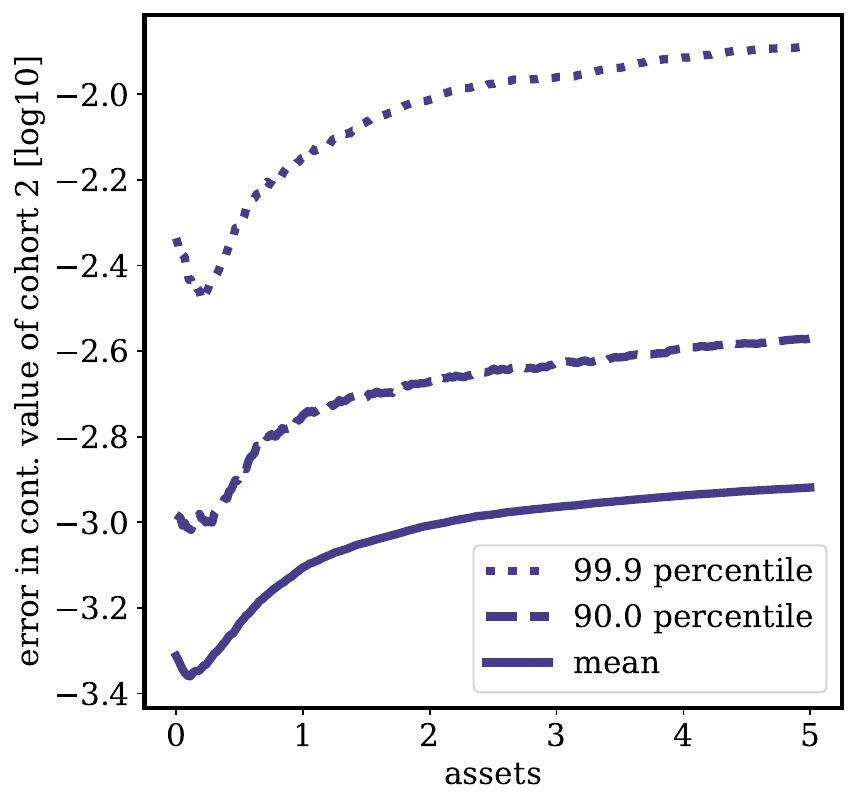}
    \includegraphics[width=0.32\linewidth]{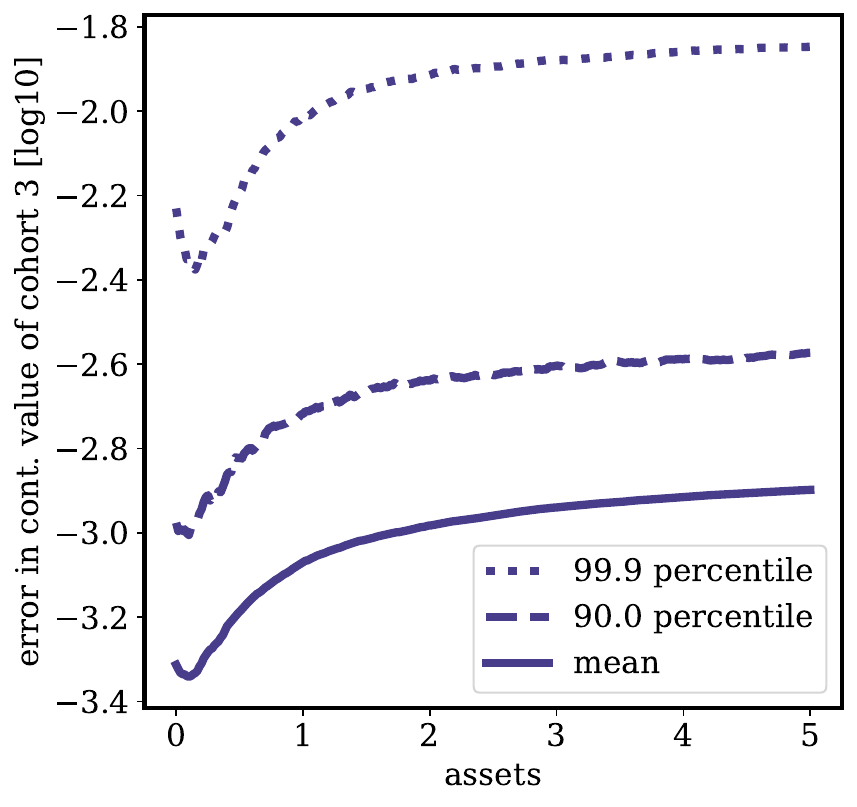}\\
    \includegraphics[width=0.32\linewidth]{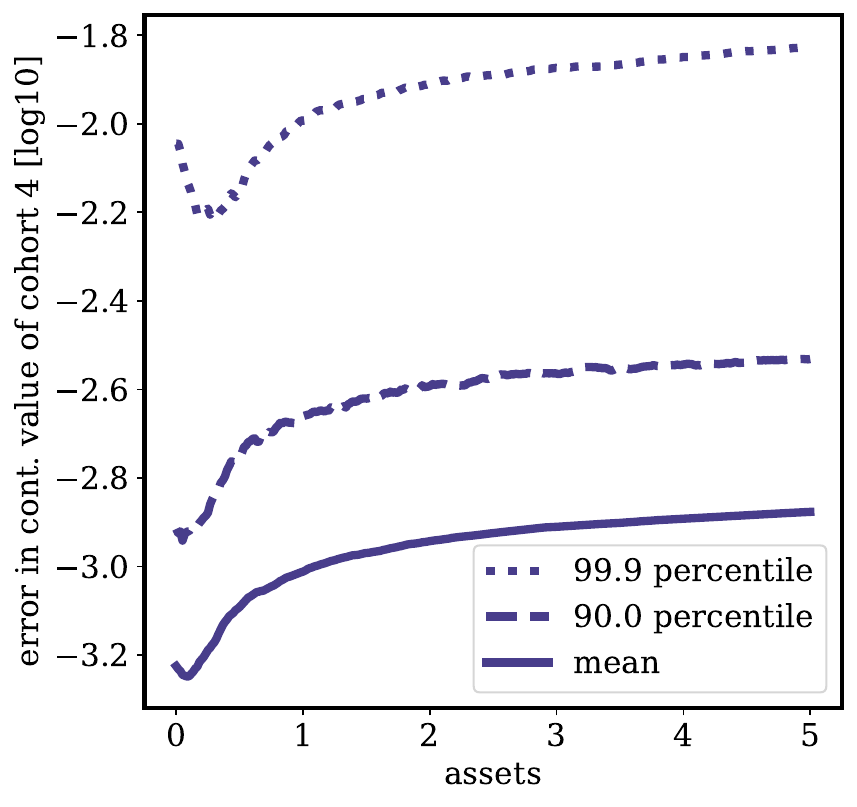}
    \includegraphics[width=0.32\linewidth]{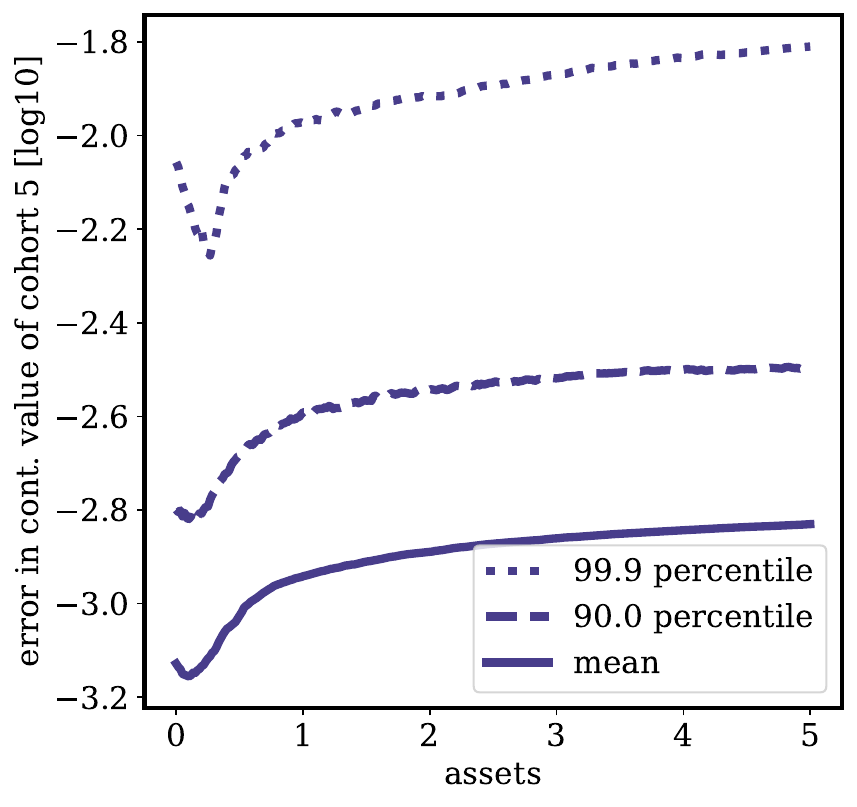}
    \includegraphics[width=0.32\linewidth]{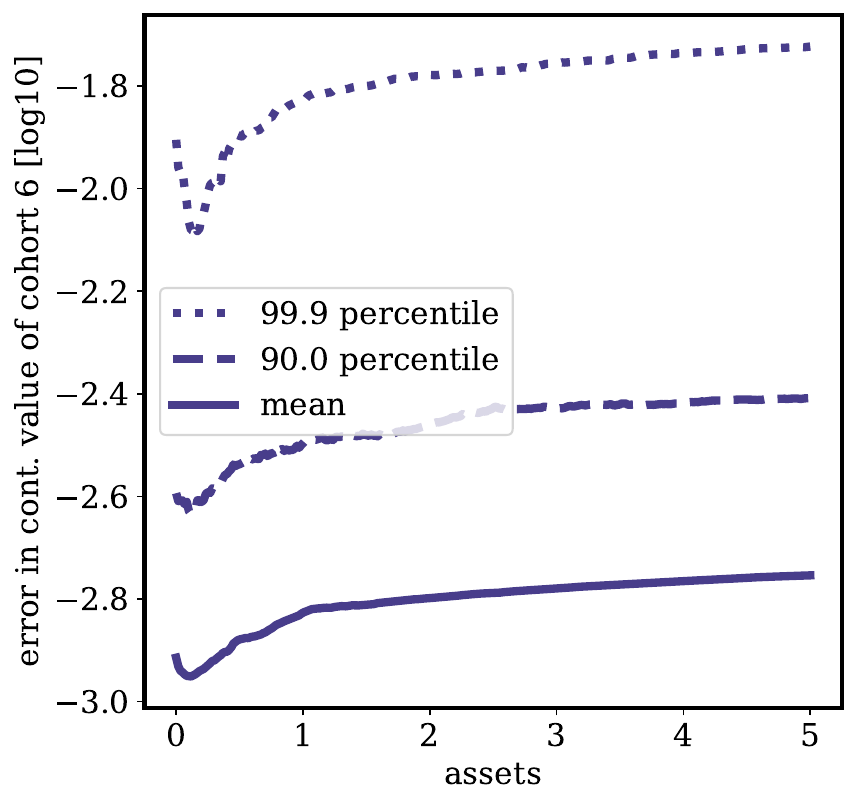}\\
    \includegraphics[width=0.32\linewidth]{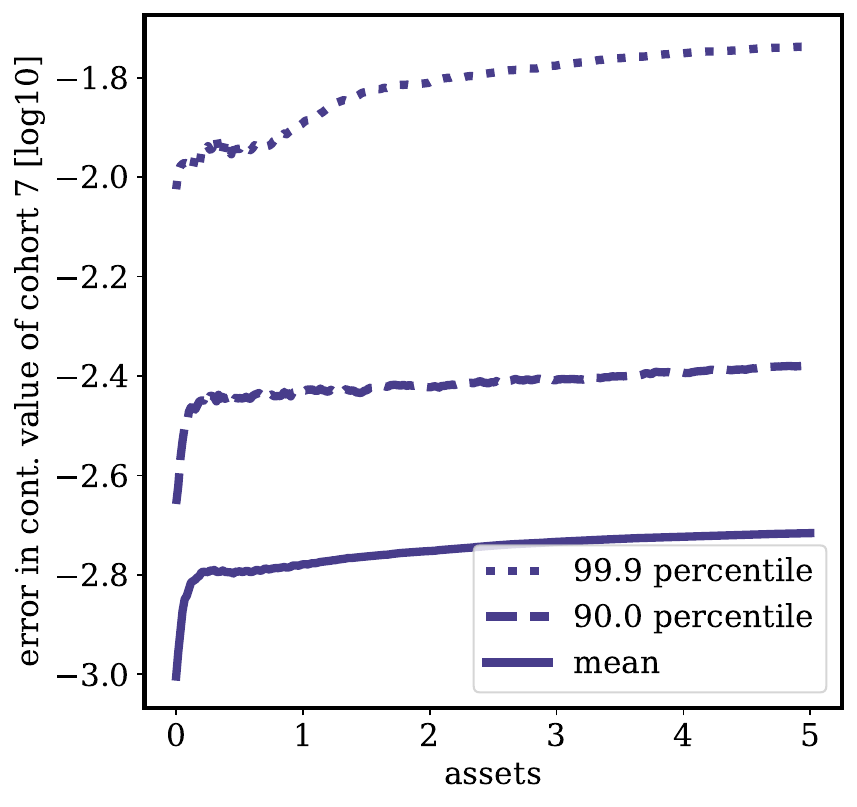}
    \includegraphics[width=0.32\linewidth]{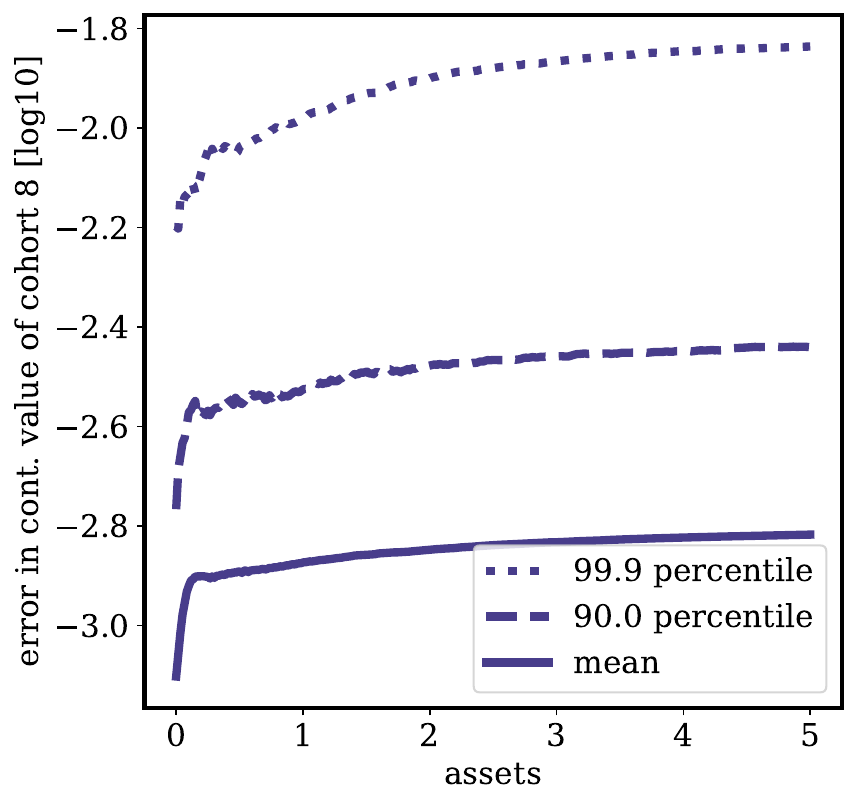}
    \includegraphics[width=0.32\linewidth]{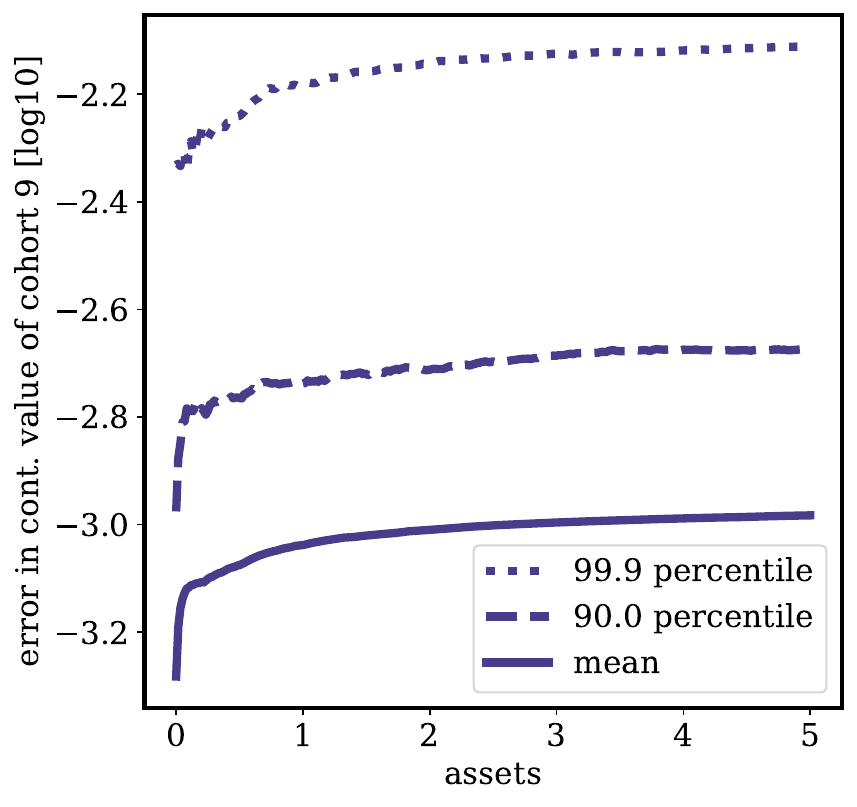}
    \caption{The distribution of (absolute) prediction errors for the continuation value by age and asset holdings. 
    \label{fig:ret_model_errors}}
\end{figure}
\begin{figure}[H]
    \centering
    \includegraphics[width=0.32\linewidth]{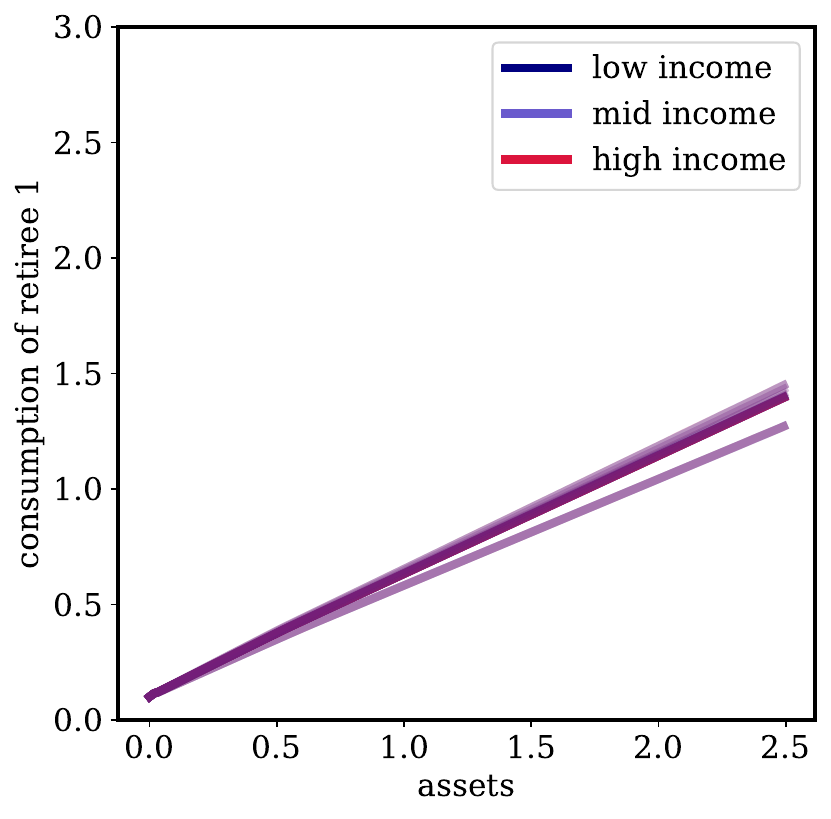}
    \includegraphics[width=0.32\linewidth]{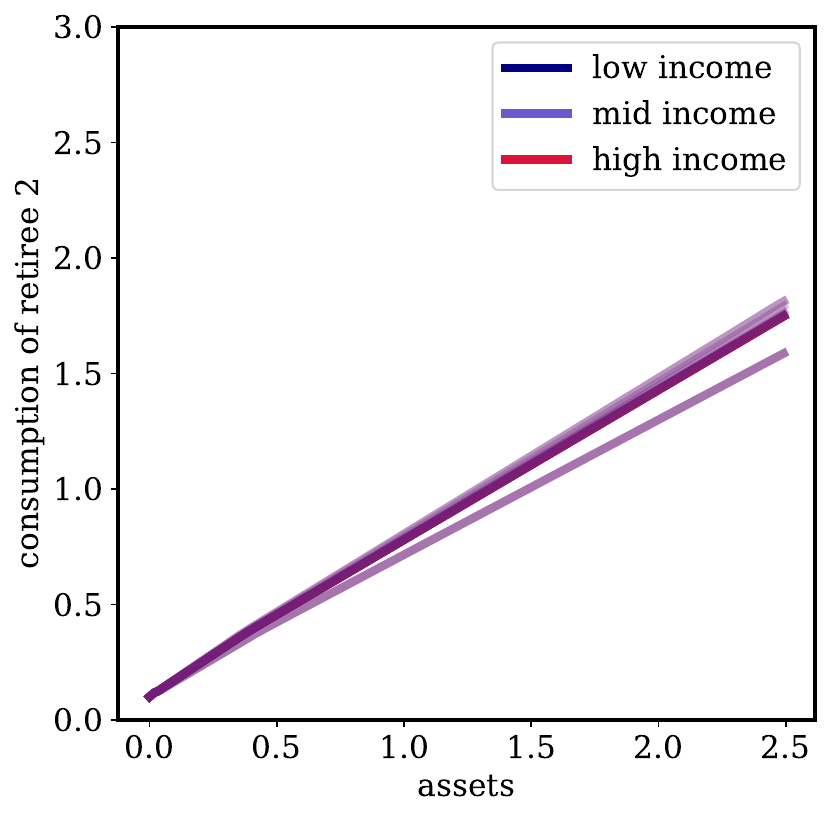}
    \includegraphics[width=0.32\linewidth]{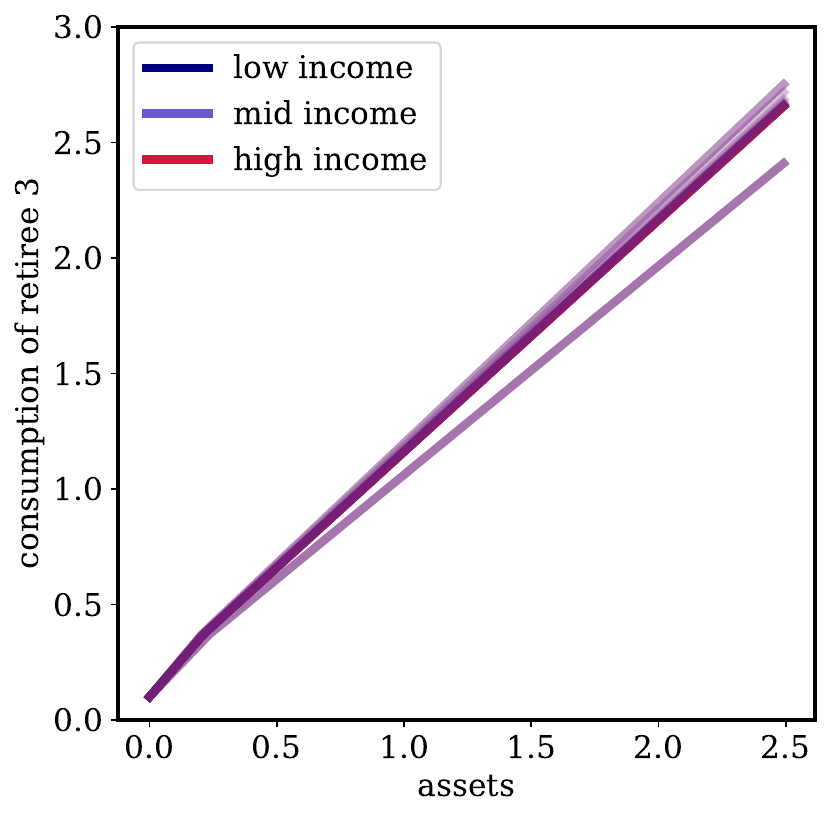}
    \caption{Consumption functions of retired cohorts. The shaded lines show the policies for different aggregate states, and the solid lines show the average taken across the sample of aggregate states.\label{fig:ret_model_retiree_consumption}}
\end{figure}
\end{appendices}

\end{document}